\documentclass[prb,  twocolumn, superscriptaddress, notitlepage]{revtex4-1}
\usepackage{amsmath,amsfonts, amssymb, amsthm, dsfont}
\usepackage{yfonts}
\usepackage{bm}
\usepackage{mathrsfs}
\usepackage{graphicx}
\usepackage{verbatim}
\usepackage{hyperref}
\usepackage{multirow}
\usepackage[normalem]{ulem}
\usepackage{sseq}
\usepackage{shuffle}
\usepackage{tikz}
	\usetikzlibrary{calc}
	\usetikzlibrary{shapes.multipart}
\hypersetup{
    colorlinks, bookmarks=true,breaklinks=true,linkcolor=magenta, citecolor=blue, linktocpage=true, urlcolor=blue
}

\newcommand{\comments}[1]{}
\newcommand{\mb}[1]{\mathbf{#1}}

\newcommand{\ag}[2]{{#1}_\mb{#2}}
\newcommand{\ga}[2]{{}^\mb{#1}{#2}}
\newcommand{\gia}[2]{{}^{\overline{\mb{#1}}}{#2}}
\newcommand{\nugh}[1]{\nu(\mb{#1})}

\newcommand{\coho}[1]{\textswab{#1}}

\def\U{\mathrm{U}(1)}
\def\H{{{\mathcal{H}}}}
\def\Z{\mathbb{Z}}
\def\RZ{\mathbb{R}/\mathbb{Z}}

\newcommand{\stkout}[1]{\ifmmode\text{\sout{\ensuremath{#1}}}\else\sout{#1}\fi}

\newcommand{\sbullet}[1][.6]{{\mathbin{\vcenter{\hbox{\scalebox{#1}{$\bullet$}}}}}}
\def\hs{h^{\sbullet}}

\makeatletter
\def\l@subsubsection#1#2{}
\makeatother

\usetikzlibrary{decorations.pathreplacing,decorations.markings}
\tikzset{middlearrow/.style={
        decoration={markings,
            mark= at position 0.55 with {\arrow{#1}} ,
        },
        postaction={decorate}
    }
}

\begin{document}

\title{Domain Wall Decorations, Anomalies and Spectral Sequences in Bosonic Topological Phases}

\author{Qing-Rui Wang}
\affiliation{Department of Physics, Yale University, New Haven, CT 06511-8499, USA}
\author{Shang-Qiang Ning}
\affiliation{Department of Physics and HKU-UCAS Joint Institute for Theoretical and Computational Physics, The University of Hong Kong, Pokfulam Road, Hong Kong, China}
\author{Meng Cheng}
\affiliation{Department of Physics, Yale University, New Haven, CT 06511-8499, USA}
\date{\today}

\begin{abstract}
	In this work we investigate the decorated domain wall construction in bosonic group-cohomology symmetry-protected topological (SPT) phases and related quantum anomalies in bosonic topological phases. We first show that a general decorated domain wall construction can be described mathematically as an Atiyah-Hirzebruch spectral sequence, where the terms on the $E_2$ page correspond to decorations by lower-dimensional SPT states at domain wall junctions. For bosonic group-cohomology SPT phases, the spectral sequence becomes the Lyndon-Hochschild-Serre (LHS) spectral sequence for ordinary group cohomology. We then discuss the physical interpretations of the differentials in the spectral sequence, particularly in the context of anomalous SPT phases and symmetry-enriched gauge theories. As the main technical result, we obtain a full description of the LHS spectral sequence concretely at the cochain level. The explicit formulae are then applied to explain Lieb-Schultz-Mattis theorems for SPT phases, and also derive a new LSM theorem for easy-plane spin model in a $\pi$ flux lattice. We also revisit the classifications of symmetry-enriched 2D and 3D Abelian gauge theories using our results.
\end{abstract}

\maketitle
\tableofcontents

\section{Introduction}
A large class of gapped quantum phases are believed to be described by topological quantum field theories (TQFT) at low energy. They are broadly divided into two categories: short-ranged entangled (SRE) states, which have unique ground state on any closed (spatial) manifold and all low-lying excitations are local; long-range entangled (LRE) states, which have topology-dependent ground state degeneracy on closed manifolds and exhibit topologically nontrivial excitations. The presence of global symmetry further enriches the landscape and leads to sharp distinctions between phases protected by the symmetry. We will often loosely refer to symmetric SRE phases as symmetry-protected topological (SPT) phases, and symmetric LRE phases as symmetry-enriched topological (SET) phases. The classification of interacting SPT phases has been studied intensively in the past decade, and by now a consistent picture for SPT phases protected by internal or spatial symmetries has been largely achieved, at least in low dimensions relevant to condensed matter systems~\cite{Chen2013, kapustin2014symmetry, Kapustin_2015, freed2019reflection, Xiong_2018, Gaiotto_2019, WangGuPRX2020, ThorngrenPRX2018}. For SET phases, a systematic mathematical framework was developed for the (2+1)d case in Ref. [\onlinecite{SET}] based on tensor category, and more recently Ref. [\onlinecite{Kong_2020}] has proposed a general theory in higher dimensions.

Even though SRE and LRE phases have very different physical properties, they are in fact closely related. On one hand, gauging a finite unitary symmetry in a symmetric SRE phase results in a topological gauge theory~\cite{Levin_2012}. The connection can be utilized in both ways: SPT phases may be characterized by fusion and braiding statistics of symmetry fluxes, while universal properties of the gauge theories can be traced back to the ungauged SPT phases. Another connection is provided by 't Hooft anomaly: on the boundary of an SPT phase, the symmetry must be implemented in an anomalous way~\cite{kapustin2014}, and as a consequence a symmetry-preserving gapped state must be LRE, i.e. topologically ordered~\cite{VishwanathPRX2013, Chen2014}. Therefore ground states of systems with 't Hooft anomaly are strongly constrained, in particular a symmetric SRE ground state is forbidden.

Recently, a new boundary phenomenon has been discovered: there exist ``anomalous'' SPT phases which can only be realized on the boundary of a topologically trivial state in one dimension higher~\cite{WangASPT}, or in an intrinsically gapless system~\cite{thorngren2020intrinsically}. Examples discussed so far all occur in fermionic systems, so one naturally wonders whether ASPT phases also exist in bosonic systems, and how they fit into the general classification scheme.
 
On the other hand, significant progress has been made in classifying SPT phases with crystalline symmetries~\cite{ThorngrenPRX2018, HuangPRB2017, shiozaki2018generalized, Song_2020}. In particular, the bulk-boundary correspondence of crystalline SPT phases shed new light on the celebrated Lieb-Schultz-Mattis theorems~\cite{LSM}, as manifestation of 't Hooft anomaly for spatial symmetries~\cite{ChengPRX2016, HuangPRB2017, JianPRB2018, ChoPRB2017}. This new perspective has naturally led to various generalizations of LSM-type constraints. While in most cases LSM theorems forbid SRE ground state, recent works~\cite{Lu_LSMSPT, Yang_LSMSPT, JiangLSM, ElseLSM} have revealed a new twist in which microscopic constraints allows SRE states, which nevertheless have to be in a nontrivial SPT phase. 

The goal of this work is to provide a unified view of the three kinds of phenomena: a) anomalies in topological gauge theories, b) anomalous SPT phases and c) LSM theorems for SPT phases.  We will see that all of them can be understood within the same framework, namely a systematic theory of decorated domain wall (DW) construction of SPT phases~\cite{ChenNC2013}, which are described mathematically using Atiyah-Hirzebruch spectral sequence for generalized cohomology, as shown in Ref. [\onlinecite{Gaiotto_2019}] (see also Ref. [\onlinecite{Xiong_2018}]). We will explain the mathematical structure through explicit constructions and further expound on the relations with anomalies in this work.  

 The basic idea is that given a global symmetry group $G$, a symmetric state can be obtained from quantum disordering a symmetry-breaking state, or proliferation of symmetry domain walls. One can imagine that domain walls and their junctions are decorated by lower-dimensional SRE phases (possibly protected by unbroken symmetries). 
In order to produce a fully gapped, symmetric state, it is important that all possible domain wall configurations can be smoothly deformed from one to another, so by local Hamiltonian terms that fluctuate the domain walls one can find a ground state which is the superposition of all domain wall configurations. The key observation is that consistency conditions for domain wall decorations can be organized by Atiyah-Hirzebruch spectral sequence.  

As will be explained below, in a general domain wall decoration one successively considers domain wall junctions with increasing codimension (codimension-$0$ is the whole system, codimension-1 is the domain wall, etc.). At each step, one needs to make sure that the decoration is consistent, which leads to an obstruction-vanishing condition. These conditions can be formulated in terms of differentials in the associated spectral sequence. Computing differentials is generally a difficult task. Recently Ref. [\onlinecite{WangGuPRX2020}] essentially derived explicit expressions for differentials for the spectral sequence that describes fermionic SPT phases. The main technical achievement of this work is to provide explicit expressions for \emph{cochain-level} differentials in the Lyndon-Hochschild-Serre (LHS) spectral sequence~\cite{Lyndon, HS}, which correspond to decorating domain walls with bosonic group-cohomology SPT phases protected by another group $A$, but $G$ is nontrivially extended by $A$.

We then provide physical interpretations of the differentials. First of all, for each domain wall decoration in $D$ dimensions, differentials compute possible inconsistencies, represented by a domain wall decoration in $(D+1)$ dimensions. If the differential is nontrivial, it means that the decoration can not be realized, unless the system is actually on the boundary of a $(D+1)$-dimensional bulk. Therefore, such obstructed decorations correspond to anomalous SPT phases, which need to be excluded from the classification of SPT phases in $D$ dimensions.

Using the explicit expressions for differentials, we find that the examples of SPT-LSM theorems for magnetic translation symmetry in the literature can indeed be understood as ASPTs, when the spatial symmetry is formally treated as internal. We also study a new example of LSM theorem, which forbids SRE ground states in a spin-$1/2$ model with easy-plane anisotropy and $\pi$-flux background. Using the tools developed in this work we prove that if the a fully symmetric topologically ordered ground state exists, then anyons must be permuted by some discrete symmetries of the system.

Now we consider a fully consistent decoration, which should describe an actual SPT state. However, when it is the image of a differential map from another decoration in one dimension lower, the SPT state is actually trivial topologically. We will argue that the differential implies that a gapped symmetry-preserving boundary has to be an anomalous SPT phase. We give an example of a bosonic anomalous SPT phase in one-dimensional spin chain with a non-on-site $\Z_4$ symmetry for illustration.

For another important application, it is well-known that gauging a normal subgroup of the symmetry group in an SPT phase turns it into a topological gauge theory, enriched by the remaining quotient symmetry. In this gauging procedure, the group extension structure determines how gauge charges transform under the quotient symmetry. Different decorations translate into patterns of ``fractionalization'' of the quotient symmetry on gauge flux excitations. When the differentials are nontrivial (i.e. the decoration is inconsistent), however, the gauging is also obstructed. In other words, the corresponding fractionalization class is not compactible with the prescribed symmetry action on gauge charges. A well-known special case of the obstructions is the 't Hooft anomaly, but there are other kinds of obstructions beyond 't Hooft anomaly.  Using the LHS spectral sequence, we revisit the classification of symmetry-enriched Abelian gauge theory in 2D and showed that the various obstructions can be understood within the framework established in Ref. [\onlinecite{SET}]. We also consider symmetry-enriched U(1) gauge theory in 3D, recovering the classification in Ref. [\onlinecite{NingU1SL}] and furthermore identifying SPT stacking trivialization missed in previous works.

\section{Domain wall decoration and spectral sequence}
\label{sec:ddw}
In this section we explain how the decorated domain wall construction naturally leads to a spectral sequence description of the SPT classification~\cite{Gaiotto_2019, Xiong_2018}. A brief introduction to spectral sequence can be found in Appendix \ref{sec:ss}.

We will consider a finite, unitary symmetry group $G$. Here we remark that $G$ always refers to a ``bosonic'' symmetry, in the sense that a local bosonic order parameter can be defined and therefore the symmetry can be broken, either spontaneously or explicitly. As briefly explained in the introduction, an SPT wavefunction can be written as a superposition of all possible $G$ domain wall configurations. Typically, this is achieved by Hamiltonian terms that fluctuate the domain walls locally.  In order for the superposition to be a SRE state, we postulate that the following conditions must be true in a symmetry-breaking state:

\begin{enumerate}
	\item For any symmetry-breaking pattern (i.e. a configuration of domain walls), the system is fully gapped and short-range entangled, while preserving any remaining global symmetry.
	\item Symmetry-breaking states corresponding to different domain wall configurations can be smoothly connected. In other words, two such states can be transformed into each other by a constant-depth local unitary circuit preserving any remaining symmetry. This is necessary to ensure that a parent Hamiltonian can exist.
\end{enumerate}
The two conditions are clearly necessary to obtain a SRE state after the symmetry is restored by fluctuating domain walls. They are implicit in previous constructions of fixed-point wavefunctions, see for example Ref. \onlinecite{WangGu2018}. Here our focus will be on general structures, independent of any particular model construction. 

We will now assume that there is an unbroken symmetry group $A$ (which could be trivial). Denote the equivalence classes of SRE phases with symmetry group $A$ (and possibly with other conditions, such as fermionic/bosonic) in $D$ dimensions by $h^{D+1}(A)$. In other words, any SRE state with symmetry $A$ (obeying whatever additional conditions imposed) is associated to a unique element in $h^{D+1}(A)$. For this section we do not need more details about $h^{D+1}(A)$, besides that it is always a discrete Abelian group, with multiplication given by stacking. For brevity we will at times just write $\hs$ instead of $\hs(A)$, since the group $A$ does not play any role in the following discussions. We will denote by $\mathcal{C}^p[G, \hs], \mathcal{Z}^p[G, \hs]$ the $\hs$-valued $p$-cochains and $p$-cocycles of $G$, and $\mathcal{H}^p[G, \hs]$ the cohomology group. 

Below we describe the decorated domain wall construction in $D=2$ as an illustration. Our presentation is schematic and the main purpose is to demonstrate how the physical data can be naturally organized mathematically into a spectral sequence (see Appendix.\ref{sec:ss} for brief introduction). The computational details are discussed in the next section and also in Appendix \ref{sec:LHS1234}. 

\begin{figure}[t!]
	\centering
	\includegraphics[width=.9\columnwidth]{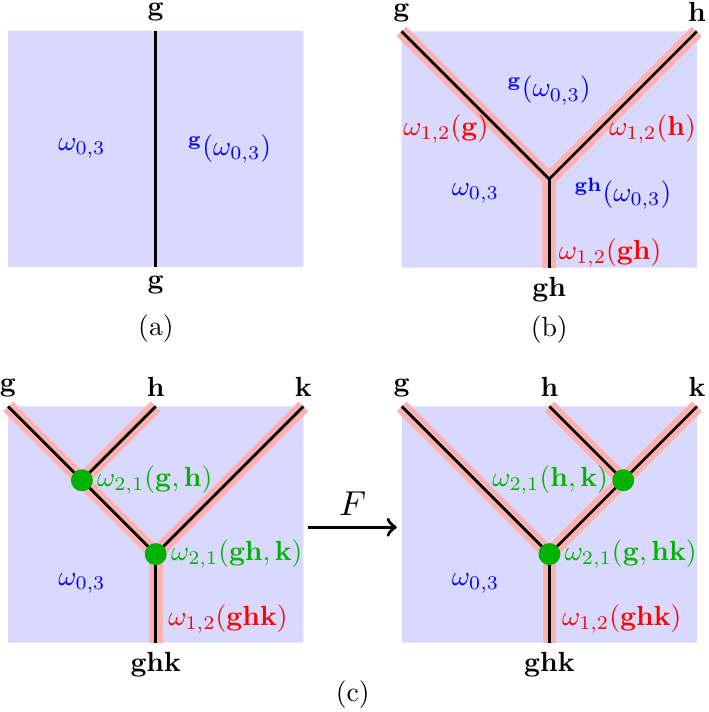}
	\caption{Domain wall decorations and obstructions in 2D. Wavefunction $\omega_{i,3-i}$ is decorated on codimension-$i$ domain walls. (a) The obstruction $O_{1,3}$ indicates that the boundary between two $G$-domains should be gapped. (b) $O_{2,2}$ means that the DW tri-junction does not harbor any projective representation of $A$. (c) $O_{3,1}$ imposes the constraint of $A$-charge conservation under the $F$ move. There is a final layer of obstruction $O_{4,0}$ from the pentagon equation of DW fusions that is not shown in the figure.}
	\label{fig:2D}
\end{figure}

To describe the decorated domain wall states, we start from the ``top'' level, where there is no domain wall, but the $G$ symmetry is already broken. There are $|G|$ different ways that the symmetry can be broken, related to each other by $G$ symmetry transformations.  For each type of domain, the ground state wavefunction belongs to an SPT phase with the unbroken symmetry $A$, the equivalence class of which is given by an element $\omega_{0,3}\in h^{3}$ [see Fig.~\ref{fig:2D}(a)]. 
It is clear that in order to have a consistent decorated domain wall wavefunction, $\omega_{0,3}$ must be ``invariant'' under the natural $G$ action on $h^{3}$ (e.g. when $G$ has a nontrivial action on $A$), so that different domains belong to the same SPT phase. As the actual wavefunctions are generally different in different domains,  the difference between the states before and after the symmetry action $(\delta_1 \omega_{0,3}) (\mb g)={}^{\mb g}(\omega_{0,3})/\omega_{0,3}$ should belong to the trivial class in $h^3$. So we have the first obstruction
\begin{align}
	O_{1,3} &= \delta_1\omega_{0,3} \in \mathcal{Z}^{1}[G,h^3],
\end{align}
which is the $G$-invariant condition for the 2D $A$-SPT phases.  Here $\delta_1$ is the usual coboundary operator acting on the $G$ elements. This obstruction condition implies that  if $O_{1,3}$ is nontrivial, the 1D domain wall between two different domains has protected gapless modes, or spontaneously breaks $A$. In contrast, the vanishing of this obstruction ensures that 1D domain walls can be fully gapped out without breaking $A$.

Next we consider 1D domain walls between different domains, labeled by elements of the symmetry group $G$. They can be viewed as (gapped) boundaries between two SPT states differed from each other by $G$ symmetry action [see Fig.~\ref{fig:2D}(a)].  The boundary wavefunction along the $\mb g$ DW is determined by the two 2D $A$-SPT states up to a 1D $A$-SPT phase, and will be denoted by $\omega_{1,2}(\mb g)$ [see red lines in Fig.~\ref{fig:2D}(b)]. In other words, the 1D $A$-SPT phase should be thought of as an torsor over different equivalence classes of the boundary state. Here ``equivalence'' is defined by adiabatic local deformation along the 1D domain wall.

In the next step, we examine the simplest junction of domain walls, namely a tri-junction of $\mb{g,h}$ and $\mb{gh}$ domain walls depicted in Fig.~\ref{fig:2D}(b).   { The tri-junction essentially provides} a way to check the consistency of group actions on the $A$-SPT states. The two sides of the junction are basically the same domain configuration, but a codimension-$1$ operator is applied to split the $\mb{gh}$ DW to $\mb{g}$ and $\mb{h}$ DWs. This codimension-$1$ operator can not create an nontrivial 1D SPT state, since such a configuration should be smoothly connected to a single domain wall by local adiabatic deformation. Equivalently, the choice of the wavefunctions of the 1D domain walls determines whether the junction harbors a 0D zero mode protected by $A$ or not. Denote this 1D SPT phase  by $O_{2,2}(\mb{g,h})\in h^2$.
Since the wavefuntion of a $\mb{g}$ domain wall is ambiguous up to 1D $A$-SPT 
$\omega_{1,2}(\mb{g})\in h^{2}$, $O_{2,2}$ is only determined up to 
\begin{equation}
	(\delta_1\omega_{1,2})(\mb{g,h})=\frac{\omega_{1,2}(\mb{g})\omega_{1,2}(\mb{h})}{\omega_{1,2}(\mb{gh})}.
	\label{}
\end{equation}
In other words, $O_{2,2}$ up to $\delta_1\omega_{1,2}$ is a function of $\omega_{0,3}$. So we have the following expression
\begin{align}
	O_{2,2} &= (\delta_2\omega_{0,3}) (\delta_1\omega_{1,2}) \in \mathcal{Z}^{2}[G,h^2].
\label{eqn:O22}
\end{align}
This is the second obstruction in the decorations. Physically  $O_{2,2}=\mathds{1}$ imposes the constraint that the DW tri-junction can not harbor any non-trivial 0D boundary states of 1D A-SPT in $h^2$, otherwise the configuration is not fully gapped.  

Here we introduce the notation $\delta_2\omega_{0,3}$, which formally represents the contribution to the obstruction on the codimension-$2$ defect junction from the decoration on the codimension-0 defect (i.e. the 2D $A$-SPT phase). The actual form of $\delta_2\omega_{0,3}$ depends on the nature of $\hs$ (e.g. fermionic or bosonic), and will be discussed in Sec. \ref{sec:lhs}. An important remark follows: $\omega_{0,3}$ is a 3-cocycle in $h^3$ as it is the top-dimension decoration. $\omega_{1,2}$ however is not necessarily a cocycle in $\H^1[G, h^2]$, since it describes the physical state on the domain wall. This is generally true for the other $\omega_{p,q}$'s with $p>0$. The only exception is that when $\omega_{0,3}$ is completely trivial, then $\delta_1\omega_{1,2}=1$ which means that $\omega_{1,2}\in \mathcal{Z}^1[G, h^2]$. On the other hand, even though $\omega_{1,2}$ itself is only a cochain in $\mathcal{C}^1[G, h^2]$, difference of two $\omega_{1,2}$'s both satisfying Eq. \eqref{eqn:O22} is a cocycle in $\mathcal{Z}^1[G, h^2]$. In other words, $\mathcal{Z}^1[G, h^2]$ (and actually, $\H^1[G, h^2]$ as $\mathcal{B}^1[G, h^2]$ can be deformed to a trivial decoration) forms a \emph{torsor} over different solutions of Eq. \eqref{eqn:O22}. Similar comments apply to other $\omega_{p,q}$ as well.

Once $O_{2,2}$ vanishes, we can move to the next level and consider the two configurations depicted in Fig.~\ref{fig:2D}(c). They are exactly the same away from the local patch,  so they can at most differ by a $0$D SRE state, classified by $h^{1}$. To go from one configuration to the other one has to apply a ``F move'', which is a locality-preserving, $A$-symmetric unitary operator creating an additional SPT state. For now denote this SPT phase by $O_{3,1}(\mb{g,h,k})\in h^{1}$.  Since the wavefuntion of the tri-junction of $\mb{g,h}$ domain wall is ambiguous up to 0D state $\omega_{2,1}(\mb{g,h})$ [see green dots in Fig.~\ref{fig:2D}(c)], $O_{3,1}$ is only determined up to 
\begin{equation}
	(\delta_1\omega_{2,1})(\mb{g,h,k})=\frac{\omega_{2,1}(\mb{h,k})\omega_{2,1}(\mb{g,hk})}{\omega_{2,1}(\mb{gh,k})\omega_{2,1}(\mb{g,h})}.
	\label{}
\end{equation}
We can split $O_{3,1}$ into terms from $\omega_{0,3}, \omega_{1,2}$ and $\omega_{2,1}$ as
\begin{equation}
	O_{3,1} = (\delta_3\omega_{0,3}) (\delta_2\omega_{1,2}) (\delta_1\omega_{2,1}) \in \mathcal{Z}^{3}[G,h^1].
\label{eqn:O31}
\end{equation}
The physical meaning of this obstruction is that the F move does not change the 0D SRE state in $h^{1}(A)$. Again here we will keep the actual form of $\delta_3, \delta_2, \cdots$ implicit, except that $\delta_1$ is the usual coboundary operator.
 
Finally, if $O_{3,1}$ vanishes, then the only known information is a phase factor in the F move (since we have gotten down to 0D quantum states) and the last consistency condition is that the F moves need to satisfy the pentagon equation. We can check whether the pentagon equation holds given all the previous decorations, and it can be violated at most by a phase factor, denoted by $O_{4,0}(\mb{g,h,k,l})$. Again we can express $O_{4,0}$ as the following form
\begin{equation}
	O_{4,0} = (\delta_4\omega_{0,3}) (\delta_3\omega_{1,2}) (\delta_2\omega_{2,1}) \in \mathcal{Z}^{4}[G,h^0],
\label{eqn:O40}
\end{equation}
originated from different decoration layers. Here $h^0$ is just $\U$. The F move itself can acquire an additional phase factor $\omega_{3,0}$. If $O_{4,0} = \delta_1\omega_{3,0}$ is a coboundary of $G$, then we can absorb the F move by another phase factor $\omega_{3,0}$ such that the pentagon equation is satisfied.

At this point we have exhausted all consistency conditions for the decorations. Physically, it means that the two conditions for a consistent decoration laied out in the beginning of this section are satisfied at all levels, and one can in principle define a SRE ground-state wavefunction for the corresponding SPT state using the data $\omega_{p,3-p}$ for $p=0,1,2,3$. 

Mathematically, the consistency conditions are the four coupled equations $O_{p,4-p}=\mathds{1}$ for $p=1,2,3,4$. For most applications, we do not need the full solutions of these equations. Instead, one may be interested the particular obstruction function arising from a certain layer of decoration, e.g. $\omega_{1,2}$. In that case, one can assume that the $\omega_{0,3}=\mathds{1}$, so $\omega_{1,2}$ is an actual cocycle in $\H^1[G, h^2]$. For clarify, we will rename it to $\nu_{1,2}$. Then one checks Eq. \eqref{eqn:O31} to see whether there is a $\omega_{2,1}\in \mathcal{C}^2[G, h^1]$ satisfying the condition. The solvability of Eq. \eqref{eqn:O31} is equivalent to the vanishing of $\delta_2\nu_{1,2}$ as a cohomology class in $\H^3[G, h^1]$. Therefore we define a \emph{cocycle-level} map, called a \emph{differential}:
\begin{equation}
	d_2: \H^1[G,h^2]\rightarrow \H^3[G, h^1].
	\label{}
\end{equation}
In this case, $d_2$ is basically the same as $\delta_2$, just regarded as a map between cohomology classes~\footnote{This is consistent because $\delta_2$ must commute with $\delta_1$.}, but this is not true for higher differentials.  

If $d_2$ of $\nu_{1,2}$ vanishes, one can find at least one solution $\omega_{2,1}^0$ for Eq. \eqref{eqn:O31}. Then one can plug $\nu_{1,2}$ and $\omega_{2,1}^0$ into Eq. \eqref{eqn:O40} to compute the obstruction function. This way one may attempt to define another differential using this procedure:
\begin{equation}
	d_3: \H^1[G, h^2]\rightarrow \H^4[G, h^0],
	\label{}
\end{equation}
which increases the degree of the cohomology group by $3$. However, the $d_3$ map is not well-defined yet: $\omega_{2,1}^0$ is one particular solution, which is determined up to a cocycle $\nu_{2,1}$ in $\mathcal{Z}^2[G, h^1]$. If we shift $\omega_{2,1}^0\rightarrow \omega_{2,1}^0\nu_{2,1}$, the obstruction function changes by $d_2\nu_{2,1}$. In other words, $d_3$ is defined only up to $d_2\nu_{2,1}$. To resolve this ambiguity, we can restrict to a subgroup of $\H^2[G, h^1]$ with vanishing $d_2$. In that case, the $d_3$ map becomes well-defined. Physically the restriction makes sense, as to discuss $d_3$ we have to assume that $d_2\nu_{1,2}$ vanishes, so naturally the same assumption should be applied to other layers of decorations as well. More formally, we define $E_3^{p,q}\subset \H^p[G, h^{q}]$ as the kernel of $d_2$, and $d_3$ is only defined on $E_3$. In this sense, $\H^p[G, h^q]$ should be called $E_2^{p,q}$ as they are the kernel of $d_1\equiv\delta_1$. By the same reasoning one can define $E_n^{p,q}$ for higher $n$ iteratively. The collection of $E_n^{p,q}$ is called the $E_n$ page. The differential map $d_n$ is only defined on the $E_n$ page, and one can further show that its image automatically sits in the $E_n$ page (of one dimension higher) as well, so we have
\begin{equation}
	d_n: E_n^{p,q}\rightarrow E_n^{p+n,q-n+1}.
	\label{}
\end{equation}
The collection of all $E_n^{p,q}$ and the differentials that map between them in fact forms an Atiyah-Hirzebruch spectral sequence.

Heuristically, as the ``page number'' $n$ increases, more and more obstructions are required to vanish and one obtains better and better ``approximations'' of the actual SPT states. Eventually, when all obstructions vanish, we arrive at the ``$E_\infty$'' page, which is the collection of all physical SPT states. A concrete example of  solving the consistency conditions to compute the differential maps in (2+1)d when $\hs$ is the group cohomology of $A=\Z_n$ can be found in Appendix \ref{app:example}.

Now let us briefly discuss the general structure of a domain wall decoration construction in $D$ dimensions.
	
Starting from the top dimension $p=0$, we consider topological junctions of $G$ domain walls of increasing codimension successively. The fact that the $p$-junction should be fully gapped requires that the obstruction function
		\begin{equation}
			O_{p,D+2-p} = \delta_p \omega_{0,D+1}\delta_{p-1}\omega_{1,D}\cdots \delta_1\omega_{p-1,D+2-p}
			\label{}
		\end{equation}
		must vanish. In other words, any gapless modes resulting from decorations on $q$-junctions with $q\leq p-1$ must cancel out on the $p$-junction.  This procedure formally continues up to $p=D+2$. Note that for $p=D+2$, $\omega_{D+1,0}$ is a phase factor associated with a certain rearrangement of topological junctions, generalizing the F move. The condition follows from checking the $(D+1)$-cocycle condition for these rearranging moves.

		We can similarly define $d_2$ as basically the $\delta_2$ operator. To define higher differentials, first one needs to define the $E_n$ page as the subgroup of $E_{n-1}$ page whose images under $d_{n-1}$ vanish, where $E_2^{p,q}=\H^p[G, h^q]$. $d_n\omega_{p,q}$ is then defined as the obstruction function $O_{n+p, D+2-n-p}$ arising from $\omega_{p,q}$ with all lower-codimension decorations set to $\mathds{1}$. The operation is well-defined on the $E_n$ page.

		So far we have focused on the construction of obstruction-free domain wall decorations. We have not yet considered the equally important question of whether the decorated domain wall state is trivial or not. This will be addressed in Sec. \ref{sec:aspt}.

Below we discuss two spectral sequences relevant for the classification of SPT phases.

\subsection{Lyndon-Hochschild-Serre spectral sequence}
\label{sec:lhs}
Suppose that $h^{D+1}(A)$ is the group-cohomology bosonic SPT phases, i.e.
\begin{equation}
	h^{D+1}(A)=\H^{D+2}[A, \Z].
	\label{}
\end{equation}
In the following we assume the symmetry group is either finite, or a compact Lie group. In this case, we have $\H^*[A, \mathbb{R}]=0$ and thus $\H^{D+2}[A, \Z]=\H^{D+1}[A, \U]$. The same is true for $A$ a free Abelian group. The Atiyah-Hirzebruch here reduces to the so-called Lyndon-Hochschild-Serre (LHS) spectral sequence~\cite{Lyndon, HS} for group cohomology.

More concretely, denote by $\tilde{G}$ the total symmetry group. Because $A$ is a normal subgroup, $\tilde{G}$ fits in the short exact sequence:
\begin{equation}
	1\rightarrow A\rightarrow \tilde{G}\rightarrow G\rightarrow 1.
	\label{eqn:extensionG}
\end{equation}
In general, the group extension can be specified by two pieces of data: a homomorphism $\rho$ from $G$ to the outer automorphism group $\mathrm{Out}(A)$ which satisfies a certain-obstruction vanishing condition, and a torsor $\nu$ over $\H^2_\rho[G, Z(A)]$ where $Z(A)$ is the center of $A$. When $A$ is Abelian, there is no distinction between $\mathrm{Out}(A)$ and $\mathrm{Aut}(A)$, and the obstruction always vanishes canonically. The multiplication rule in $\tilde G$ is given by
\begin{equation}
a_\mb{g}\times b_\mb{h} = [a \cdot \rho(\mb{g})(b) \cdot \nu(\mb{g,h})]_{\mb{gh}},
\end{equation}
with $a,b\in A$ and $\mb{g,h}\in G$.

When $\tilde{G}=A\times G$, one has the K\"unneth formula:
\begin{equation}
	\H^{d}[G, \U]=\bigotimes_{\substack{p+q=d\\ p,q\geq 0}} \H^{p}[G, \H^{q}[A, \U]].
	\label{}
\end{equation}
This is basically the $E_2$ page of the LHS spectral sequence, and all differentials are trivial.   In general, however, explicit expressions for differentials appear to be unknown in literature except $d_2$, though partial results have been obtained in previous works~\cite{kapustin2014, Tachikawa_2020, NingU1SL}.

In Appendix \ref{sec:LHS1234} we provide a complete and explicit description of the LHS spectral sequence for $\H^{d}[\tilde{G}, \U]$ for $d=1,2,3,4$, at the cochain level, and present the conjectured general structure for higher $d$. We find that a general $d$-cocycle in $\mathcal{Z}^d[\tilde{G}, \U]$ can always be constructed from cochains $F_{p,d-p}\in E_0^{p,d-p}$ for $0\le p\le d$ at different positions of the spectral sequence. These cochains satisfy the following obstruction-vanishing conditions:
\begin{align}
E_0^{0,d+1}&:\quad (\delta_0F_{0,d}) = 1,\\
E_0^{1,d}&:\quad (\delta_1F_{0,d})(\delta_0F_{1,d-1}) = 1,\\
E_0^{2,d-1}&:\quad (\delta_2F_{0,d})(\delta_1F_{1,d-1})(\delta_0F_{2,d-2}) = 1,\\
&\quad\quad\vdots\\
E_0^{d,1}&:\quad (\delta_dF_{0,d})(\delta_{d-1}F_{1,d-1})\cdots(\delta_0F_{d,0}) = 1,\\
E_0^{d+1,0}&:\quad (\delta_{d+1}F_{0,d})(\delta_{d}F_{1,d-1})\cdots(\delta_1F_{d,0}) = 1.
\end{align}
Here $\delta_i$'s are the cochain-level differentials, whose explicit expressions can be found in Eq.~\eqref{diFpq} of Appendix \ref{App:diff}. For example, the zeroth and the first differentials $\delta_0=d_A$ and $\delta_1=\pm d_G$ are just the usual differentials with respect to $A$ and $G$, respectively. Higher differentials are more complicated, so we refer the interested readers to Appendix \ref{sec:LHS1234} for details. As an example, the second differential has the following general form
\begin{align}
&\quad(\delta_2F_{p,q})(a_1,...,a_{q-1},\mb{g}_1,...,\mb{g}_{p+2})\\\nonumber
&=\left[ \iota_{(-1)^q\nu(\mb{g_1,g_2})} \left({}^{1_{\mb{g}_1\mb{g}_2}}F_{p,q}\right)^{\cdot,...,\cdot,1_{\mb{g}_3},...,1_{\mb{g}_{p+2}}} \right]^{a_1,...,a_{q-1}}.
\end{align}
Here, the notation ${}^{\ag{1}{g}}\!F$ is defined in Eq.~\eqref{hF}. And $\iota$ is the slant product operation, the definition of which can be found in Eq.~\eqref{eqn:slant}. If $\delta_2$ lands on the $p$ axis (i.e., $q=1$), the above equation is simplified to
\begin{align}
\quad(\delta_2F_{p,1})(\mb{g}_1,...,\mb{g}_{p+2})
=\frac{1}{\left({}^{1_{\mb{g}_1\mb{g}_2}}F_{p,q}\right)^{\nu(\mb{g}_1,\mb{g}_2),1_{\mb{g}_3},...,1_{\mb{g}_{p+2}}}},
\end{align}
which is basically Theorem 4 of the original LHS spectral sequence paper Ref.~\onlinecite{HS}.

These equations can be thought of as the concrete mathematical representations of the physical obstruction-vanishing conditions discussed in Sec. \ref{sec:ddw} for domain wall decorations. With these cochain-level equations, one can further extract the ``cocycle-level'' differentials $d_i$'s. As a special case, the second cocycle-level differential $d_2$ is in fact equal to $\delta_2$ when viewed as operations on cocycles of the $E_2$ page.  For an example of computations of higher differentials, in Appendix \ref{app:example} we derive all the differentials for $A=\Z_n$ and $d=3$ when the extension is central.

In Appendix \ref{sec:lyndon} we provide an algorithm to extract the cochains at different positions of the LHS spectral sequence from a general cocycle, and give explicit expressions in low dimensions.

An easy collorary of our results is that when $A$ is Abelian and the short exact sequence splits (i.e. $\tilde{G}=A\rtimes G$), every cochain-level differentials $\delta_n$ with $n\geq 2$ vanish and $E_\infty=E_2$. 

\subsection{Spectral sequence for fermionic SPT phases}
Let us give another example of Atiyah-Hirzebruch spectral sequence in the classification of fermionic SPT phases~\cite{Kapustin_2015, Thorngren2018, WangGuPRX2020}.  The symmetry group of a fermionic system fits into the following short exact sequence:
\begin{equation}
	1\rightarrow \Z_2^f\rightarrow G\rightarrow G_b\rightarrow 1
	\label{}
\end{equation}
where $G_b$ is the symmetry group that act on bosonic operators, and $\Z_2^f$ is the conservation of the total fermion parity.   A general fermionic SPT state can be constructed from decorating fermionic invertible topological phases on $G_b$ domain walls. Therefore we set $h^{D+1}$ to be the group of fermionic invertible phases, denoted by $h_F^{D+1}$.  Up to spatial dimension $D=4$, they are given by
\begin{center}
	\begin{tabular}{|c|ccccc|}
		\hline
		$D$ & $0$ & $1$ & $2$ & $3$ & $4$\\
		\hline
		$h_F^{D+1}$ & $\Z_2$ & $\Z_2$ & $\Z$ & $\Z_1$ & $\Z_1$\\
		\hline
\end{tabular}
\end{center}
Here the $0$-dimensional SPT state is just a single fermion. In 1D the nontrivial phase is the Majorana chain, and in 2D the generating one is a $(p_x+ip_y)$ superconductor. 

Then decoration on codimension-$p$ junctions of $G_b$ is given by $\H^{p}[G_b, h^{D+1-p}_F]$. Explicit expressions for differentials up to $D=3$ are obtained in Ref. \onlinecite{WangGu2018} (see Ref. \onlinecite{Thorngren2018} for related results). Interestingly, in this case even when the extension is trivial, i.e. $G=G_b\times \Z_2^f$, the differentials are still nontrivial \cite{WangGu2018}.

\section{Applications}

\subsection{Anomalous SPT phases}
\label{sec:aspt}
\begin{figure}[t!]
	\centering
	\includegraphics[width=\columnwidth]{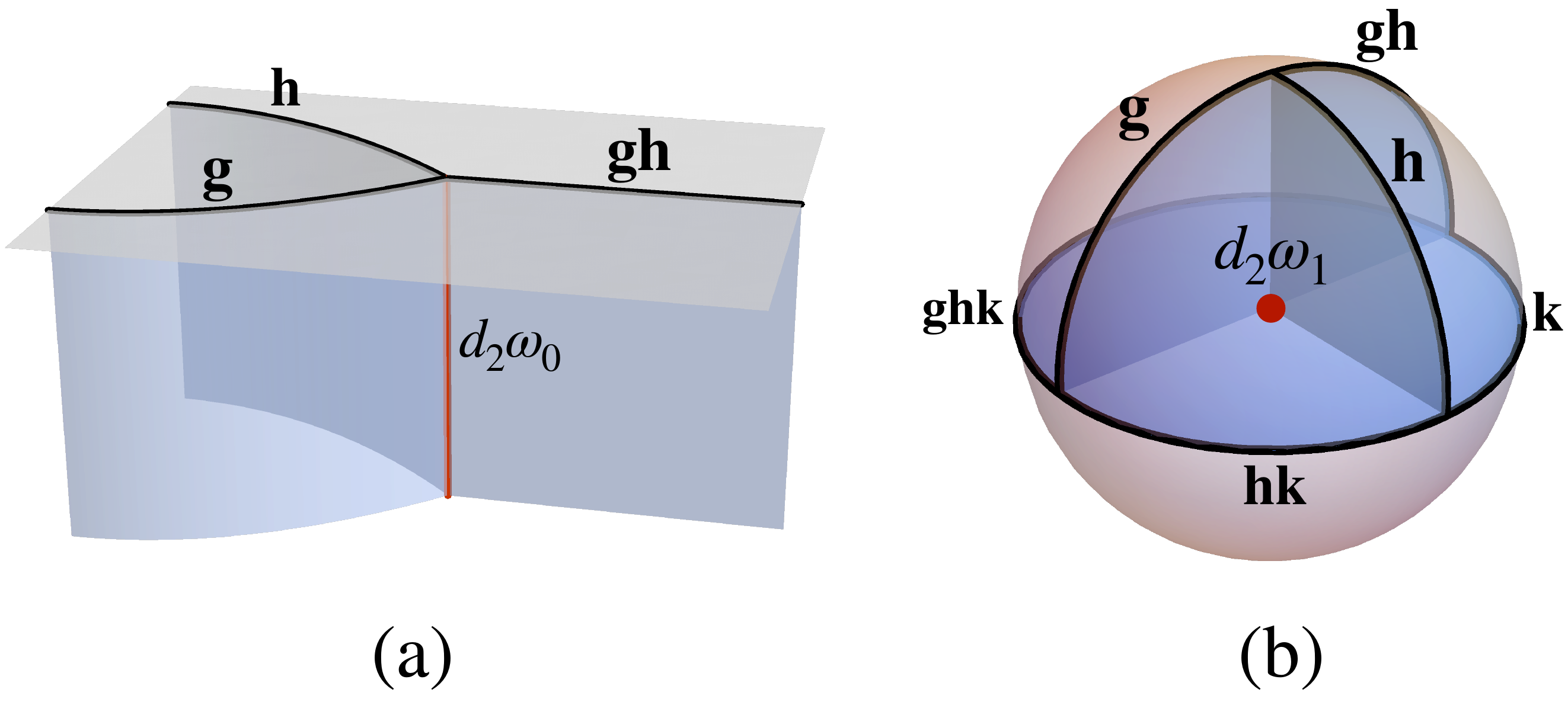}
	\caption{Illustration of $d_2$ obstructions and anomalous SPT phase.}
	\label{fig:aspt}
\end{figure}

As discussed in Sec. \ref{sec:ddw}, obstructions for a consistent domain wall decoration are represented by cohomology classes in $\H^{p}[G, h^{D+2-p}]$ for codimension-$p$ junctions. For example, we have seen that the top level decoration, i.e. $\omega_0$, may cause an obstruction represented by a class $d_2\omega_0\in \H^2[G, h^{D}(A)]$. $d_2\omega_0$ can be understood as boundary states of a $(D-1)$-dimensional SPT phase decorated on a codimension-2 junction in a $(D+1)$-dimensional ``bulk''. In other words, $d_2\omega_0$ gives a DW decoration of the bulk. In Fig. \ref{fig:aspt} we illustrated two examples, for $E_2^{2,2}$ and $E_2^{3,1}$, where obstructions in DW decoration on the surface are ``compensated'' by decorations in the bulk, so that the boundary can be made non-degenerate. For example, in Fig. \ref{fig:aspt}(a), the boundary decoration is obstructed on the 0D junction, which is neutralized by a compensating 1D SPT state in the bulk. Similarly, in Fig. \ref{fig:aspt}(b), the boundary decoration can be viewed as stacking the two sides of the F move together to eliminate open DWs, and the obstruction, a 0D SPT state characterized by $d_2\omega_1$, is neutralized by the corresponding bulk decoration. In other words, these ``obstructed'' decorations can only be consistently realized as boundary of a higher-dimensional SPT state. Thus they can be thought of as ``anomalous'' SPT phases. On the other hand, the bulk in this case must be a trivial SPT phase, since its boundary can form a SRE state preserving all symmetries.

In summary, any nontrivial obstruction class in the decorated DW construction give rise to an anomalous SPT phase. Mathematically, these obstructions are computed using the same differential maps of the spectral sequence.

One may wonder what if one adiabatically deforms the bulk to a trivial product state, which should always be possible because the bulk is topologically trivial. Naively once the bulk is completely disentangled, one is left with the boundary SRE state, contradicting the claimed anomalous nature. The resolution is that the disentangling transformation necessarily removes the boundary SRE state as well. Let us describe the required transformation for a decorated DW construction of a 1D SPT state in $D=1$, with an obstruction class given by $d_2: E_2^{0,2}\rightarrow E_2^{2,1}$. The splitting of $\mb{gh}$ DW into $\mb{g}$ and $\mb{h}$ DWs produces an additional 0D SPT state (i.e. a ``charge''), which is cancelled by the codimension-2 decoration in the 2D bulk. This bulk-boundary relation immediately suggests the following disentangling transformation: first fixing a DW configuration in the bulk. Inside the $\mb{h}$ domain, we adiabatically ``pump'' out a 1D SPT state parametrized by $^{\mb{h}}\omega_0$, and bring the 1D SPT close to the DWs. It is easy to see that along each DW the two 1D SPT states can be annihilated, which also neutralizes the 0D SPT state at the junction. On a disk, the pumping can be intuitively thought as ``pushing'' the edge SPT states inside, towards the junction. For each bulk DW configuration, this kind of pumping can be performed to remove the decoration and it should be evident that the process preserves the residual symmetry. We believe that the pumping can be realized globally by a symmetric adiabatic transformation on the whole bulk wavefunction, even though we do not know how to explicitly write it down. 

This discussion also makes it clear that the differential maps always have dual interpretations: consider $d_n: E_n^{p,q}\rightarrow E_n^{p+n, q-n+1}$. On one hand, it is the obstruction map for the SPT phase described by the DW decoration in $E_n^{p,q}$. Non-anomalous decorations must belong to the kernel of the differential map. On the other hand, it is also the trivialization map for SPT phases described by DW decoration in $E_n^{p+n, q-n+1}$ (in one dimension higher), and the image of the map gives trivial SPT phase.

An alternative perspective on ASPT is to discard the bulk altogether and view the ``anomaly'' as the property of a non-on-site symmetry. To be concrete, suppose we have a $D$-dimensional bosonic lattice model, with a global symmetry $\tilde{G}$. Denote the Hilbert space of the lattice model by $\mathcal{H}$. The symmetry group is represented on the Hilbert space by locality-preserving unitary operators $\{U_\mb{g}\}_{\mb{g}\in \tilde{G}}$, which satisfy $U_\mb{g}U_\mb{g}=U_\mb{gh}$. For simplicity, let us assume that $U_\mb{g}$ is a local unitary circuit with a constant depth.  Generally, it is believed that one can associate $\{U_\mb{g}\}_{\mb{g}\in \tilde{G}}$ uniquely to an element in $\H^{D+2}[\tilde{G}, \U]$~\cite{ElseNayak}. When this element is nontrivial, the symmetry has a 't Hooft anomaly, forbidding the existence of any $\tilde{G}$-symmetry preserving SRE state in this Hilbert space. 

Now suppose $\tilde{G}$ is an extension of $G$ by $A$, where $A$ is the normal subgroup. If the anomaly class of $U_\mb{g}$'s is trivial, then a symmetry-preserving SRE state is possible. Given the group extension structure, it should be possible to locate the anomaly class in the $E_2$ page of LHS spectral sequence, i.e. the ``bulk'' can have a certain domain wall decoration pattern, trivialized by differentials. In that case, the $D$-dimensional system must realize an ASPT phase, whose differential gives the bulk domain wall decoration. Below we present a concrete example in (1+1)d, where a non-on-site internal symmetry enforces the system to realize an ASPT phase. Very similar phenomena arise for spatial symmetries as SPT-LSM theorems discussed below in Sec. \ref{sec:SPTLSM}.

We consider a spin chain with $A=\Z_2=\{1,a\}$ and $G=\Z_2=\{1, g\}$ in (1+1)d. Suppose that $g^2=a$, so $G$ is extended nontrivially by $\Z_2$ to $\Z_4$. There are no nontrivial SPT phase in (1+1)d with $\Z_4$ symmetry, but there is an anomalous one. In terms of the LHS spectral sequence, the only nontrivial term is $E_2^{1,1}=\H^1[G, \H^1[A, \U]]=\Z_2$. Denote the 1-cocycle by $F^{a,g}=\pm 1$ (the other ones are normalized to 1). The $d_2$ differential gives a 3-cocycle in $\H^3[G, \U]$:
\begin{equation}
	F^{g,g,g}=F^{\nu(g,g),g}=F^{a,g}.
	\label{}
\end{equation}
Thus the nontrivial class with $F^{a,g}=-1$ is $d_2$ obstructed.

Let us now turn to the lattice model. Consider a 1D chain with Ising spins $\sigma_j$ on sites and $\tau_{j+\frac12}$ on bonds.  The symmetry group $A$ is generated by $U_a=\prod_{j}\tau_{j+\frac12}^x$, and the $G$ symmetry is generated by the following unitary operator:
\begin{equation}
	U_g=\prod_j \sigma_j^x\cdot e^{\frac{i\pi}{2}\sum_j \frac{1-\sigma_j^z\sigma_{j+1}^z}{2}}\cdot  e^{\frac{\pi i}{4}\sum_j(\tau^x_{j+1/2}-1)}.
	\label{}
\end{equation}
Under periodic boundary condition, it is easy to see that $U_g^2=U_a$, so we have the desired group extension structure. Besides the last factor, $U_g$ is essentially the anomalous $\Z_2$ symmetry transformation on the boundary of a 2D $\Z_2$ SPT phase~\cite{Levin_2012}.

To realize a SRE state preserving both $U_g$ and $U_a$, first we define a subspace by the following constraint:
\begin{equation}
	\tau_{j+\frac12}^x = \sigma_{j}^z\sigma_{j+1}^z.
	\label{eqn:Z2constraint}
\end{equation}
Physically, this is the $U_a$ charge decoration on $U_g$ domain wall, corresponding to a nontrivial class in $\H^1[\Z_2, \H^1[\Z_2,\U]]$. In this subspace, the symmetry $U_g$ simplifies to $\prod_j \sigma_j^x$. Define the projector to the subspace as $P$:
\begin{equation}
	P=\prod_j \frac{1+\tau_{j+\frac12}^x\sigma_j^z\sigma^z_{j+1}}{2}.
	\label{}
\end{equation}
We choose the Hamiltonian to be
\begin{equation}
	H=-\sum_{j}\tau_{j-\frac12}^z \sigma_j^x\tau_{j+\frac12}^zP.
	\label{}
\end{equation}
The term is invariant under $U_a$, and under $U_g$ as well in the projected space.  The ground state satisfies
\begin{equation}
	\tau_{j+\frac12}^x\sigma_{j}^z\sigma_{j+1}^z=\tau_{j-\frac12}^z \sigma_j^x\tau_{j+\frac12}^z=1.
	\label{}
\end{equation}
We recognize that with periodic boundary condition, the ground state is the well-known 1D cluster state. It is often described as a 1D SPT state protected by the $\Z_2\times\Z_2$ symmetry group generated by $U_a$ and $U_g'=\prod_j \sigma_{j}^x$. On an open chain starting with site 1 and ending at site $L$, one finds zero mode operators $\sigma_1^z$ and $\sigma_1^x\tau_{\frac32}^z$ localized at the site $1$, which anticommute with each other and span a two-dimensional projective representation of $\Z_2\times\Z_2$. In our case, we define the restricted symmetry transformations as follows:
\begin{equation}
	\begin{gathered}
		\tilde{U}_a=\sigma_1^z \prod_{j=0}^{L-1}\tau_{j+1/2}^x \sigma_L^z,\\
		\tilde{U}_g=\prod_{j=1}^L \sigma_j^x\cdot e^{\frac{i\pi}{2}\sum_{j=1}^{L-1} \frac{1-\sigma_j^z\sigma_{j+1}^z}{2}}\cdot  e^{\frac{\pi i}{4}\sum_{j=1}^{L-1}(\tau^x_{j+1/2}-1)}.
	\end{gathered}
	\label{}
\end{equation}
Notice that we introduce two additional $\sigma^z$ operators at the ends of the chain in $\tilde{U}_a$, since the restriction is ambiguous up to local operators at the ends. The restricted symmetry operators satisfy all the group relations: $\left( \tilde{U}_a \right)^2=\mathds{1}, \left( \tilde{U}_g \right)^2=\tilde{U}_a$. With these symmetry transformations, both $\sigma_1^z$ and $\sigma_1^x\tau_{3/2}^z$ are forbidden by $\tilde{U}_g$, and therefore the ground state degeneracy is protected.

Intuitively, the constraint in Eq. \eqref{eqn:Z2constraint} must be enforced to get rid of the non-on-site part of $U_g$ in a suitable subspace, and therefore a non-degenerate ground state must be an anomalous SPT phase. The constraint basically binds $\tau$ charges to $\sigma$ domain walls. Without this particular decoration, the system either breaks the symmetry or forms a gapless state. For example, one may let the $\tau$ spins to form an Ising paramagnet, i.e. $\tau_{j+1/2}^x=1$ for every $j$ by adding a large on-site magnetic field $-h\sum_j \tau_{j+1/2}^x$. Within this low-energy subspace, the $U_g$ transformation effectively generates a $\Z_2$ group with 't Hooft anomaly, which is an example of the emergent anomaly discussed in Ref. [\onlinecite{Metlitski_2018}]. The $\sigma$ spins can form a Luttinger liquid preserving the effective $\Z_2$ symmetry.


\subsection{Symmetry-enriched gauge theory}
In a consistent decorated DW state, one can gauge the symmetry group $A$ to arrive at a $A$ gauge theory. The global symmetry becomes $G$ after $A$ is gauged, so the result is a $G$-enriched $A$ gauge theory~\cite{MesarosPRB2013, Tachikawa_2020}. This is believed to be the most general construction of a $A$ gauge theory enriched by $G$ symmetry, as long as the enrichment is compactible with the gauge structure.

Let us examine the symmetry-enriched phenomena in more details. In the gauge theory, there are two types of fundamental excitations: gauge charges, which are point-like excitations labeled by irreducible representations of the gauge group, and gauge fluxes, $(D-2)$-dimensional objects labeled by conjugacy classes.  The group extension determines how the gauge charges transform under the global symmetry $G$: the automorphism of $G$ on $A$ specifies how different types of gauge charges are permuted by $G$ (and at the same time determine how different kinds of fluxes are permuted), and the $2$-cocycle indicates that gauge charges transform projectively under $G$. 

On the other hand, each $E_2^{p,q}$ with $p\neq 0, q\neq 0$ describes a particular way that the $G$ symmetry acts on gauge flux excitations, including changing the topological type of the excitations, or symmetry transformation being ``fractionalized'' on the excitations. The latter is most familiar in 2D, and similar phenomena have also been investigated for loop excitations in 3D topological gauge theories~\cite{cheng2015symmetry, ChenPRB2016, ChenPRR2021}. Below in Sec. \ref{sec:segex} we will carefully examine the physical interpretations of these terms for Abelian gauge theories in (2+1)d and U(1) gauge theory in (3+1)d.

The differentials $d_n: E_n^{p,D+1-p}\rightarrow E_n^{p+n,D+2-p-n}$ represent obstructions in gauging the symmetry group $A$. In other words, the gauged theory is ``obstructed'' (with the given action on gauge charges). More specifically, a non-vanishing obstruction typically implies that the symmetry action given by $E_2^{p,q}$ is incompactible with the underlying fusion and braiding structure of the topological order, when the group extension fixes how the symmetry acts on the charge sector.  In addition, $d_n: E_n^{D+2-n, n-1}\rightarrow E_n^{D+2,0}$ leads to 't Hooft anomaly of the symmetry-enriched gauge theory. However there are a whole ladder of obstructions. We study the 2D case for general Abelian group $A$, and 3D case for $A=\U$ below in Sec. \ref{sec:U1}, and unpack the physical meaning of the obstructions.

On the other hand, if certain element in $E_n^{p,D+1-p}$ is in the image of the $d_n$ differential, according to the discussion in, the domain wall decoration after the symmetry is ungauged can be trivialized. As a result, the corresponding symmetry action on flux excitations in the gauged theory is also trivial.



\section{Examples}

In this section we discuss a few examples to illustrate the applications of LHS spectral sequence.

\subsection{SPT-LSM theorems}
\label{sec:SPTLSM}
In recent years, the celebrated Lieb-Schultz-Mattis theorem has found many generalizations and refinements~\cite{Watanabe_2015, ChengPRX2016,  PoPRL2017, ChengPRB2019, Lu_2020, Lu_LSMSPT, Yang_LSMSPT, ElsePRB2020}. A common theme underlying these developments is the realization that LSM theorem can be understood as the manifestation of 't Hooft anomaly of a particular class of crystalline SPT phases, the boundary of which hosts a projective representation per unit cell~\cite{ChengPRX2016, JianPRB2018, ChoPRB2017}. Here we will focus on a class of LSM theorems which allows SRE ground state~\cite{Lu_2020, Lu_LSMSPT, Yang_LSMSPT}, but nevertheless must be in a nontrivial SPT phase, thus referred to as ``SPT-LSM'' theorems. While these theorems have been understood in the theoretical framework of crystalline SPT phases~\cite{ElsePRB2019, Song_2020, shiozaki2018generalized}, here we will ``rederive'' some of the bosonic SPT-LSM theorems using the LHS spectral sequence.

More specifically, we will treat the relevant crystalline symmetry formally as an internal one and identify the SPT-LSM theorems from a certain $E_2$ page corresponding to certain decoration with projective representations of the actual internal symmetry. This is motivated by the crystalline correspondence principle in Ref. [\onlinecite{ThorngrenPRX2018}], which establishes the equivalence between classification of topological phases with a crystalline symmetry group $G$ and that with an internal symmetry group $G$. Notice that however it is not clear whether the spectral sequence that classifies bosonic crystalline SPT phases~\cite{ElsePRB2019, Song_2020, shiozaki2018generalized} and the Atiyah-Hirzebruch spectral sequence studied here for internal symmetry are completely equivalent (at the level of spectrums, not just classifications). Therefore the computations performed below should be regarded as being conjectural, and suggest that the crystalline correspondence principle may actually extend to the whole spectral sequence.

We will focus on SPT-LSM theorems with ``magnetic'' translation symmetry.  Let us consider $G=\Z^D$, extended by an Abelian group $A$:
\begin{align}\label{A_ZD}
1 \rightarrow A \rightarrow \tilde G \rightarrow \Z^D \rightarrow 1.
\end{align}
$G$ here represents the lattice translation symmetry in $D$-spatial dimensions, and $A$ is the internal symmetry group. When $D=2$ and $A=\U$, $\tilde{G}$ is the usual magnetic translation symmetry group. To describe projective representation of $A$ per unit cell, we need to consider $E_2^{D,2} = \H^D[\Z^D,\H^2[A,\U]]$. Roughly speaking, this is because a unit cell can be formally viewed as a particular codimension-$D$ junction of translation symmetry defects. For example, in 2D the commutator of primitive translations along the two Bravis basis vectors geometrically encloses exactly one unit cell. We will fix a $\alpha\in E_2^{D,2}$, which is uniquely determined by the $A$ projective representation. We look for SPT-LSM theorems in the strongest form, that is, nontrivial SPT ground states protected by $A$ alone are enforced when the symmetries are preserved. 

In order to have a SPT-LSM theorem, the ``decoration'' should be trivializable. In particular, in order to enforce a ``strong'' SPT ground state, we consider the following differential:
\begin{equation}
	d_D: E_D^{0,D+1}\rightarrow E_D^{D,2}.
	\label{}
\end{equation}
Suppose there exists a $\omega_0\in E_2^{0,D+1}$ such that $d_D\omega_0=\alpha$. To obtain an SPT-LSM theorem, it is necessary to check that $d_n\omega_0$ vanish for $2\leq n<D$. In addition, it is also desirable that $\nu$ does not have other trivializations, to get the most stringent constraint possible.  If there exists such a $\omega_0$ with all the requirements satisfied, then a SRE ground state must be a nontrivial $A$ SPT phase described by $\omega_0$.  Note that higher differentials of $E_D^{0,D+1}$ vanish automatically since $\H^{k}[\Z^D, *]=\Z_1$ for $k>D$. 

For $D=2$, the above criteria are all trivially satisfied since $d_2$ is the very first obstruction. Applying the first term in formula (\ref{dFabkl}), we have
\begin{align}
&\quad(d_2\omega_0)(a,b,\ag{1}{k},\ag{1}{l})\\\nonumber
&=\frac{{}^{\ag{1}{kl}}\!\!\left[\omega_0\!\left(\gia{kl}{a},\gia{kl}{\nugh{k,l}},\gia{kl}{b}\right)\right]}{{}^{\ag{1}{kl}}\!\!\left[\omega_0\!\left(\gia{kl}{a},\gia{kl}{b},\gia{kl}{\nugh{k,l}}\right)\right] \times {}^{\ag{1}{kl}}\!\!\left[\omega_0\!\left(\gia{kl}{\nugh{k,l}},\gia{kl}{a},\gia{kl}{b}\right)\right]},
\end{align}
where $\nu\in\H^2[\Z^2,A]$ specifies the central extension of the group. For simplicity, let us consider the special case where $G=\Z^2$ has trivial action on $A$, and $\nu(T_x,T_y)=g\in A$, and $\nu(\mb{k,l})=0$ otherwise. It represents the magnetic translation symmetry with $T_xT_yT_x^{-1}T_y^{-1}=g$ ($g$ is a fixed element in $A$). Then $(d_2\omega_0)(a,b,T_x,T_y)$ has a simpler expression $\frac{\omega_0(a,g,b)}{\omega_0(a,b,g) \omega_0(g,a,b)}$. If it is cohomologous to a nontrivial cocycle $\alpha(a,b)$ in $E_2^{2,2}=\H^2[\Z^2,\H^2[A,\U]]=\H^2[A,\U]$ representing the projective representation in the unit cell, then we have an SPT-LSM theorem. In this way, we rederived and generalized the SPT-LSM theorem in Ref.~\onlinecite{Yang_LSMSPT}.

Now we consider $D=3$, and suppose that $\H^3[A, \U]=\Z_1$. An SPT-LSM theorem is given by the first differential term of Eq.~(\ref{dFabklm})
\begin{equation}
	d_3: E_3^{0,4}\rightarrow E_3^{3,2}.
	\label{}
\end{equation}
If $\H^3[A, \U]$ vanishes, both differentials $d_2: E_2^{0,4}\rightarrow E_2^{2,3}$ and $d_2: E_2^{1,3}\rightarrow E_2^{3,2}$ vanish automatically. An example of 3D SPT-LSM theorem was given in Ref. \onlinecite{JiangLSM}, with $G=\Z^3$ and $A=\U\times\Z_2^\mathsf{T}$, with translations acting as charge conjugation.

\subsection{2D topological phases with magnetic translation symmetry}
\label{sec:set-magnetic}
We now apply the LHS spectral sequence to the problem of 2D symmetry-enriched topological (SET) phase with magnetic translation symmetry~\cite{Lu_2020, manjunath2020classification}. As in the previous section, the system has an internal symmetry group $A$ and lattice translation $G=\Z^2$, but the primitive translations $T_x$ and $T_y$ satisfy
\begin{equation}
	T_xT_yT_x^{-1}T_y^{-1}=a,
	\label{}
\end{equation}
where $a\in A$ represents an internal symmetry transformation. Apparently, the internal symmetry must all commute with $a$, i.e. $a$ is a central element. For simplicity, let us assume that the internal symmetry is an Abelian group. The total symmetry group $\tilde G$ is then a central extension of $\Z^2$ by $A$ [see Eq.~(\ref{A_ZD})]. It is specified by $\nu\in\H^2[\Z^2,A]=A$. If a general element of the translation group is represented as $T_x^{m_x}T_y^{m_y}$ where $m_x,m_y\in\Z$, then we can choose 
\begin{equation}
	\nu(T_x^{m_x}T_y^{m_y}, T_x^{n_x}T_y^{n_y})=a^{m_xn_y}.
	\label{}
\end{equation}

In the topological phase, anyons can carry fractionalized quantum numbers of the global symmetry.
Assuming that no anyons are permuted by the symmetry, symmetry fractionalization is classified by $\H^2[\tilde{G}, \cal{A}]$, where $\cal{A}$ is the group of Abelian anyons. There are three terms in the $E_2$ page:
\begin{equation}
	\begin{split}
		E_2^{0,2}&=\H^2[A, \cal{A}], \\
		E_2^{1,1}&=\H^1[\Z^2, \H^1[A, \mathcal{A}]]={(\H^1[A, \cal{A}])}^2, \\
		E_2^{2,0}&=\H^2[\Z^2, \cal{A}]=\cal{A}.
	\end{split}
	\label{eqn:E2magnetic}
\end{equation}
Ref. \onlinecite{ChengPRX2016} gave the physical interpretations of the various terms when $a=1$: $E_2^{0,2}$ is just the fractionalization of $A$ symmetry itself on anyons. $E_2^{1,1}$ means anyons carry ``dipole moment'' of $A$, i.e. as one separates a pair of anyons apart the total $A$ charge changes. The $E_2^{2,0}$ is the background anyon charge in each unit cell. For magnetic translation the results can be understood similarly. In particular, $E_2^{2,0}$ now describes the background charge in a unit cell of the original lattice (not the magnetic unit cell). Notice that however now they are subject to obstruction and trivialization differentials. 

The only obstruction differential to be considered is
\begin{equation}
	d_2: E_2^{0,2}\rightarrow E_2^{2,1}.
	\label{}
\end{equation}
Suppose $\omega_0$ is a 2-cocycle in $\H^2[A, \cal{A}]$. Using Eq. \eqref{dF21}, we find $d_2\omega_0\in E_2^{2,1}=\H^2[\Z^2, \H^1[A, \cal{A}]]$ given by:
\begin{equation}
	(d_2\omega_0)(b, \mb{g,h})=\frac{\omega_0(b, \nu(\mb{g,h}))}{\omega_0(\nu(\mb{g,h}),b)}.
	\label{}
\end{equation}
Here $b\in A$. For fixed $\mb{g,h}$, $(d_2\omega_0)(, \mb{g,h})$ gives a homomorphism from $A$ to $\cal{A}$. To check whether $d_2\omega_0$ is a nontrivial cocycle in $E_2^{2,1}$, we consider the following invariant:
\begin{equation}
	\frac{(d_2\omega_0)(b, T_x, T_y)}{(d_2\omega_0)(b, T_y, T_x)}=\frac{\omega_0(b, a)}{\omega_0(a,b)}.
	\label{}
\end{equation}
Therefore $\omega_0$ is $d_2$-obstructed if and only if there exists $b\in A$ such that $\omega_0(a,b)\neq \omega_0(b,a)$. What this condition means heuristically is that the group element $a$ must be central (i.e. commuting with all other elements in $A$) even when acting on an individual anyon. For $A=\U$ or a cyclic group, this obstruction always vanishes.

Now let us consider the trivialization differentials. The only nontrivial one is
\begin{equation}
	d_2: E_2^{0,1}\rightarrow E_2^{2,0}.
	\label{}
\end{equation}
Here $E_2^{0,1}=\H^1[A, \cal{A}]$, i.e. a group homomorphism $\varphi$ from $A$ to $\cal{A}$. Under $d_2$, one finds $(d_2\varphi)(\mb{g,h})=\varphi\big(\nu(\mb{g,h})\big)$. In particular, for $G=\Z^2$ we only need to check the $T_x, T_y$ commutator:
\begin{equation}
	\frac{(d_2\varphi)(T_x, T_y)}{(d_2\varphi)(T_y, T_x)} = \varphi(a).
	\label{}
\end{equation}
Namely, we exclude from $E_2^{2,0}=\cal{A}$ the elements that can be written as $\varphi(a)$ for some group homomorphism from $A$ to $\cal{A}$.  Such trivializations do not occur for $A=\U$ since $\H^1[\U, \cal{A}]$ is trivial. 

Neither the obstruction nor the trivialization affects $E_2^{1,1}$, as the physical effect only involves the translation along a certain direction. We therefore conclude that for $A=\U$, the $E_2$ page already converges to $E_\infty$, and we obtain a complete classification of symmetry fractionalization from Eq. \eqref{eqn:E2magnetic}.

\subsubsection{Magnetic LSM theorem}
Now we study an example of LSM theorem with magnetic translation symmetry. Consider a spin system with $\mathrm{O}(2)=\U_{S_z}\rtimes \Z_2$ symmetry.  It is often sufficient to consider the (maximal) finite subgroup $\Z_2^2=\{1, Z, X, ZX\}$. Here $Z$ is the $\pi$ spin rotation for the $\U_{S_z}$ subgroup of $\mathrm{O}(2)$, and $X$ is the $\pi$ rotation around $x$ axis.We will assume that on the lattice there is a $\pi$ flux of $\U_{S_z}$ in each unit cell. In other words, the lattice translations satisfy
\begin{equation}
	T_xT_yT_x^{-1}T_y^{-1}=Z.
	\label{}
\end{equation}
The 2-cocycle $\nu\in\H^2[\Z^2,\Z_2^2]=\Z_2^2$ characterizing the central extension takes values $\nu(T_x,T_y)=Z$ and $1$ for all other arguments. Each unit cell contains a projective representation of O(2), i.e. a spin-$1/2$. The setup is quite similar to the 2D SPT-LSM theorem discussed in Sec.~\ref{sec:SPTLSM}, but we will show that this is an actual LSM system. 

One way to prove this result is to apply the filling constraint in Ref. [\onlinecite{Lu_2020}]. Focusing on the $\U_{S_z}$ subgroup, the system is half-filled with a background $\pi$ flux. Therefore, if the ground state is SRE, the Hall conductance must be an odd integer, which is not allowed for bosonic systems. 

Alternatively, we can use the LHS spectral sequence. In fact, it is sufficient to consider the $\Z_2\times\Z_2$ subgroup of $\mathrm O(2)$. Physically, a would-be SPT phase must have a $\Z_2\times\Z_2$ projective representation bound to a $Z$ flux. However, from the classification of $\Z_2\times\Z_2$ SPT phases in 2D, we know that no such SPT states exist. In fact, the cocycle $\omega_0\in E_2^{03}=\H^3[\Z_2^2,\U]=\Z_2^3$ for 2D $\Z_2\times\Z_2$ SPT can be parametrized as 
\begin{equation}
	\omega_0(a,b,c)=e^{i\pi\sum_{1\le i\le j\le 2}p_{ij}a_ib_jc_j},
	\label{}
\end{equation}
with $p_{ij}=0,1$. From Eq.~(\ref{dFabkl}), the differential $d_2$ of $\omega_0$ is $(d_2\omega_0)(a,b,\ag{1}{k},\ag{1}{l}) = \frac{\omega_0(a,\nu(\mb{k,l}),b)}{\omega_0(a,b,\nu(\mb{k,l})) \omega_0(\nu(\mb{k,l}),a,b)}$. It is easy to check that they are all trivial in $E_2^{2,2}=\H^2[\Z^2,\H^2[\Z_2^2,\U]]=\Z_2$. So the anomaly does not match, and we have an LSM theorem.

Now let us consider what kind of SET phases can be realized. We will show that quite surprisingly, in order to saturate the anomaly, certain symmetry transformations (among translations and $X$) must permute anyons. A key observation is that the spin-$1/2$ representation of O(2) can not be lifted to a projective representation of the full group $\tilde{G}$. If the translation symmetry is not magnetic, the spin-$1/2$ representation must be ``screened'' in the ground state by a background anyon charge which also carries the spin-$1/2$. However, such mechanism does not generalize to the present case. 

We now present a detailed calculation of the anomaly, assuming no anyons are permuted. For technical reasons, we focus on the $A=\Z_2^2$ subgroup of O(2), as the anomaly remains essentially the same. The classification for symmetry fractionalization $\H^2[\tilde{G}, \cal{A}]$ has already been computed from the LHS spectral sequence in Sec. \ref{sec:set-magnetic}. Here we adopt the results:
\begin{itemize}
	\item $E_2^{0,2}$ is characterized by the three invariants
\begin{equation}
	I_Z=\coho{w}(Z,Z), I_X=\coho{w}(X,X), I_{ZX}=\frac{\coho{w}(Z,X)}{\coho{w}(X,Z)}.
	\label{}
\end{equation}
Physically, $\coho{w}(Z,Z)$ determines which anyons carry fractional $Z$ charge, and similarly for $\coho{w}(X,X)$. $I_{ZX}$ determines which anyons carry a two-dimensional projective representations of $\Z_2^2$, i.e. the ``spin-$1/2$'' representation.   We also notice that for projective represetations of O(2) we must have $I_Z=I_{ZX}$. It can be proven by considering the two generating classes: one of them has only nontrivial $I_X$, and $I_Z, I_{ZX}$ both trivial. The other generator can be lifted a projective representation of SO(3), which satisfies $I_Z=I_{ZX}$.  According Sec. \ref{sec:set-magnetic}, $I_{ZX}$ must be trivial in order for the $d_2$ obstruction to vanish. In other words, no anyons carry the two-dimensional projective representation of $\Z_2\times\Z_2$.  In conclusion, the non-obstructed class has both $I_Z$ and $I_{ZX}$ trivial.
\item $E_2^{1,1}$ is characterized by four invariants, $\gamma_{Z/X}(T_{x/y})$. The invariant $\gamma_{a}(\mb{g})$ is defined as 
	\begin{equation}
		\gamma_a(\mb{g})=\frac{\coho{w}(a,\mb{g})}{\coho{w}(\mb{g},a)}.
		\label{}
	\end{equation}
	One can show that all of them must be self-dual Abelian anyons because $Z^2=X^2=1$. In addition, since $Z$ is actually the $\pi$ rotation of $\U_{S_z}$ and $E_2^{1,1}$ is actually completely trivial for $G=\U$, $\gamma_{Z}(T_{x/y})$ must vanish in order to be lifted to O(2).
\item $E_2^{2,0}$ is given by $\cal{A}$, but module all self-dual Abelian anyons due to a $d_2$ trivialization from $E^{0,1}$.
\end{itemize}

The $\H^4$ anomaly class, denoted by $O$ below, can be computed using the formula derived in Refs. \onlinecite{Chen2014} and \onlinecite{SET}. We can use Eqs.~(\ref{Lyn_w04}), (\ref{Lyn_w13}) and (\ref{Lyn_w22}) to extract $E_2^{0,4}, E_2^{1,3}$ and $E_2^{2,2}$. Since we are interested in those fractionalization classes that can be lifted back to O(2), we set $\gamma_Z(T_{x/y})=\mb{0}$. Recall that the obstructio-vanishing condition requires $I_Z=I_{ZX}=\mb{0}$. Let us focus on the most relevant component, $O^{2,2}$ in $E_2^{2,2}$, can be characterized by the following invariant:
\begin{equation}
	I=\frac{O^{2,2}(Z,X,T_x, T_y)O^{2,2}(X,Z, T_y, T_x)}{O^{2,2}(Z,X, T_y, T_x)O^{2,2}(X,Z, T_x, T_y)}.
	\label{}
\end{equation}
The magnetic LSM anomaly requires $I=-1$. A tedious but straightforward calculation yields $I=1$ for the fractionalization classes in our case. Therefore, it is impossible to saturate the magnetic LSM anomaly.


If we do not demand that the symmetry can be enhanced to O(2), the anomaly can be realized simply by a $\Z_2$ gauge theory. More specifically, denote the electric and magnetic gauge charges by $e$ and $m$, respectively. One can show that the fractionalization class
\begin{equation}
	\gamma_Z(T_x)=e,\quad \gamma_X(T_y)=m,
	\label{}
\end{equation}
correctly produces the LSM anomaly.


\subsection{Symmetry-enriched Abelian gauge theory in 2D and 3D}
\label{sec:segex}

\subsubsection{(2+1)d Abelian gauge theory}
Let us apply the LHS spectral sequence to (2+1)d Dijkgraaf-Witten topological gauge theory with an Abelian gauge group $A$ and a symmetry group $G$. It will be assumed that $G$ is a finite group below, although the results can be easily adopted to the continuous case. 

Topological excitations in a $A$ gauge theory are dyons $(\hat{\chi},a)$ where $a\in A$ labels gauge flux, and gauge charges labeled by a character $\hat{\chi}\in \hat{A}$. Fusion rules and braiding statistics are determined by the cohomology twist $\alpha\in\H^3[A, \U]$. The resulting topological order is denoted by $\mathcal{Z}^\alpha(A)$ (known as the twisted quantum double of $A$). We will assume that there is no type-III 3-cocycle~\cite{propitius1995}, since in that case the topological order becomes non-Abelian. We will discuss how to connect the results from the LHS spectral sequence to the classification framework established in Ref. [\onlinecite{SET}]. For simplicity, we assume that the extension is central so $G$ does not act on $A$. In the language of Ref. [\onlinecite{SET}], $G$ action does not change the types of gauge charges.

The $E_2$ page has three terms, $E_2^{0,3}, E_2^{1,2}$ and $E_2^{2,1}$. The first term $E_2^{0,3}=\H^3[A, \U]$ is just the cohomology twist $\alpha$ in the gauge theory. Notice that it is subject to three obstruction differentials 
\begin{equation}
	\begin{split}
	d_2: E_2^{0,3}\rightarrow E_2^{2,2},\\
	d_3: E_3^{0,3}\rightarrow E_3^{3,1},\\	
	d_4: E_4^{0,3}\rightarrow E_4^{4,0}.
	\end{split}
\end{equation}
In fact, similar structures already appeared in 2D SPT-LSM theorems, where the $d_2$-``obstructed'' SPT phase can only exist when the $G$ symmetry is implemented with a mixed anomaly with $A$. In the gauge theory context, the physical interpretation is that it is inconsistent to have the $G$ action on gauge charges as specified by the group extension in the twisted gauge theory. Let us illustrate this by an example of the $d_3$ obstruction, which first appeared in Ref. \onlinecite{Thorngren2015}. Let $A=\Z_n$, and the 3-cocycle
\begin{equation}
	\alpha(a,b,c)=\frac{p}{n^2} a(b+c-[b+c]_n).
	\label{}
\end{equation}
Here $p=0,1,\dots, n-1$ and $a,b,c\in \Z/n\Z$.  The $d_2$ map on $\alpha$ is trivial for $E_2^{2,2}=0$. The $d_3$ map to $\H^3[G, \H^1[A, \U]]$ is (see Appendix \ref{app:example} for the derivation)
\begin{equation}
	(d_3\alpha)(a, \mb{g,h,k})=-\frac{2p}{n}a(\beta_n\tilde{\nu})(\mb{g,h,k}),
	\label{}
\end{equation}
where $\tilde{\nu}$ is an integral lift of $\nu$, and $\beta_n\tilde{\nu}=\frac{1}{n}d\tilde{\nu}$ is the Bockstein homomorphism of $\nu$.  If $-2p\nu$ is trivial as a cohomology class in $\H^2[G, A]$, then for the integral lift there exists a 2-cochain $\mu$ in ${\cal{C}}^2[G, \Z]$ and a 1-cochain $\lambda$ in ${\cal{C}}^1[G, \Z]$ such that $d\lambda+n\mu=-2p\tilde{\nu}$, then 
\begin{equation}
	(d_3\alpha)(a, \mb{g,h,k})=a(\beta_n\mu)(\mb{g,h,k}).
	\label{}
\end{equation}
This is actually a coboundary $d\omega \in \mathcal{B}^3[G, \H^1[A, \U]]$ if we define $\omega(a, \mb{g,h})=a\frac{\mu(\mb{g,h})}{n}\in \mathcal{Z}^2[G, \H^1[A, \U]]$. The converse is also true, that is if $-2p\nu$ is nontrivial in $\H^2[G, A]$, then $d_3\alpha$ is a nontrivial class in $\H^3[G, \H^1[A, \U]]$.

We can actually understand what goes wrong more intuitively: in this twisted $\Z_n$ gauge theory, the gauge flux $(0,1)$ satisfies the following fusion rules: $(0,1)^n=([2p]_n,0)$. On the other hand, the $[2p]_n$ gauge charge carries a projective representation of $G$ characterized by the 2-cocycle $2p\nu$. If $2p\nu$ is a nontrivial cohomology class, it is impossible to take a ``$n$-th'' root of the representation, so the $G$ representation on a unit flux is ill-defined. In summary, the group extension structure is inconsistent with the fusion rules.

Finally, when $d_3$ vanishes, $d_4: E_4^{0,3}\rightarrow E_4^{4,0}$ contributes to the 't Hooft anomaly. An explicit expression of $d_4$ for $A=\Z_n$ is derived in Appendix \ref{app:example}:
\begin{equation}
	d_4\alpha=-\frac{p}{n}\tilde{\nu}\cup_1\beta_n\tilde{\nu}+\frac{p}{n^2}\tilde{\nu}\cup\tilde{\nu}.
	\label{}
\end{equation}
Here $\tilde{\nu}$ is an integral lift of $\nu$, valued in $0,1,\cdots, n-1$.

Next we have a term $E_2^{1,2}=\H^1[G, \H^2[A, \U]]$. The physical meaning is the following: consider gapped invertible domain walls in the $A$ gauge theory, obtained by embedding a 1D $A$-SPT state before gauging. After gauging it means that gauge fluxes have additional charge attached when passing through the domain wall. To see why this is the case, consider the topological response of a 1D SPT phase labeled by $\omega\in \H^2[A, \U]$. When put on a ring with a flux $a\in A$ threaded, the global charge of the system under $b\in A$ changes by $\hat{i}^\omega_a(b)=\omega(b,a)/\omega(a,b)$~\cite{ZaletelJS2014}. Now if the 1D SPT state is embedded in the 2D gauge theory, the additional gauge charge must be carried away by the gauge flux to ensure charge neutrality. Therefore we have the transformation:
\begin{equation}
	(\hat{\chi}, a)\rightarrow (\hat{\chi}\times \hat{i}^\omega_a, a)
	\label{eqn:permutation}
\end{equation}
as the particle passes through the wall.
In other words, each element of $\H^2[A, \U]$ corresponds to a nontrivial permutation in the MTC $\cal{Z}^\alpha(A)$. Notice that these permutations form a subgroup of the topological symmetry group $\text{Aut}\big(\cal{Z}^\alpha(A)\big)$which preserves the gauge structure (i.e. the electric charge sector remains invariant). $E_2^{1,2}$ is then a homomorphism from $G$ to this subgroup, as expected from the general classification~\cite{SET}. 

Let us discuss the obstruction differential $d_2: E_2^{1,2}\rightarrow E_2^{3,1}$. The $E_2^{1,2}$ cocycle is a homomorphism $\rho: G\rightarrow \H^2[A,\U]$:
\begin{equation}
	(d_2\rho)(a, \mb{h,k,l})=\frac{\rho_\mb{l}(a, \nu(\mb{h,k}))}{\rho_\mb{l}(\nu(\mb{h,k}),a))}.
	\label{eqn:d2E12}
\end{equation}
The $\mb{l}$ symmetry acts on a flux $a\in A$ by attaching a charge $\hat{i}_a^\omega$. The projective representation of this charge is characterized by the factor set $\hat{i}_a^{\rho_\mb{l}}(\nu(\mb{h,k}))$ for $\mb{h,k}\in G$, which is exactly Eq. \eqref{eqn:d2E12}. Therefore the $d_2$ represents an obstruction in assigning consistent fractionalization to fluxes, due to potential conflict between the permutations by $\rho$ and the projective represetations carried by gauge charges. As an example, let us assume $G=G_1\times G_2$. Suppose $\nu$ can be completely restricted to $G_2$, and $\rho_\mb{g}$ is nontrivial only when $\mb{g}\in G_1$. If under a transformation $\mb{g}_1\in G_1$, a certain flux is attached a charge carrying a nontrivial projective representation under $G_2$, this is physically impossible as the projective representation of the flux under $G_2$ (by assumption, it is invariant under $G_2$) should not change under symmetry transformations. This inconsistency can be detected by the $d_2$ map through slant product.

The next term, $E_2^{2,1}=\H^2[G, \H^1[A, \U]]=\H^2[G, \hat{A}]$ describes symmetry fractionalization on gauge fluxes. 
 $d_2: E_2^{2,1}\rightarrow E_2^{4,0}$ contributes to the 't Hooft anomaly.
 There is also the trivialization map $d_2: E_2^{0,2}\rightarrow E_2^{2,1}$. More specifically, for $\omega\in \H^2[A, \U]$,
\begin{equation}
	(d_2\omega)(a, \mb{h,k})= \frac{\omega\big(a, \nu(\mb{h,k})\big)}{\omega\big(\nu(\mb{h,k}),a\big)}.
	\label{eqn:E02d2}
\end{equation}
As we have discussed for the obstruction map of $E_2^{1,2}$, for an $a$ flux, a charge $\hat{i}^\omega_a$ is attached, which carries a projective representation exactly given by Eq. \eqref{eqn:E02d2}. Thus the trivialization map simply indicates whether the projective representations on fluxes can be removed by relabeling.

Finally, collecting all contributions to $E_2^{4,0}$ ($d_2$ from $E_2^{2,1}$, $d_3$ from $E_3^{1,2}$, and $d_4$ from $E_4^{0,3}$), one obtains a formula for the 't Hooft anomaly of the symmetry-enriched gauge theory. 

Together, we see that terms on the $E_2$ page describe how gauge fluxes are permuted under the symmetry group $G$, as well the fractionalization on the fluxes. The obstruction and trivialization maps can be interpreted in terms of the consistency between the symmetry actions and the underlying topological order, as well as the 't Hooft anomaly.

\subsubsection{(3+1)d U(1) gauge theory}
\label{sec:U1}
We now consider the classification of (3+1)d U(1) gauge theory enriched by $G$. Recently the formal classification was established in Ref. [\onlinecite{NingU1SL}] (see also Ref. [\onlinecite{HsinU1SL}]), and here we will see how the results can be reproduced using the LHS spectral sequence. Suppose $G$ preserves the gauge structure (i.e. it does not mix electric and magnetic charges), then as discussed in we can use the LHS spectral sequence to compute the classification. As many details can be found in Ref. [\onlinecite{NingU1SL}], here we will be brief.

The nontrivial terms on the $E_2$ page are $E_2^{1,3}, E_2^{3,1}$ and $E_2^{4,0}$.  More concretely:
\begin{equation}
	\begin{gathered}
	E_2^{1,3}=\H^1[G, \H^3[\U, \U]]=\H^1[G, \Z]\\
	E_2^{3,1}=\H^3[G, \H^1[\U, \U]]=\H^3[G, \Z]\\
	E_2^{4,0}=\H^4[G, \U].
	\end{gathered}
	\label{}
\end{equation}
Physically, $E_2^{1,3}$ describes 2D bosonic integer quantum Hall states decorated on $G$ domain walls. Equivalently, $G$ implements a $T^2$ duality transformation on magnetic charges. Under $G$ magnetic charges may transform projectively, which is described by $E_2^{3,1}$. Finally $\H^4[G, \U]$ describes stacking a $G$ SPT state.

Now we consider the differentials. Many of them vanish automatically, and we focus on the possibly nontrivial ones. First the obstruction maps:
\begin{equation}
	\begin{split}
	d_2: 
		E_2^{3,1}&\rightarrow E_2^{5,0}\\
	d_3: 
		E_2^{1,3}&\rightarrow E_2^{4,1}\\
	d_4: 
		E_2^{1,3}&\rightarrow E_2^{5,0}\\
	\end{split}
	\label{eqn:diffU1}
\end{equation}

The differentials $d_2$ and $d_4$ ending in $E_2^{5,0}$ give the 't Hooft anomaly, which was also obtained in Ref. \onlinecite{NingU1SL}. The $d_4:E_4^{1,3}\rightarrow E_4^{4,1}$ was an obstruction class to $G$ domain wall decorations (by 2D U(1) SPT phases), and it was termed ``deconfinement obstruction'' in  Ref. [\onlinecite{NingU1SL}]. Explicit expressions for the differentials in Eq. \eqref{eqn:diffU1} were obtained in Ref. [\onlinecite{NingU1SL}].

Next we turn to the trivialization differentials:
\begin{equation}
	\begin{split}
	d_2: 
		E_2^{2,1}&\rightarrow E_2^{4,0}\\
	d_3: 
		E_3^{0,3}&\rightarrow E_3^{3,1}\\
	d_4: 
		E_4^{0,3}&\rightarrow E_4^{4,0}.
	\end{split}
	\label{}
\end{equation}
The explicit expressions for these trivialization differentials are calculated in Appendix.~\ref{app:example} (for a similar example of $A=\Z_n$). If $G$ is unitary and has trivial action on U(1), the differentials can be simplified: 
\begin{equation}
	\label{dU1}
\begin{aligned}
(d_2\omega_{2,1})(\mb{g,h,k,l})&=
-\nu(\mb{g,h})n(\mb{k,l})
\\
(d_3\omega_{0,3})(a,\mb{h,k,l})&=
-2 k a \cdot (\beta\nu)(\mb{h,k,l})
\\
(d_4\omega_{0,3})(\mb{g,h,k,l})&=
- k [\nu(\mb{ghk,l}) \cdot (\beta\nu)(\mb{g,h,k}) \\
&\quad+ \nu(\mb{g,hkl}) \cdot (\beta\nu)(\mb{h,k,l})]\\
&\quad+ \frac{k}{2\pi}\nu(\mb{g,h})\nu(\mb{k,l}),
\end{aligned}
\end{equation}
where $\nu(\mb{g,h})\in [0,2\pi)$ is the symmetry extension 2-cocycle from $G$ to $\U$ in $\H^2[G,\U]$. 
In the first equation, the cocycle $\omega_{2,1}$ in $E_2^{2,1}=\H^2[G,\H^1[\U,\U]]$ is parametrized by a $\Z$-valued 2-cocycle $n\in \H^2[G,\Z]$.
In the last two equations, the cocycle $\omega_{0,3}$ at $E_2^{0,3}=\H^3[\U,\U]=\Z$ is chosen to be
\begin{align}
\omega_{0,3}(a,b,c)=\frac{k}{2\pi} a (b+c-[b+c]_{2\pi}),
\end{align}
with $a,b,c\in [0,2\pi) \cong \U$ and $k\in\Z$. And $\beta\nu$ is the Bockstein homomorphism of $\nu$ defined as
\begin{align}
(\beta\nu)(\mb{g,h,k}) = \frac{\nu(\mb{h,k}) - \nu(\mb{gh,k}) + \nu(\mb{g,hk}) - \nu(\mb{g,h})}{2\pi}.
\end{align}

Both $d_2$ and $d_4$ lead to trivialization of SPT stacking~\cite{{WangPRX2016},{ZouPRB2018}}. The $d_2$ trivialization was discussed in Ref. [\onlinecite{HsinU1SL}].  The $d_4$ trivialization appears to be unknown before. Since the involved data is a $\H^3[\U,\U]$ cocycle, we conjecture that physically it corresponds to a duality $T^2$ transformation.

Finally, the $d_3$ map trivializes monopole symmetry fractionalization. The physical meaning is very clear: if the projective representation carried by the monopole is $-2k\nu$, then a duality transformation $T^{2k}$ makes the monopole completely neutral.

\section{Acknowledgement}
Q.-R. W. would like to thank Yang Qi for helpful discussions on diagonal approximations. M. C. and Q.-R. W acknowledge support from Alfred P. Sloan Foundation and the NSF under grant No. DMR-1846109. S.-Q. Ning acknowledges support from  Research Grant Council of Hong Kong (ECS 21301018) and URC, HKU (Grant No. 201906159002).

\bibliography{lhs.bib}

\appendix

\onecolumngrid

\section{Spectral sequence: a practical introduction}
\label{sec:ss}
Here we give a quick and operational definition of the (cohomology) spectral sequence, using the LHS spectral sequence as an example as it is most relevant for our work.  

A spectral sequence consists of a family of Abelian groups $E_r^{p,q}$ where $r,p,q$ are non-negative integer. The collection of all $E_r^{p,q}$ with the fixed $r$  are called the $E_r$ \emph{page}. Within each page, there exist a set of group homomorphisms, called differentials, that map elements of one Abelian group $E_r^{p,q}$  to another Abelian group $E_r^{p+n, q-n+1}$. We denote such a homomorphism by $d_r^{p,q}$ where the subscript $r$ indicates that it is defined on the $E_r$ page,  and the superscripts $p,q$  imply that its preimage is in $E_r^{p,q}$. Such a homomorphism can be represented by the following map: 
\begin{align}
d_r^{p,q}: E_r^{p, q}\rightarrow E_r^{p+r,q-r+1}
\end{align}
We can sometimes  omit the superscript, and just use $d_r$ for simplicity when there is no ambiguity. We require that the differentials  should satisfy $d_r^2=0$ (i.e. $d_r^2$ maps into the group identity).   In addition, we have the following isomorphisms
\begin{align}
	E_{r+1}^{p,q}\simeq \frac{\mathrm{ker}(d_r^{p,q})}{\mathrm{img} (d_r^{p-n, q-n+1})}.
\label{eqn:iso_ss}
\end{align}

The family of $\{ E_r^{p,q}\}$ can be constructed iteratively, namely, knowing the $E_r$ page (i.e., $E_r^{p,q}$ for all $p,q$ with  given $r$), one can construct the next parge, i.e., $E_{r+1}$-page via the isomorphism (\ref{eqn:iso_ss}).  

Let us take the LHS spectral sequence as an example.  Given a short exact sequence of groups
\begin{equation}
	1\rightarrow A\rightarrow \tilde{G}\rightarrow G\rightarrow 1,
	\label{}
\end{equation}
the LHS spectral sequence computes the group cohomology of $\tilde{G}$ in terms of the group cohomology of the normal subgroup $A$ and the quotient group $G=\tilde G/A$. We denote the set of $i$-cochains, $i$-cocycles, and $i$-coboundaries  of an group $G_0$  with  coefficients in $M$ by ${\mathcal C}^i[ G_0, M]$, ${\mathcal Z}^i[ G_0, M]$, and ${\mathcal B}^i[ G_0, M]$ repsectively. 

\begin{figure}[h]
   \centering
   \includegraphics[width=\textwidth]{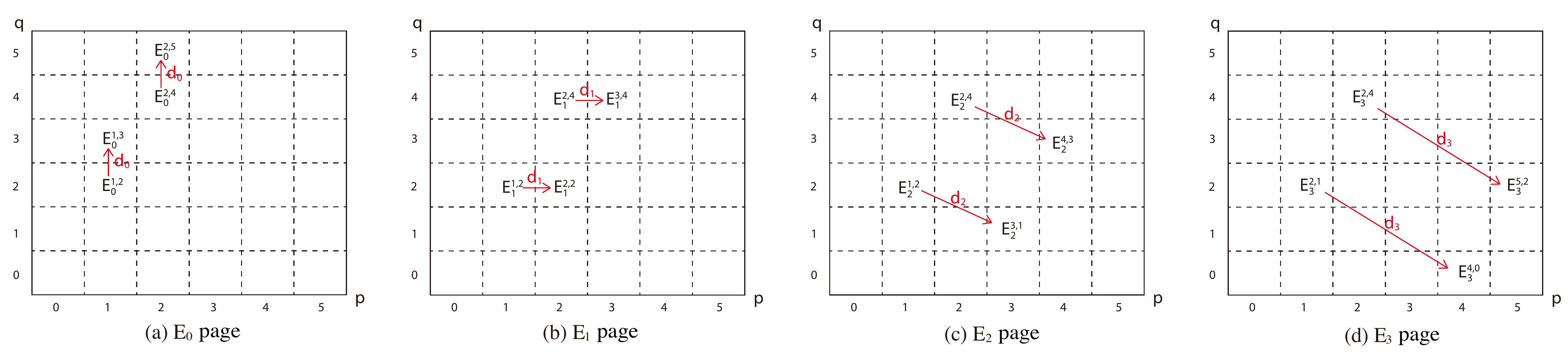} 
   \caption{Examples for illustrating  how $d_n$ differentials map within $E_n$ page.}
   \label{fig:Er_differential}
\end{figure}

We begin with defining $E_0$-page of the LHS spectral sequence, which is just the group of  cochains  ${\mathcal C}^p[G, {\mathcal C}^q[A, M]]$. The $d_0$ differential is given by Eq. \eqref{eqn:d0_def}, which maps a cochain in $E_0^{p,q}={\mathcal C}^p[G, {\mathcal C}^q[A, M]]$ to a cochain in $E_0^{p,q+1}={\mathcal C}^p[G, {\mathcal C}^{q+1}[A, M]]$. See Fig.\ref{fig:Er_differential}(a) for illustration. The kernel of $d_0^{p,q}$ is just ${\mathcal C}^p[G, {\mathcal Z}^q[A, M]]$, while the  image of $d_0^{p,q-1}$ is just ${\mathcal C}^p[G, \mathcal B^q[A, M]]$, hence the $E_1$ page is given by
\begin{align}
	E_1^{p,q}=\frac{\mathrm{ker}(d_0^{p,q})}{\mathrm{img}(d_0^{p,q-1})}=\frac{{\mathcal C}^p[G, {\mathcal Z}^q[A, M]]}{{\mathcal C}^p[G, \mathcal B^q[A, M]]}={\mathcal C}^p[G, \H^q[A, M]].
\end{align}
Then we see $E_1$ page is a subgroup of $E_0$ page (modulo additional equivalence relations), that is,
\begin{align}
E_0\supset E_1. \label{eqn:sequence_0}
\end{align}
If we denote the elements of $E_0^{p,q}$ by $w_0^{p,q}$,  the more precise meaning of (\ref{eqn:sequence_0}) is that 
the elements of $E_1^{p,q}$ are equivalence classes of elements in $E_0^{p,q}$ that satisfy the condition $d_0^{p,q}w_0^{p,q}=0$, or more compactly, 
\begin{align}
d_0 w_0=0,
\end{align}
with the equivalence relation given by $w_0\sim w_0d_0c_{-1}$, where $c_{-1}\in {\mathcal C}^p[G, {\mathcal C}^{q-1}[A,M]]$.

 Now on the $E_1$ page given by $E_1^{p,q}={\mathcal C}^p[G, \H^q[A,M]]$,  the differential $d_1$ is defined  by Eq. \eqref{eqn:d1_def}. Namely $d_1$ maps a cochain in ${\mathcal C}^p[G, \H^q[A,M]]$ to a cochain in ${\mathcal C}^{p+1}[G, \H^q[A,M]]$.  The kernel of $d_1^{p,q}$ is nothing but cocycles ${\mathcal Z}^{p}[G, \H^q[A, M]]$, and the image of $d_1^{p-1, q}$ is nothing but the coboundaries $\mathcal B^p[G, \H^q[A,M]]$. We illustrate $d_1$ in Fig.\ref{fig:Er_differential}(b). Then we define the $E_2$ page by
\begin{align}
	E_2^{p,q}=\frac{\mathrm{ker}(d_1^{p,q})}{\mathrm{img}(d_1^{p-1,q})}=\frac{{\mathcal Z}^{p}[G, \H^q[A, M]]}{\mathcal B^p[G, \H^q[A,M]]}=\H^p[G, \H^q[A, M]],
\end{align}
 Then we see $E_2$ page is a subset of $E_1$-page (modulo equivalence relations), that is,
\begin{align}
E_0\supset E_1\supset E_2.
\label{eqn:sequence_1}
\end{align}

More concretely, elements of $E_2$ page are those elements in $E_0$-page that satisfy the following conditions
\begin{subequations}
\begin{align}
d_0 w_0&=0\label{eqn:d0_cond0},\\
d_1w_0&=0\label{eqn:d1_cond0},
\end{align}
\end{subequations}
where the first condition (\ref{eqn:d0_cond0}) ensures that it is in $E_1$ page, and the second (\ref{eqn:d1_cond0}) ensures it is in $E_2$ page. 

On the $E_2$ page, we have the $d_2$ differentials
\begin{align}
d_2^{p,q}: E^{p,q}_2\rightarrow E^{p+2, q-1}_2,
\end{align}
or more explicitly,
\begin{align}
d_2^{p,q}: \H^p[G,\H^q[A,M]]\rightarrow \H^{p+2}[G,\H^{q-1}[A,M]].
\end{align}
The explicit expression of $d_2$ is given by Eq.~\eqref{diFpq} (with $i=2$). We illustrate  how the $d_2$ differential maps within the $E_2$ page in Fig.\ref{fig:Er_differential}(c).

Following the pattern, the $E_3$ page can be constructed from the $E_2$ page:
\begin{align}
	E_3^{p,q}=\frac{\mathrm{ker}(d_2^{p,q})}{\mathrm{img}(d_{2}^{p-2,q+1})}.
\end{align}
Within $E_3$ page, we can also define the $d_3$ differentials, as given by Eq.~\eqref{diFpq}(with $i=3$). We illustrate some examples about how the $d_3$ differential maps within $E_3$ page in Fig. \ref{fig:Er_differential}d. We also see that $E_3$ page is a subset of $E_2$ and hence of $E_1$ page. Then the inclusion relation Eq.~\eqref{eqn:sequence_1} can be extended to include $E_3$-page, given by
\begin{align}
E_0\supset E_1\supset E_2 \supset E_3.\label{eqn:sequence_2}
\end{align}
More precisely the above relation sequence implies that the elements of $E_3$ page are those  elements in $E_0$ page that satisfy  the following conditions
\begin{subequations}
\begin{align}
&d_0 w_0=0\label{eqn:d0_cond1},\\
&d_1w_0=0\label{eqn:d1_cond1},\\
&d_2w_0=0\label{eqn:d2_cond1}.
\end{align}
\end{subequations}

We can continue this operational definition for arbitary $E_r$-page. We have the relation sequence
\begin{align}
E_0\supset E_1\supset E_2 \supset E_3 \supset \cdots \supset  E_r.
\label{eqn:sequence_3}
\end{align}
 Starting from $E_0$ page, the elements that belong to $E_{r+1}$-page must satisfy the following $r+1$ conditions
\begin{equation}
\begin{split}
&d_0 w_0=0\\
&d_1w_0=0\\
&d_2w_0=0\\
&\quad\quad \vdots\\
&d_{r} w_0=0
\end{split}
\label{eqn:sequence_cond_general}
\end{equation}
If for a large enough $r$ the condition $d_{r}w_{0}=0$ is satisfied over the entire $E_r$, then the elements in $E_r$ automatically become the elements in $E_{r+1}$ which means that we have $E_{r+1}=E_r$, and all the higher pages are the same. We say that the sequence stablizes at $E_r$, or we have reached the $E_\infty$ page. For the LHS spectral sequence, the $E_\infty$ is isomorphic to the group cohomology $\H^n[\tilde G, M]$ as a set. In order to fully recover the group $\H^n[\tilde{G}, M]$ one also needs to understand the group multiplication structure on $E_\infty$, which is called the extension problem.

Now we briefly discuss the relation between  group cohomology of $\H^n[\tilde G, M]$ and  $E^{p,q}$.
For a fixed $n$, a general cocycle in $\H^n[\tilde G, M]$ can be constructed from cochains in $E_0^{p,n-p}$, and to satisfy the cocycle condition these cochains must satisfy conditions which schematically take the form of Eq.~\eqref{eqn:sequence_cond_general}. However, the actual expressions involve ``cochain-level'' differentials instead of differentials mapping between cohomology classes. In the following sections (\ref{App:H1}--\ref{App:H4}), we will show how a $n$-cocycle can be decomposed into $E_0^{p,n-p}$ cochains for $n\leq 4$, and work out the cochain-level differentials explicitly for general $n$. Together these results provide an explicit description of the LHS spectral sequence.

\section{LHS spectral sequence for degree-$1,2,3$ and $4$}
\label{sec:LHS1234}

The goal of this section is to obtain the explicit expressions for $n$-cocycles and (cochain-level) differentials in the LHS spectral sequence. Although only the cases with degree $n=1,2,3,4$ will be discussed in full detail, the method can be generalized to higher degrees order by order. And based on the computations at degree $n\leq 4$, we conjecture a unified description of the cochain-level differentials using the slant products and diagonal approximations.

For simplicity, we assume that $A$ is Abelian. It is not difficult to generalize all the results to the non-Abelian case. We will present our results using the cohomology coefficient $\U$. To obtain the results for the coefficient of other Abelian group, we can simplify replace the multiplication in $\U$ by the corresponding group multiplication.


In the following part of this appendix, we will first review some useful notations and concepts such as slant product and diagonal approximation in Sec.~\ref{ssec:notation}. The obstruction conditions, cochain-level differentials $\delta_i$ and cocycle-level differentials $d_i$ in the LHS spectral sequence are summarized in Sec.~\ref{App:diff}. After that, we will show the details of calculating the explicit expressions for $n$-cocycles ($n=1,2,3,4$) and differentials on each page of the LHS spectral sequence. In Sec.~\ref{app:example}, we will show how to solve the obstruction conditions and obtain the differentials in the example of $A=\Z_n$ with trivial $G$ action in degree 3.

\subsection{Notations and conventions}
\label{ssec:notation}

\subsubsection{Symmetry actions}

There are two group actions in our discussions. One is an ingredient in the definition of group $\tilde G$, i.e., a map from $G$ to $\operatorname{Aut}(A)$ which specifies an action of $G$ on $A$. If $A$ is non-Abelian, this map is not necessarily a group homomorphism. In this appendix, we will always assume that $A$ is Abelian, and the non-Abelian generalization is straightforward. We will use the notation $\ga{g}{a}$ ($a\in A, g\in G$) to denote this action. In terms of the group multiplication in $\tilde G$, we have $\ga{g}{a} = \ag{1}{g}\cdot a\cdot (\ag{1}{g})^{-1}$. Another one is the action of $\tilde G$ on the cohomology coefficient, which makes the coefficient into a $\tilde G$ module. For our purpose, we mainly use the coefficient $\U$ in our paper, although all the formulas and results below also apply to other coefficient. The most common action in physics is complex conjugation if the element in $\tilde G$ is time reversal. We will abuse the notation such as ${}^{\ag{a}{g}}(F^{b,c,d})$ to indicate this action of $\ag{a}{g}$ on $F^{b,c,d}\in\U$.

From the above two symmetry actions, we can define a natural action of $G$ on $\H^q[A,\U]$, which appears in the definition of $E_2^{p,q}=\H^p[G,\H^q[A,\U]]$ \cite{Lyndon,HS}. Following Ref.~\onlinecite{HS}, we use the convention that $G$ acts on $\H^q[A,\U]$ as
\begin{align}\label{Gaction}
({}^{\mb{g}} \omega)(a_1,a_2,...,a_q) := {}^{\ag{1}{g}}\! \left[\omega\!\left(\gia{g}{a_1},\gia{g}{a_2},...,\gia{g}{a_q}\right)\right],
\end{align}
where $\omega \in \H^q[A,\U]$ is a $q$-cocycle of $A$. On the right hand side, the $\overline{\mb{g}}=\mb{g}^{-1}$ action on $a$ in the arguments of $\omega$ is the action of $G$ on $A$ in the definition of $\tilde G$. And the $\ag{1}{g}$ action is the $\tilde G$ action on the coefficient $\U$.

Both of two symmetry actions appears in the usual differentials of the LHS spectral sequence. The simplest $\delta_0$ and $\delta_1$ are the differentials with respect to the group $A$ and $G$ respectively:
\begin{align}
\delta_0F_{p,q} &= d_A F_{p,q},\\
\delta_1F_{p,q} &= \left(d_G F_{p,q}\right)^{(-1)^{q}}.
\end{align}
To be more precise, the differentials $d_AF_{p,q}\in E^{p,q+1}$ and $d_GF_{p,q}\in E^{p+1,q}$ are
\begin{align}\nonumber
(d_A F_{p,q})(a_1,...,a_{q+1},\mb{g}_1,...,\mb{g}_p)
&=
{}^{a_1}\!\left(F^{a_2,...,a_{q+1},1_{\mb{g}_1},...,1_{\mb{g}_p}}\right)
\times \prod_{i=1}^q \left(F^{a_1,...,a_ia_{i+1},...,a_{q+1},1_{\mb{g}_1},...,1_{\mb{g}_p}}\right)^{(-1)^i}\\
&\quad
\times \left(F^{a_1,...,a_{q},1_{\mb{g}_1},...,1_{\mb{g}_p}}\right)^{(-1)^{q+1}}, \label{eqn:d0_def}
\\\nonumber
(d_G F_{p,q})(a_1,...,a_{q},{\mb{g}_1},...,{\mb{g}_{p+1}}) &= {}^{1_{\mb{g}_1}}\!\left(F^{\gia{g_\mathrm 1}{a_1},...,\gia{g_\mathrm 1}{a_q},1_{\mb{g}_2},...,1_{\mb{g}_{p+1}}}\right)
\times \prod_{i=1}^p \left(F^{a_1,...,a_q,1_{\mb{g}_1},...,1_{\mb{g}_i\mb{g}_{i+1}},...,1_{\mb{g}_{p+1}}}\right)^{(-1)^i}\\
&\quad\times \left(F^{a_1,...,a_q,1_{\mb{g}_1},...,1_{\mb{g}_{p}}}\right)^{(-1)^{p+1}}.\label{eqn:d1_def}
\end{align}
We note that the first terms in both $d_A$ and $d_G$ have an $\tilde G$ action on the $\U$ coefficient. And the first $q$ variables $a_i$ ($1\le i\le q$) of the first term in $d_G$ are acted $\overline {\mb{g}_1}$. This is the $G$ action on $A$ in the definition of $\tilde G$.

For convenience, we define the $G$ action on $F\in E_0^{p,q}=\mathcal{C}^p[G,\mathcal{C}^q[A,\U]]$ to be
\begin{align}\label{hF}
\left({}^{\ag{1}{h}}\!F\right)^{a_1,...,a_q,1_{\mb{g}_1},...,1_{\mb{g}_p}}
:=
{}^{\ag{1}{h}}\!\left( F^{\gia{h}{a_1},...,\gia{h}{a_q},1_{\mb{g}_1},...,1_{\mb{g}_p}} \right),
\end{align}
where all the $A$ part variables $a_1,...,a_q$ of $F$ are acted by $\overline{\mb{h}}=\mb{h}^{-1}$. On the other hand, all the $G$ part variables $1_{\mb{g}_1},...,1_{\mb{g}_p}$ of $F$ are fixed without $G$-action by $\mb{h}$.

\subsubsection{Slant product}

Slant product in algebraic topology is a paring between chains and cochains \cite{Spanier1989}. As it produces lower degree cochains, slant product is important in dimension reduction constructions of both bosonic and fermionic SPT phases \cite{propitius1995,Cheng_2018,Tantivasadakarn_2017}. The simplest example of slant product is a map sending a 1-chain and an $n$-cochain to an $(n-1)$-cochain as alternating product:
\begin{align}\label{eqn:slant}
(\iota_b \omega_n) (a_1,...,a_{n-1}) := \omega_n(a_1,...,a_{n-1},b) \times \omega_n(a_1,...,a_{n-2},b,a_{n-1})^{-1} \times ...\times \omega_n(b,a_1,...,a_{n-1})^{(-1)^{n-1}}.
\end{align}
Given a 2-chain $(b_1,b_2)$, we can similarly map an $n$ cochain to an $(n-2)$-cochain:
\begin{align} \label{eqn:slant2}
[\iota_{(b_1,b_2)} \omega_n] (a_1,...,a_{n-2}) 
&:= \prod_{0\le i\le j\le n-2} \left[\omega_n(a_1,...,a_{i},b_1,a_{i+1},...,a_j,b_2,a_{j+1},...,a_{n-2})\right]^{(-1)^{i+j}}\\\nonumber
&=\omega_n(a_1,...,a_{n-2},b_1,b_2) \times \omega_n(a_1,...,a_{n-3},b_1,a_{n-2},b_2)^{-1} \times ...\times \omega_n(b_1,b_2,a_1,...,a_{n-2}).
\end{align}
In general, we can define the shuffle product \cite{eilenberg1953} for two chains $a=(a_1,...,a_m)$ and $b=(b_1,...,b_n)$. The result $a \shuffle b$ is a sum over the $\frac{(m+n)!}{m!n!}$ $(m+n)$-chains of interleaving $a$ and $b$. We use the convention that the first chain $(a_1,...,a_m,b_1,...,b_n)$ has plus sign, and all other terms have alternating signs according to the parity of the permutation. Using shuffle product, the slant product is defined as
\begin{align}
[\iota_{(b_1,...,b_n)}\omega_{m+n}](a_1,...,a_m)
&:=
\omega_{m+n}[(a_1,...,a_m) \shuffle (b_1,...,b_n)]\\\nonumber
&=
\omega_{m+n}(a_1,...,a_m,b_1,...,b_n)
\times [\omega_{m+n}(a_1,...,a_{m-1},b_1,a_m,b_2,...,b_n)]^{-1}\\\nonumber
&\quad\times...\times
[\omega_{m+n}(b_1,...,b_n,a_1,...,a_m)]^{(-1)^{mn}}.
\end{align}
Given an $n$-chain $b$, the slant product maps an $(m+n)$-cochain $\omega_{m+n}$ to an $m$-cochain $\iota_b\omega_{m+n}$.

\subsubsection{Diagonal approximation and higher cup product}
\label{sec:HigherCup}

The diagonal approximation maps one chain to a summation of tensor product of two chains. The zeroth order diagonal approximation (Alexander-Whitney map) is defined as:
\begin{align}\label{Delta0}
\Delta_0(g_1,...,g_n) := \sum_{p=1}^n (g_1,...,g_p)\otimes \left[{}^{g_1...g_p}(g_{p+1},...,g_n)\right].
\end{align}
When pairing the above expression with two cochains $\nu_m$ and $\nu_n$, it produces the cup product
\begin{align}
(\nu_m \cup \nu_n)(g_1,...,g_{m+n}) = \nu_m(g_1,...,g_m) \times {}^{g_1...g_m}[\nu_n(g_{m+1},...,g_{m+n})].
\end{align}
The right-hand-side of the cup product has only one term $p=m$ in Eq.~(\ref{Delta0}), as other terms in the diagonal approximation do not match the degrees of $\nu_m$ and $\nu_n$.

The higher order diagonal approximation is a chain homotopy measuring the failure of commutativity of lower order diagonal approximations \cite{bredon2013}. In this paper, we only use the first order diagonal approximation defined as
\begin{align}\label{Delta1}
\Delta_1(g_1,...,g_n) := \sum_{p=1}^n \sum_{i=0}^{p-1} (-1)^{(p-i)(n-p)} (g_1,...,g_i,g_{i+1}...g_{i+n-p+1},g_{i+n-p+2},...,g_n) \otimes [{}^{g_1...g_i}(g_{i+1},...,g_{i+n-p+1})].
\end{align}
Similar to the zeroth diagonal approximation and cup product, the first order diagonal approximation is related to cup-1 product \cite{steenrod1947}
\begin{align}\nonumber
&\quad(\nu_m \cup_1 \nu_n)(g_1,...,g_{m+n-1})\\
&= \sum_{i=0}^{m-1} (-1)^{(m-i)(n+1)} \nu_m(g_1,...,g_i,g_{i+1}...g_{i+n},g_{i+n+1},...,g_{m+n-1}) \times {}^{g_1...g_i}[\nu_n(g_{i+1},...,g_{i+n})].
\end{align}
Due to the degrees of $\nu_m$ and $\nu_n$, only the $p=m$ terms in Eq.~(\ref{Delta1}) appear in cup-1 product.

For convenience, we list some of the useful cup-1 products:
\begin{align}\label{n2cup1n2}
(\nu_2\cup_1\nu_2)(\mb{g,h,k})&=
\nu_2(\mb{gh,k})\times\nu_2(\mb{g,h})
-\nu_2(\mb{g,hk})\times \ga{g}{\nu_2(\mb{h,k})},
\\\label{n3cup1n2}
(\nu_3\cup_1\nu_2)(\mb{g,h,k,l})&=
-\nu_3(\mb{gh,k,l})\times\nu_2(\mb{g,h})
+\nu_3(\mb{g,hk,l})\times\ga{g}{\nu_2(\mb{h,k})}
-\nu_3(\mb{g,h,kl})\times\ga{gh}{\nu_2(\mb{k,l})},
\\\nonumber\label{n4cup1n2}
(\nu_4\cup_1\nu_2)(\mb{g,h,k,l,m})&=
\nu_4(\mb{gh,k,l,m})\times\nu_2(\mb{g,h})
-\nu_4(\mb{g,hk,l,m})\times\ga{g}{\nu_2(\mb{h,k})}\\
&\quad
+\nu_4(\mb{g,h,kl,m})\times\ga{gh}{\nu_2(\mb{k,l})}
-\nu_4(\mb{g,h,k,lm})\times\ga{ghk}{\nu_2(\mb{l,m})}.
\end{align}
They are related to the following first order diagonal approximations by replacing $\times$ by $\otimes$:
\begin{align}
(\nu_2\otimes\nu_2)(\Delta_1(\mb{g,h,k}))&=
\nu_2(\mb{gh,k})\otimes\nu_2(\mb{g,h})
-\nu_2(\mb{g,hk})\otimes \ga{g}{\nu_2(\mb{h,k})},
\\
(\nu_3\otimes\nu_2)(\Delta_1(\mb{g,h,k,l}))&=
-\nu_3(\mb{gh,k,l})\otimes\nu_2(\mb{g,h})
+\nu_3(\mb{g,hk,l})\otimes\ga{g}{\nu_2(\mb{h,k})}
-\nu_3(\mb{g,h,kl})\otimes\ga{gh}{\nu_2(\mb{k,l})},
\\\nonumber
(\nu_4\otimes\nu_2)(\Delta_1(\mb{g,h,k,l,m}))&=
\nu_4(\mb{gh,k,l,m})\otimes\nu_2(\mb{g,h})
-\nu_4(\mb{g,hk,l,m})\otimes\ga{g}{\nu_2(\mb{h,k})}\\
&\quad
+\nu_4(\mb{g,h,kl,m})\otimes\ga{gh}{\nu_2(\mb{k,l})}
-\nu_4(\mb{g,h,k,lm})\otimes\ga{ghk}{\nu_2(\mb{l,m})}.
\end{align}
Using Eqs.~(\ref{n2cup1n2}) - (\ref{n4cup1n2}), we can derive the expressions for iterative cup-1 product $((\nu_2\cup_1\nu_2)\cup_1...)\cup_1\nu_2$ with $n$ $\nu_2$'s. It is a summation of $n!$ terms if we expand the iterative cup-1 of $\nu_2$. For example, $(\nu_2\cup_1\nu_2)\cup_1\nu_2$ has $3!=6$ terms:
\begin{align}\nonumber\label{cup1_3nu}
((\nu_2\cup_1\nu_2)\cup_1\nu_2)(\mb{g,h,k,l})
&=
-\nu_2(\mb{ghk,l})\times\nu_2(\mb{gh,k})\times\nu_2(\mb{g,h})+\nu_2(\mb{gh,kl})\times\ga{gh}{\nu_2(\mb{k,l})}\times\nu_2(\mb{g,h})\\
&\quad
+\nu_2(\mb{ghk,l})\times\nu_2(\mb{g,hk})\times\ga{g}{\nu_2(\mb{h,k})}-\nu_2(\mb{g,hkl})\times\ga{g}{\nu_2(\mb{hk,l})}\times\ga{g}{\nu_2(\mb{h,k})}\\\nonumber
&\quad
-\nu_2(\mb{gh,kl})\times\nu_2(\mb{g,h})\times\ga{gh}{\nu_2(\mb{k,l})}+\nu_2(\mb{g,hkl})\times\ga{g}{\nu_2(\mb{h,kl})}\times\ga{gh}{\nu_2(\mb{k,l})}.
\end{align}
Similarly, we can apply the first order diagonal approximations iteratively as $(\Delta_1\otimes\operatorname{id.}\otimes...\otimes\operatorname{id.})...(\Delta_1\otimes\operatorname{id.})\Delta_1$. The higher cup products and diagonal approximations have the following correspondence:
\begin{align}
\nu_2\cup_1\nu_2
\quad&\longleftrightarrow\quad
(\nu_2\otimes\nu_2)(\Delta_1),\\
(\nu_2\cup_1\nu_2)\cup_1\nu_2
\quad&\longleftrightarrow\quad
(\nu_2\otimes\nu_2\otimes\nu_2)[(\Delta_1\otimes\operatorname{id.})\Delta_1],\\
((\nu_2\cup_1\nu_2)\cup_1\nu_2)\cup_1\nu_2
\quad&\longleftrightarrow\quad
(\nu_2\otimes\nu_2\otimes\nu_2\otimes\nu_2)[(\Delta_1\otimes\operatorname{id.}\otimes\operatorname{id.})(\Delta_1\otimes\operatorname{id.})\Delta_1],\\\nonumber
&\,\,\,\,\, ...
\end{align}
We will find later that the three above formulas of iterative cup-1 product or first order diagonal approximation would appear in the differentials $\delta_3,\delta_4$ and $\delta_5$ in the LHS spectral sequence.

\subsection{Summary of obstructions and differentials}
\label{App:diff}

We note that the differentials $\delta_i$ we will obtain later are defined for all cochains (not only cocycles). Hence, we will call them the \emph{cochain-level differentials} $\delta_i: E_0^{p,q} \rightarrow E_0^{p+i,q-i+1}$, in contrast to the \emph{cocycle-level differentials} $d_i: E_i^{p,q} \rightarrow E_i^{p+i,q-i+1}$.

In the following, we will first show the general form of the obstruction conditions. Then we will summarize the expressions for the cochain-level differentials $\delta_i$ and the cocycle-level differentials $d_i$.

\subsubsection{Solving obstruction conditions}
\label{sec:obs}

In the later sections of this appendix, we will obtain the obstruction conditions of $n$-cocycle $F$ for $n=1,2,3,4$. In general, the obstruction conditions of the $n$-cocycle $F$ can be summarized as
\begin{align}\label{d0F0n}
E_0^{0,n+1}&:\quad (\delta_0F_{0,n}) = 1,\\\label{d0F1n-1}
E_0^{1,n}&:\quad (\delta_1F_{0,n})(\delta_0F_{1,n-1}) = 1,\\
E_0^{2,n-1}&:\quad (\delta_2F_{0,n})(\delta_1F_{1,n-1})(\delta_0F_{2,n-2}) = 1,\\
&\quad\quad...\\
E_0^{n,1}&:\quad (\delta_nF_{0,n})(\delta_{n-1}F_{1,n-1})...(\delta_0F_{n,0}) = 1,\\\label{d1Fn0}
E_0^{n+1,0}&:\quad (\delta_{n+1}F_{0,n})(\delta_{n}F_{1,n-1})...(\delta_1F_{n,0}) = 1.
\end{align}
Here, $F$ is decomposed into cochains $F_{i,n-i}\in E_0^{i,n-i}$ for $0\le i\le n$ at different positions of the LHS spectral sequence. And $\delta_i$'s are the cochain-level differentials, with the explicit expressions shown in the next subsection. We note that all the $F_{i,n-i}$ are cochains and all the obstruction functions are cochain-level equations.

To work out the expression for the $n$-cocycle $F$, we have to solve the obstructions one after another. The first obstruction Eq.~\eqref{d0F0n}, for example, means that $F_{0,n}$ is a cocycle with respect to $A$ due to $d_0=d_A$. After choosing a solution $F_{0,n}$ of it, we can now try to solve the second obstruction Eq.~\eqref{d0F1n-1} for another cochain $F_{1,n-1}$. If we obtain a set of cochains $\{F_{i,n-i}\}$ satisfying all these obstruction equations, we can use them to construct an explicit $n$-cocycle $F$.

\subsubsection{Cochain-level differentials $\delta_i$}

Let us first show some examples of the cochain-level differentials $\delta_i$, which appear in the obstruction equations. The differential $\delta_2$ in LHS spectral sequence is basically the slant product of $\nu=\nu_2$, which is the 2-cocycle in $\H^2[G,A]$ specifying the short exact sequence of $\tilde G$. For example, when acting on $F_{0,1}\in E_0^{0,1}$, the result $\delta_2F_{0,1}\in E_0^{2,0}$ is [see Eq.~(\ref{dF20})]
\begin{align}
\delta_2F_{0,1} = \iota_{[-\nugh{g,h}]}\left({}^{\ag{1}{gh}}\!F\right)^\cdot
=\frac{1}{\left({}^{\ag{1}{gh}}\!F\right)^{\nu(\mb{g,h})}}
=\frac{1}{{}^{\ag{1}{gh}}\!{\left[F^{\gia{gh}{\nu(\mb{g,h})}}\right]}},
\end{align}
where we used the action of $G$ on $F$ in Eq.~(\ref{Gaction}) in the last step.

The differential $\delta_3$ is roughly the slant product of $(\nu_2\otimes\nu_2)(\Delta_1)$. For example, $\delta_3F_{0,2}\in E_0^{3,0}$ in Eq.~(\ref{dF30}) is
\begin{align}\nonumber
\delta_3F_{0,2} &= \iota_{(\nu\otimes\nu)(-\Delta_1(\mb{g,h,k}))}\left({}^{\ag{1}{ghk}}\!F\right)^{\cdot,\cdot}
=\left({}^{\ag{1}{ghk}}\!F\right)^{(\nu\otimes\nu) [(\mb{g,hk})\otimes\ga{g}{(\mb{h,k})} -(\mb{gh,k})\otimes(\mb{g,h})]}\\
&
=\frac{\left({}^{\ag{1}{ghk}}\!F\right)^{\nu(\mb{g,hk}),\ga{g}{\nu(\mb{h,k})}}}{\left({}^{\ag{1}{ghk}}\!F\right)^{\nu(\mb{gh,k}),\nu(\mb{g,h})}}
=\frac{{}^{\ag{1}{ghk}}\!\!\left[F^{\gia{ghk}{\nu(\mb{g,hk})},\gia{hk}{\nu(\mb{h,k})}}\right]}{{}^{\ag{1}{ghk}}\!\!\left[F^{\gia{ghk}{\nu(\mb{gh,k})},\gia{ghk}{\nu(\mb{g,h})}}\right]}
=
\frac{\vcenter{\hbox{\includegraphics[page=1]{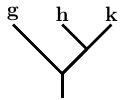}}}}{\vcenter{\hbox{\includegraphics[page=2]{Fig_combined.pdf}}}}
.
\end{align}
It has $2!=2$ terms just as $\nu_2\cup_1\nu_2$ in Eq.~(\ref{n2cup1n2}). In the last step, we used binary trees to represent $F$ terms. For example, the tree in the numerator represents $\left({}^{\ag{1}{ghk}}\!F\right)^{\nu(\mb{g,hk}),\ga{g}{\nu(\mb{h,k})}}$. It has two vertices from top to bottom. The first vertex is understood as $\ga{g}{\nu(\mb{h,k})}$ with two coming edges labelled by $\mb{h}$ and $\mb{k}$. The second one is $\nu(\mb{g,hk})$ with coming edges $\mb{g}$ and $\mb{hk}$. They are put into the variables of $\left({}^{\ag{1}{ghk}}\!F\right)^{\cdot,\cdot}$ from right to left.

Similarly, the fourth differential $\delta_4$ is roughly the slant product of $(\nu_2\otimes\nu_2\otimes\nu_2) [(\Delta_1\otimes\operatorname{id.})\Delta_1]$ [see also $(\nu_2\cup_1\nu_2)\cup_1\nu_2$ in Eq.~(\ref{cup1_3nu})]. For instance, $\delta_4F_{0,3}\in E_0^{4,0}$ in Eq.~(\ref{dFghkl}) is
\begin{align}\nonumber
\delta_4F_{0,3}
&= \iota_{(\nu\otimes\nu\otimes\nu)[(-(\Delta_1\otimes\operatorname{id.})\Delta_1)(\mb{g,h,k,l})]} \left({}^{\ag{1}{ghkl}}\!F\right)^{\cdot,\cdot,\cdot}\\
&=
\frac{\left({}^{\ag{1}{ghkl}}\!F\right)^{\nu(\mb{ghk,l}),\nu(\mb{gh,k}),\nu(\mb{g,h})}}{\left({}^{\ag{1}{ghkl}}\!F\right)^{\nu(\mb{gh,kl}),\ga{gh}{\nu(\mb{k,l})},\nu(\mb{g,h})}}
\frac{\left({}^{\ag{1}{ghkl}}\!F\right)^{\nu(\mb{g,hkl}),\ga{g}{\nu(\mb{hk,l})},\ga{g}{\nu(\mb{h,k})}}}{\left({}^{\ag{1}{ghkl}}\!F\right)^{\nu(\mb{ghk,l}),\nu(\mb{g,hk}),\ga{g}{\nu(\mb{h,k})}}}
\frac{\left({}^{\ag{1}{ghkl}}\!F\right)^{\nu(\mb{gh,kl}),\nu(\mb{g,h}),\ga{gh}{\nu(\mb{k,l})}}}{\left({}^{\ag{1}{ghkl}}\!F\right)^{\nu(\mb{g,hkl}),\ga{g}{\nu(\mb{h,kl})},\ga{gh}{\nu(\mb{k,l})}}}\\
&=
\frac{\vcenter{\hbox{\includegraphics[page=3]{Fig_combined.pdf}}}}{\vcenter{\hbox{\includegraphics[page=4]{Fig_combined.pdf}}}}
\frac{\vcenter{\hbox{\includegraphics[page=5]{Fig_combined.pdf}}}}{\vcenter{\hbox{\includegraphics[page=6]{Fig_combined.pdf}}}}
\frac{\vcenter{\hbox{\includegraphics[page=7]{Fig_combined.pdf}}}}{\vcenter{\hbox{\includegraphics[page=8]{Fig_combined.pdf}}}}
.
\end{align}
The three fractions correspond to the three terms of $\nu_3\cup_1\nu_2$ in Eq.~(\ref{n3cup1n2}). Each fraction can be further expressed as two terms using $\nu_3=\nu_2\cup_1\nu_2$ from Eq.~(\ref{n2cup1n2}). In the last step, we again used binary trees to represent $3!=6$ $F$ terms. The final result is similar to the six $F$ terms appearing in the $\H^4[G,\U]$ obstruction function of 2+1D SET classifications \cite{Chen2014,SET}. It is not a surprise since we can also use LHS spectral sequence to understand symmetry-enriched gauge theories as discussed in the main text of the paper.

The differential $\delta_5$ is more complicated. It is the slant product of $(\nu_2\otimes\nu_2\otimes\nu_2\otimes\nu_2)[(\Delta_1\otimes\operatorname{id.}\otimes\operatorname{id.})(\Delta_1\otimes\operatorname{id.})\Delta_1]$, which has $4!=24$ terms. From Eq.~(\ref{dFghklm}), $\delta_5F_{0,4}\in E_0^{5,0}$ is given by
\begin{align}\label{d5F04}
\delta_5F_{0,4}
&=
\iota_{(\nu\otimes\nu\otimes\nu\otimes\nu)[-((\Delta_1\otimes\operatorname{id.}\otimes\operatorname{id.})(\Delta_1\otimes\operatorname{id.})\Delta_1)(\mb{g,h,k,l,m})]}{\tilde F}^{\cdot,\cdot,\cdot,\cdot}\\\nonumber
&=
\frac{{\tilde F}^{\nu(\mb{ghkl,m}),\nu(\mb{ghk,l}),\nu(\mb{gh,k}),\nu(\mb{g,h})}
{\tilde F}^{\nu(\mb{gh,klm}),\ga{gh}{\nu(\mb{kl,m})},\ga{gh}{\nu(\mb{k,l})},\nu(\mb{g,h})}
{\tilde F}^{\nu(\mb{ghk,lm}),\nu(\mb{gh,k}),\ga{ghk}{\nu(\mb{l,m})},\nu(\mb{g,h})}}
{{\tilde F}^{\nu(\mb{ghk,lm}),\ga{ghk}{\nu(\mb{l,m})},\nu(\mb{gh,k}),\nu(\mb{g,h})}
{\tilde F}^{\nu(\mb{ghkl,m}),\nu(\mb{gh,kl}),\ga{gh}{\nu(\mb{k,l})},\nu(\mb{g,h})}
{\tilde F}^{\nu(\mb{gh,klm}),\ga{gh}{\nu(\mb{k,lm})},\ga{ghk}{\nu(\mb{l,m})},\nu(\mb{g,h})}}\\\nonumber
&\quad\times
\frac{{\tilde F}^{\nu(\mb{ghk,lm}),\ga{ghk}{\nu(\mb{l,m})},\nu(\mb{g,hk}),\ga{g}{\nu(\mb{h,k})}}
{\tilde F}^{\nu(\mb{ghkl,m}),\nu(\mb{g,hkl}),\ga{g}{\nu(\mb{hk,l})},\ga{g}{\nu(\mb{h,k})}}
{\tilde F}^{\nu(\mb{g,hklm}),\ga{g}{\nu(\mb{hk,lm})},\ga{ghk}{\nu(\mb{l,m})},\ga{g}{\nu(\mb{h,k})}}}
{{\tilde F}^{\nu(\mb{ghkl,m}),\nu(\mb{ghk,l}),\nu(\mb{g,hk}),\ga{g}{\nu(\mb{h,k})}}
{\tilde F}^{\nu(\mb{g,hklm}),\ga{g}{\nu(\mb{hkl,m})},\ga{g}{\nu(\mb{hk,l})},\ga{g}{\nu(\mb{h,k})}}
{\tilde F}^{\nu(\mb{ghk,lm}),\nu(\mb{g,hk}),\ga{ghk}{\nu(\mb{l,m})},\ga{g}{\nu(\mb{h,k})}}}
\\\nonumber
&\quad\times
\frac{{\tilde F}^{\nu(\mb{ghkl,m}),\nu(\mb{gh,kl}),\nu(\mb{g,h}),\ga{gh}{\nu(\mb{k,l})}}
{\tilde F}^{\nu(\mb{g,hklm}),\ga{g}{\nu(\mb{hkl,m})},\ga{g}{\nu(\mb{h,kl})},\ga{gh}{\nu(\mb{k,l})}}
{\tilde F}^{\nu(\mb{gh,klm}),\nu(\mb{g,h}),\ga{gh}{\nu(\mb{kl,m})},\ga{gh}{\nu(\mb{k,l})}}}
{{\tilde F}^{\nu(\mb{gh,klm}),\ga{gh}{\nu(\mb{kl,m})},\nu(\mb{g,h}),\ga{gh}{\nu(\mb{k,l})}}
{\tilde F}^{\nu(\mb{ghkl,m}),\nu(\mb{g,hkl}),\ga{g}{\nu(\mb{h,kl})},\ga{gh}{\nu(\mb{k,l})}}
{\tilde F}^{\nu(\mb{g,hklm}),\ga{g}{\nu(\mb{h,klm})},\ga{gh}{\nu(\mb{kl,m})},\ga{gh}{\nu(\mb{k,l})}}}
\\\nonumber
&\quad\times
\frac{{\tilde F}^{\nu(\mb{gh,klm}),\ga{gh}{\nu(\mb{k,lm})},\nu(\mb{g,h}),\ga{ghk}{\nu(\mb{l,m})}}
{\tilde F}^{\nu(\mb{ghk,lm}),\nu(\mb{g,hk}),\ga{g}{\nu(\mb{h,k})},\ga{ghk}{\nu(\mb{l,m})}}
{\tilde F}^{\nu(\mb{g,hklm}),\ga{g}{\nu(\mb{h,klm})},\ga{gh}{\nu(\mb{k,lm})},\ga{ghk}{\nu(\mb{l,m})}}}
{{\tilde F}^{\nu(\mb{ghk,lm}),\nu(\mb{gh,k}),\nu(\mb{g,h}),\ga{ghk}{\nu(\mb{l,m})}}
{\tilde F}^{\nu(\mb{g,hklm}),\ga{g}{\nu(\mb{hk,lm})},\ga{g}{\nu(\mb{h,k})},\ga{ghk}{\nu(\mb{l,m})}}
{\tilde F}^{\nu(\mb{gh,klm}),\nu(\mb{g,h}),\ga{gh}{\nu(\mb{k,lm})},\ga{ghk}{\nu(\mb{l,m})}}}\\\nonumber
&=
\frac{\vcenter{\hbox{\includegraphics[page=9]{Fig_combined.pdf}}}
\vcenter{\hbox{\includegraphics[page=11]{Fig_combined.pdf}}}
\vcenter{\hbox{\includegraphics[page=13]{Fig_combined.pdf}}}}
{\vcenter{\hbox{\includegraphics[page=10]{Fig_combined.pdf}}}
\vcenter{\hbox{\includegraphics[page=12]{Fig_combined.pdf}}}
\vcenter{\hbox{\includegraphics[page=14]{Fig_combined.pdf}}}}
\frac{\vcenter{\hbox{\includegraphics[page=15]{Fig_combined.pdf}}}
\vcenter{\hbox{\includegraphics[page=17]{Fig_combined.pdf}}}
\vcenter{\hbox{\includegraphics[page=19]{Fig_combined.pdf}}}}
{\vcenter{\hbox{\includegraphics[page=16]{Fig_combined.pdf}}}
\vcenter{\hbox{\includegraphics[page=18]{Fig_combined.pdf}}}
\vcenter{\hbox{\includegraphics[page=20]{Fig_combined.pdf}}}}
\frac{\vcenter{\hbox{\includegraphics[page=21]{Fig_combined.pdf}}}
\vcenter{\hbox{\includegraphics[page=23]{Fig_combined.pdf}}}
\vcenter{\hbox{\includegraphics[page=25]{Fig_combined.pdf}}}}
{\vcenter{\hbox{\includegraphics[page=22]{Fig_combined.pdf}}}
\vcenter{\hbox{\includegraphics[page=24]{Fig_combined.pdf}}}
\vcenter{\hbox{\includegraphics[page=26]{Fig_combined.pdf}}}}
\frac{\vcenter{\hbox{\includegraphics[page=27]{Fig_combined.pdf}}}
\vcenter{\hbox{\includegraphics[page=29]{Fig_combined.pdf}}}
\vcenter{\hbox{\includegraphics[page=31]{Fig_combined.pdf}}}}
{\vcenter{\hbox{\includegraphics[page=28]{Fig_combined.pdf}}}
\vcenter{\hbox{\includegraphics[page=30]{Fig_combined.pdf}}}
\vcenter{\hbox{\includegraphics[page=32]{Fig_combined.pdf}}}}
.
\end{align}
where we used the abbreviation $\tilde F:=\left({}^{\ag{1}{ghklm}}F\right)$. The four fractions of the above equation correspond to the four terms of $\nu_4\cup_1\nu_2$ in Eq.~(\ref{n4cup1n2}). Each fraction itself is then a product of six $F$ terms using $\nu_4=(\nu_2\cup_1\nu_2)\cup_1\nu_2$.

Form the above expressions for several lower degree differentials, it is natural to expect that $\delta_iF_{0,i-1}\in E_0^{i,0}$ ($i\ge 2$) has the following expression
\begin{align}
(\delta_iF_{0,i-1})(\mb{g}_1,...,\mb{g}_i)
=
\iota_{(\nu\otimes...\otimes\nu)[-((\Delta_1\otimes\operatorname{id.}\otimes...\otimes\operatorname{id.})...(\Delta_1\otimes\operatorname{id.})\Delta_1)(\mb{g}_1,...,\mb{g}_i)]} \left({}^{1_{\mb{g}_1...\mb{g}_i}}\!F\right)^{\cdot,...,\cdot},
\end{align}
where there are $(i-1)$ $\nu$'s and $(i-2)$ $\Delta_1$'s in the subscript of slant product. When $(\mb{g}_1,...,\mb{g}_i)$ is fixed, the slant product sends a $(i-1)$-cochain $F_{0,i-1}$ of $A$ to a 0-cochain of $A$ in $E_i^{i,0}$. This map is usually called transgression in mathematics, as it maps a $A$ cochain to a $G$ cochain.

The above result can be easily generalized to $\delta_iF_{0,q}\in E_0^{i,q-i+1}$ for $q> i-1$: the slant product will send a $q$-cochain $F_{0,q}$ to a $(q-i+1)$-cochain of $A$. So we have
\begin{align}\nonumber\label{diF0q}
&\quad (\delta_iF_{0,q})(a_1,...,a_{q-i+1},\mb{g}_1,...,\mb{g}_i)\\
&=
\left[\iota_{(\nu\otimes...\otimes\nu)[(-1)^{q-i}((\Delta_1\otimes\operatorname{id.}\otimes...\otimes\operatorname{id.})...(\Delta_1\otimes\operatorname{id.})\Delta_1)(\mb{g}_1,...,\mb{g}_i)]} \left({}^{1_{\mb{g}_1...\mb{g}_i}}\!F\right)^{\cdot,...,\cdot}\right]^{a_1,...,a_{q-i+1}}.
\end{align}
For $q<i-1$, the differential $\delta_i$ is trivial as $E_0^{i,q-i+1}=0$.

We can further generalize the result to $\delta_iF_{p,q}\in E_0^{p+i,q-i+1}$ for $p\ge 0$ and $q\ge i-1$ (note that $\delta_iF_{p,q}=0$ if $p<0$ or $q<i-1$). From the examples calculated in later sections of this appendix, we can summarize the most general differential $\delta_iF_{p,q}\in E_0^{p+i,q-i+1}$ as
\begin{align}\nonumber\label{diFpq}
&\quad (\delta_iF_{p,q})(a_1,\dots,a_{q-i+1},\mb{g}_1,\dots,\mb{g}_{p+i})\\
&=
\left[\iota_{(\nu\otimes...\otimes\nu)[(-1)^{q-i}((\Delta_1\otimes\operatorname{id.}\otimes...\otimes\operatorname{id.})...(\Delta_1\otimes\operatorname{id.})\Delta_1)(\mb{g}_1,...,\mb{g}_i)]} \left({}^{1_{\mb{g}_1...\mb{g}_i}}\!F\right)^{\cdot,...,\cdot,1_{\mb{g}_{i+1}},...,1_{\mb{g}_{p+i}}}\right]^{a_1,...,a_{q-i+1}}.
\end{align}
We have $p+i$ group elements $\mb{g}_{1},...,\mb{g}_{p+i}$ of $G$ in the argument of $\delta_iF_{p,q}$. The first $i$ of them appear in the subscript of slant product just as the $p=0$ result Eq.~(\ref{diF0q}). The last $p$ of them are put into the last $p$ variables $1_{\mb{g}_{i+1}},...,1_{\mb{g}_{p+i}}$ in the argument of $F_{p,q}$, which are fixed throughout the slant product.

Let us consider some simple examples for differentials other than transgressions. Here, we list three examples of $\delta_2F_{2,2}\in E_0^{4,1}$, $\delta_3F_{0,3}\in E_0^{3,1}$ and $\delta_3F_{1,3}\in E_0^{4,1}$:
\begin{align}
(\delta_2F_{2,2})(a,\mb{h,k,l,m})\label{d2F22}
&=
\left[\iota_{\nu(\mb{h,k})} \left({}^{\ag{1}{hk}}F\right)^{\cdot,\cdot,\ag{1}{l},\ag{1}{m}}\right]^a
=
\frac{\left({}^{\ag{1}{hk}}F\right)^{a,\nu(\mb{h,k}),\ag{1}{l},\ag{1}{m}}}{\left({}^{\ag{1}{hk}}F\right)^{\nu(\mb{h,k}),a,\ag{1}{l},\ag{1}{m}}},\\
(\delta_3F_{0,3})(a,\mb{h,k,l})
&=
\left[\iota_{(\nu\otimes\nu)[\Delta_1(\mb{h,k,l})]} \left({}^{\ag{1}{hkl}}\!F\right)^{\cdot,\cdot,\cdot}\right]^{a}
=
\frac{\left[\iota_{\nugh{hk,l},\nugh{h,k}} \left({}^{\ag{1}{hkl}}\!F\right)^{\cdot,\cdot,\cdot}\right]^a}{\left[\iota_{\nugh{h,kl},\ga{h}{\nugh{k,l}}} \left({}^{\ag{1}{hkl}}\!F\right)^{\cdot,\cdot,\cdot}\right]^a},\\
(\delta_3F_{1,3})(a,\mb{h,k,l,m})
&=
\left[\iota_{(\nu\otimes\nu)(\Delta_1(\mb{h,k,l}))} \left({}^{\ag{1}{hkl}}F\right)^{\cdot,\cdot,\cdot,\ag{1}{m}}\right]^a
=
\frac{\left[\iota_{\ga{}{\ga{}{\nu(\mb{hk,l})},\nu(\mb{h,k})}}\left({}^{\ag{1}{hkl}}F\right)^{\cdot,\cdot,\cdot,\ag{1}{m}}\right]^{a}}{\left[\iota_{\ga{}{\nu(\mb{h,kl})},\ga{h}{\nu(\mb{k,l})}}\left({}^{\ag{1}{hkl}}F\right)^{\cdot,\cdot,\cdot,\ag{1}{m}}\right]^{a}}.
\end{align}
They are three special cases of Eqs.~(\ref{diF0q}) and (\ref{diFpq}). For $\delta_2F_{2,2}$ in Eq.~(\ref{d2F22}), the slant product of $\nu(\mb{h,k})$ only involves the first two variables of $\left({}^{\ag{1}{hk}}F\right)^{\cdot,\cdot,\ag{1}{l},\ag{1}{m}}$, with the last two variables $\ag{1}{l}$ and $\ag{1}{m}$ fixed. This is true for all $\delta_iF_{p,q}$ with $p>0$.

\subsubsection{Cocycle-level differentials $d_i$}

As summarized in Appendix.~\ref{sec:obs}, all the obstruction conditions obtained in this appendix are expressed using the cochain-level differentials $\delta_i$. One natural question is how to relate them to the usual cocycle-level differentials $d_i: E_i^{p,q} \rightarrow E_i^{p+i,q-i+1}$ reviewed in Appendix.~\ref{sec:ss}.

The answer is also related to solving the obstruction conditions Eqs.~\eqref{d0F0n}-\eqref{d1Fn0} one by one. To compute the cocycle-level differential $d_iF_{p,q}$ ($p+q=n$), we can set the first $p$ cochains of degree-$n$ to be trivial:
\begin{align}
F_{0,n}=F_{1,n-1}=...=F_{p-1,q+1}=1.
\end{align}
With these conditions, the first $p$ obstruction equations in Eqs.~\eqref{d0F0n}-\eqref{d1Fn0} become trivial. The nontrivial equations are
\begin{align}
E_0^{p,q+1}&: \quad (\delta_0F_{p,q})=1,\\
E_0^{p+1,q}&: \quad (\delta_1F_{p,q})(\delta_0F_{p+1,q-1})=1,\\
E_0^{p+2,q-1}&: \quad (\delta_2F_{p,q})(\delta_1F_{p+1,q-1})(\delta_0F_{p+2,q-2})=1,\\
&\quad\quad...\\\label{diFpq_}
E_0^{p+i,q-i+1}&: \quad (\delta_i F_{p,q})(\delta_{i-1} F_{p+1,q-1})...(\delta_0F_{p+i,q-i})=1,\\
&\quad\quad...\\
E_0^{n,1}&: \quad (\delta_{n-p} F_{p,q})(\delta_{n-p-1} F_{p+1,q-1})...(\delta_0F_{n,0})=1,\\
E_0^{n+1,0}&: \quad (\delta_{n-p+1} F_{p,q})(\delta_{n-p} F_{p+1,q-1})...(\delta_1F_{n,0})=1.
\end{align}
For a given $F_{p,q}$, we have to solve the above obstruction conditions one after another. After obtaining all the cochains $F_{p',n-p'}$ for $p\le p'\le p+i-1$ satisfying the first $i$ equations, the cocycle-level differential $d_iF_{p,q}$ is defined to be
\begin{align}
d_i F_{p,q} := (\delta_i F_{p,q})(\delta_{i-1} F_{p+1,q-1})...(\delta_1F_{p+i-1,q-i+1}).
\end{align}
Besides the cochain-level differential $\delta_iF_{p,q}$ for the first term in the obstruction condition Eq.~\eqref{diFpq_}, it contains many other lower degree cochain-level differentials. All of them may contribute to the cocycle-level differential $d_i F_{p,q}$.

In the last section \ref{app:example} of this appendix, we show how to obtain the cocycle-level differentials for the example of $A = \Z_n$ in degree 3. The explicit example illustrates that the lower degree cochain-level differentials contribute to the cocycle-level differential.

\subsection{LHS spectral sequence for degree-1}
\label{App:H1}

Let us consider the 1-cocycle of $\tilde G$ in LHS spectral sequence. Since $\H[\tilde G,\U]$ is the 1-dimensional representation of $\tilde G$, we use the following ansatz of 1-cocycle:
\begin{align}\label{F1}
F^{\ag{a}{g}}
=
\underbrace{
{}^{\ag{1}{g}}{\left(F^{\gia{g}{a}}\right)}
}_{E_0^{0,1}}
\times
\underbrace{
F^{\ag{1}{g}}
}_{E_0^{1,0}}
.
\end{align}
The two terms can be understood as located at $E_0^{0,1}$ and $E_0^{1,0}$ in the LHS spectral sequence, respectively. From this expression and the the definition of group cohomology differential, $dF$ can be calculated directly as
\begin{align}\nonumber\label{dF1}
(dF)^{\ag{a}{g},\ag{b}{h}}
&=
\frac{{}^{\ag{a}{g}}\!\left(F^{\ag{b}{h}}\right) \times F^{\ag{a}{g}}}{F^{\ag{a}{g}\times\ag{b}{h}}}\\\nonumber
&=
{}^{\ag{\nu(\mb{g,h})}{gh}}\!\left[(dF)^{\gia{gh}{a},\gia{h}{b}}\right]
\times
{}^{\ag{1}{gh}}\!\left[(dF)^{\gia{gh}{\nu(\mb{g,h})},\gia{gh}{a}\gia{h}{b}}\right]
\times
{}^{\ag{1}{g}}\!\left[\left(dF\right)^{\gia{g}{a},\ag{1}{h}}\right]
\times
\left(dF\right)^{\ag{1}{g},\ag{1}{h}}\\
&=
\underbrace{
{}^{\nu(\mb{g,h})}\!\left[\left({}^{\ag{1}{gh}}\!(dF)\right)^{a,\ga{g}{b}}\right]
\times
\left[{}^{\ag{1}{gh}}\!(dF)\right]^{\nu(\mb{g,h}),a\ga{g}{b}}
}_{E_0^{0,2}}
\times
\underbrace{
\left[{}^{\ag{1}{g}}\!\left(dF\right)\right]^{a,\ag{1}{h}}
}_{E_0^{1,1}}
\times
\underbrace{
\left(dF\right)^{\ag{1}{g},\ag{1}{h}}
}_{E_0^{2,0}}.
\end{align}
The three terms in the above equation correspond to terms in $E_0^{0,2}$, $E_0^{1,1}$ and $E_0^{2,0}$, respectively.

\begin{figure}[ht]
\begin{sseq}[grid=crossword, entrysize=8mm, labelstep=1]{0...2}{0...2}
\ssmoveto{0}{1} \ssdrop[]{F_{0,1}}
\ssmoveto{1}{0} \ssdrop[]{F_{1,0}}
\ssmoveto{0}{2} \ssdropbull \ssdroplabel[RD]{\delta_0}
\ssmoveto{1}{1} \ssdropbull
\ssmoveto{2}{0} \ssdropbull
\ssmoveto{0}{1} \ssarrow{0}{1}
\end{sseq}
\quad\quad
\begin{sseq}[grid=crossword, entrysize=8mm, labelstep=1]{0...2}{0...2}
\ssmoveto{0}{1} \ssdrop[]{F_{0,1}}
\ssmoveto{1}{0} \ssdrop[]{F_{1,0}}
\ssmoveto{0}{2} \ssdropbull
\ssmoveto{1}{1} \ssdropbull \ssdroplabel[LU,color=red]{\delta_1} \ssdroplabel[RD]{\delta_0}
\ssmoveto{2}{0} \ssdropbull
\ssmoveto{1}{0} \ssarrow{0}{1}
\ssmoveto{0}{1} \ssarrow[color=red]{1}{0}
\end{sseq}
\quad\quad
\begin{sseq}[grid=crossword, entrysize=8mm, labelstep=1]{0...2}{0...2}
\ssmoveto{0}{1} \ssdrop[]{F_{0,1}}
\ssmoveto{1}{0} \ssdrop[]{F_{1,0}}
\ssmoveto{0}{2} \ssdropbull
\ssmoveto{1}{1} \ssdropbull \ssdroplabel[D,color=blue]{\delta_2}
\ssmoveto{2}{0} \ssdropbull \ssdroplabel[LD,color=red]{\delta_1}
\ssmoveto{0}{1} \ssarrow[color=blue]{2}{-1}
\ssmoveto{1}{0} \ssarrow[color=red]{1}{0}
\end{sseq}
\caption{LHS spectral sequence for degree-1.}
\label{fig:LHS1}
\end{figure}

From Eq.~(\ref{dF1}), it is easy to see that the cocycle condition $(dF)^{\ag{a}{g},\ag{b}{h}}=1$ for 1-cochain of $\tilde G$ is equivalent to
\begin{align}\label{obs1}
(dF)^{\ag{a}{g},\ag{b}{h}}=1\quad (\forall \ag{a}{g},\ag{b}{h}\in \tilde G)
\ \Longleftrightarrow\ 
\begin{cases}
(dF)^{a,b} =1 & (\forall a,b\in A),\\
(dF)^{a,\ag{1}{h}} =1 & (\forall a\in A, \ \ \forall \mb{h}\in G),\\
(dF)^{\ag{1}{g},\ag{1}{h}} =1 & (\forall \mb{g,h}\in G).
\end{cases}
\end{align}
The ``$\Rightarrow$'' direction is obvious, since all the right-hand-side equations are special cases of the left-hand-side equation. The ``$\Leftarrow$'' direction is also true, for $(dF)^{\ag{a}{g},\ag{b}{h}}$ can be expressed as the product of some $(dF)^{a,b}$, $(dF)^{a,\ag{1}{h}}$ and $(dF)^{\ag{1}{g},\ag{1}{h}}$ as in Eq.~(\ref{dF1}). The equations on the right-hand-side of Eq.~(\ref{obs1}) can be understood as conditions at different locations in the LHS spectral sequence (see the three figures in Fig.~\ref{fig:LHS1}). We expect that the equation at $E_0^{i,2-i}$ has the form that the product of several different differentials landing at $E_0^{i,2-i}$ is 1:
\begin{align}\label{E02}
E_0^{0,2}&:\quad (\delta_0 F_{0,1}) = 1,\\\label{E11}
E_0^{1,1}&:\quad (\delta_1 F_{0,1})(\delta_0 F_{1,0}) = 1,\\\label{E20}
E_0^{2,0}&:\quad (\delta_2 F_{0,1})(\delta_1F_{1,0}) = 1.
\end{align}
In fact, we can unpack the right-hand-side of Eq.~(\ref{obs1}) by using Eq.~(\ref{F1}) and the definition of differential. The explicit obstruction functions are
\begin{align}\label{dF02}
\left(dF\right)^{a,b}
&=
(d_A F^\cdot)^{a,b}
=
\underbrace{\frac{{}^{a}\!\left(F^b\right) \times F^a}{F^{ab}}}_{\delta_0F_{0,1}}
= 1,\\\label{dF11}
\left(dF\right)^{a,\ag{1}{h}}
&=
\frac{1}{(d_GF^a)^{\ag{1}{h}}} \times \left(d_AF^{\ag{1}{h}}\right)^a
=
\underbrace{\frac{F^{a}}{{}^{\ag{1}{h}}\!\left(F^{\gia{h}{a}}\right)}}_{\delta_1F_{0,1}}
\underbrace{\frac{{}^a\!\left(F^{\ag{1}{h}}\right)}{F^{\ag{1}{h}}}}_{\delta_0F_{1,0}}
=1,\\\label{dF20}
\left(dF\right)^{\ag{1}{g},\ag{1}{h}}
&=
\frac{1}{\iota_{\nugh{g,h}}\left({}^{\ag{1}{gh}}\!F\right)^\cdot}
\times (d_GF^\cdot)^{\ag{1}{g},\ag{1}{h}}
=
\underbrace{\frac{1}{{}^{\ag{1}{gh}}\!\left[F^{\gia{gh}{\nu(\mb{g,h})}}\right]}}_{\delta_2F_{0,1}}
\underbrace{\frac{{}^{\ag{1}{g}}\!\left(F^{\ag{1}{h}}\right) \times F^{\ag{1}{g}}}{F^{\ag{1}{gh}}}}_{\delta_1F_{1,0}}
=1.
\end{align}
Therefore, the 1-cochain $F^{\ag{a}{g}}$ in Eq.~(\ref{F1}) is a 1-cocycle of $\tilde G$ if and only if $F^a$ and $F^{\ag{1}{g}}$ satisfy Eqs.~(\ref{dF02})-(\ref{dF20}). These equations are obstruction functions at cochain levels. We have to solve them one by one to obtain the explicit 1-cocycle of $\tilde G$.

\subsection{LHS spectral sequence for degree-2}
\label{App:H2}

The ansatz for the 2-cocycle of $\tilde G$ is
\begin{align}\label{F2}
F^{\ag{a}{g},\ag{b}{h}} =
\underbrace{{}^{\ag{\nu(\mb{g,h})}{gh}}\!\left[F^{\gia{gh}{a},\gia{h}{b}}\right]
\times {}^{\ag{1}{gh}}\!\left[F^{\gia{gh}{\nu(\mb g,\mb h)},\gia{gh}{a}\gia{h}{b}}\right]}_{E^{0,2}}
\times \underbrace{{}^{\ag{1}{g}}{\left[F^{\gia{g}{a},\ag{1}{h}}\right]}}_{E^{1,1}}
\times \underbrace{F^{\ag{1}{g},\ag{1}{h}}}_{E^{2,0}}.
\end{align}
This expression comes from the idea of \emph{categorification}. We can think of 1-cocycle $F$ in Eq.~(\ref{F1}) as object, and $dF$ in Eq.~(\ref{dF1}) as morphism. To obtain the expression of 2-cocycle Eq.~(\ref{F2}), we simply replace the 2-coboundary morphism $dF$ in Eq.~(\ref{dF1}) by a new 2-cocycle object $F$. The result is in some sense similar to gauge fixing, as the 2-cocycle is decomposed into terms located at different positions $E_0^{i,2-i}$ of the LHS spectral sequence. The new 2-cocycle object $F$ in Eq.~(\ref{F2}) should satisfy new morphism equation in one higher dimension. Using the definition of group cohomology differential, $dF$ can be calculated directly as
\begin{align}\nonumber\label{dF2}
(dF)^{\ag{a}{g},\ag{b}{h},\ag{c}{k}}
&=
\bigg\{
{}^{\nu(\mb{g,h})\nu(\mb{gh,k})}\! \left[\left({}^{\ag{1}{ghk}}\! (dF)\right)^{a,\ga{g}{b},\ga{gh}{c}}\right]
\times {}^{\nugh{gh,k}}\! \left[\iota_{\nu(\mb{g,h})}\! \left({}^{\ag{1}{ghk}}(dF)\right)^{\cdot,a\ga{g}{b},\ga{gh}{c}}\right]\\\nonumber
&\quad
\underbrace{
\quad\times
{}^{\nu(\mb{g,hk})}\! \left\{\left[\iota_{[\ga{g}{\nu(\mb{h,k})}]} \left({}^{\ag{1}{ghk}}\! (dF)\right)^{\cdot,\cdot,\ga{g}{b}\ga{gh}{c}}\right]^a\right\}
\times
\left[\iota_{(\nu\otimes\nu)[-\Delta_1(\mb{g,h,k})]} \left({}^{\ag{1}{ghk}}\! (dF)\right)^{\cdot,\cdot,a\ga{g}{b}\ga{gh}{c}}\right]
\bigg\}
}_{E_0^{0,3}}
\\
&\quad\times
\underbrace{
{}^{\ag{\nugh{g,h}}{gh}}{\left[(dF)^{\gia{gh}{a},\gia{h}{b},\ag{1}{k}}\right]}
\times {}^{\ag{1}{gh}}{\left[(dF)^{\gia{gh}{\nugh{g,h}},\gia{gh}{a}\gia{h}{b},\ag{1}{k}}\right]}
}_{E_0^{1,2}}
\times
\underbrace{
{}^{\ag{1}{g}}\!\left[(dF)^{\gia{g}{a},\ag{1}{h},\ag{1}{k}}\right]
}_{E_0^{2,1}}
\times
\underbrace{
(dF)^{\ag{1}{g},\ag{1}{h},\ag{1}{k}}
}_{E_0^{3,0}}.
\end{align}
We note that the four terms in the above equation correspond to terms in $E_0^{0,3}$, $E_0^{1,2}$, $E_0^{2,1}$ and $E_0^{3,0}$, respectively.

\begin{figure}[ht]
\begin{sseq}[grid=crossword, entrysize=8mm, labelstep=1]{0...3}{0...3}
\ssmoveto{0}{2} \ssdrop[]{F_{0,2}}
\ssmoveto{1}{1} \ssdrop[]{F_{1,1}}
\ssmoveto{2}{0} \ssdrop[]{F_{2,0}}
\ssmoveto{0}{3} \ssdropbull \ssdroplabel[RD]{d_0}
\ssmoveto{1}{2} \ssdropbull
\ssmoveto{2}{1} \ssdropbull
\ssmoveto{3}{0} \ssdropbull
\ssmoveto{0}{2} \ssarrow{0}{1}
\end{sseq}
\quad\quad
\begin{sseq}[grid=crossword, entrysize=8mm, labelstep=1]{0...3}{0...3}
\ssmoveto{0}{2} \ssdrop[]{F_{0,2}}
\ssmoveto{1}{1} \ssdrop[]{F_{1,1}}
\ssmoveto{2}{0} \ssdrop[]{F_{2,0}}
\ssmoveto{0}{3} \ssdropbull
\ssmoveto{1}{2} \ssdropbull \ssdroplabel[LU,color=red]{d_1} \ssdroplabel[RD]{d_0}
\ssmoveto{2}{1} \ssdropbull
\ssmoveto{3}{0} \ssdropbull
\ssmoveto{0}{2} \ssarrow[color=red]{1}{0}
\ssmoveto{1}{1} \ssarrow{0}{1}
\end{sseq}
\quad\quad
\begin{sseq}[grid=crossword, entrysize=8mm, labelstep=1]{0...3}{0...3}
\ssmoveto{0}{2} \ssdrop[]{F_{0,2}}
\ssmoveto{1}{1} \ssdrop[]{F_{1,1}}
\ssmoveto{2}{0} \ssdrop[]{F_{2,0}}
\ssmoveto{0}{3} \ssdropbull
\ssmoveto{1}{2} \ssdropbull \ssdroplabel[D,color=blue]{d_2}
\ssmoveto{2}{1} \ssdropbull \ssdroplabel[LD,color=red]{d_1} \ssdroplabel[RD]{d_0}
\ssmoveto{3}{0} \ssdropbull
\ssmoveto{0}{2} \ssarrow[color=blue]{2}{-1}
\ssmoveto{1}{1} \ssarrow[color=red]{1}{0}
\ssmoveto{2}{0} \ssarrow{0}{1}
\end{sseq}
\quad\quad
\begin{sseq}[grid=crossword, entrysize=8mm, labelstep=1]{0...3}{0...3}
\ssmoveto{0}{2} \ssdrop[]{F_{0,2}}
\ssmoveto{1}{1} \ssdrop[]{F_{1,1}}
\ssmoveto{2}{0} \ssdrop[]{F_{2,0}}
\ssmoveto{0}{3} \ssdropbull
\ssmoveto{1}{2} \ssdropbull \ssdroplabel[LD,color=cyan]{d_3}
\ssmoveto{2}{1} \ssdropbull \ssdroplabel[LD,color=blue]{d_2}
\ssmoveto{3}{0} \ssdropbull \ssdroplabel[LD,color=red]{d_1}
\ssmoveto{0}{2} \ssarrow[color=cyan]{3}{-2}
\ssmoveto{1}{1} \ssarrow[color=blue]{2}{-1}
\ssmoveto{2}{0} \ssarrow[color=red]{1}{0}
\end{sseq}
\caption{LHS spectral sequence for degree-2.}
\label{fig:LHS2}
\end{figure}

Similar to the discussions for 1-cocycles in the previous subsection, the cocycle condition for $F^{\ag{a}{g},\ag{b}{h}}$ as a 2-cocycle of $\tilde G$ is equivalent to
\begin{align}\label{obs2}
(dF)^{\ag{a}{g},\ag{b}{h},\ag{c}{k}}=1\quad (\forall \ag{a}{g},\ag{b}{h},\ag{c}{k}\in \tilde G)
\ \Longleftrightarrow\ 
\begin{cases}
(dF)^{a,b,c} =1 & (\forall a,b,c\in A),\\
(dF)^{a,b,\ag{1}{k}} =1 & (\forall a,b\in A,\ \ \forall \mb{k}\in G),\\
(dF)^{a,\ag{1}{h},\ag{1}{k}} =1 & (\forall a\in A,\ \ \forall \mb{h,k}\in G),\\
(dF)^{\ag{1}{g},\ag{1}{h},\ag{1}{k}} =1 & (\forall \mb{g,h,k}\in G).
\end{cases}
\end{align}
This equivalence is a consequence of Eq.~(\ref{dF2}), which decompose the general $(dF)^{\ag{a}{g},\ag{b}{h},\ag{c}{k}}$ as the product of some $(dF)^{a,b,c}$, $(dF)^{a,b,\ag{1}{k}}$, $(dF)^{a,\ag{1}{h},\ag{1}{k}}$ and $(dF)^{\ag{1}{g},\ag{1}{h},\ag{1}{k}}$. From the LHS spectral sequence, we expect that the right-hand-side equations in Eq.~(\ref{obs2}) have the form:
\begin{align}\label{E03}
E^{0,3}&:\quad (\delta_0 F_{0,2}) = 1,\\\label{E12}
E^{1,2}&:\quad (\delta_1 F_{0,2})(\delta_0 F_{1,1}) = 1,\\\label{E21}
E^{2,1}&:\quad (\delta_2 F_{0,2})(\delta_1F_{1,1})(\delta_0F_{2,0}) = 1,\\\label{E30}
E^{3,0}&:\quad (\delta_3 F_{0,2})(\delta_2F_{1,1})(\delta_1F_{2,0}) = 1.
\end{align}
After substituting Eq.~(\ref{F2}) to the right-hand-side of Eq.~(\ref{obs2}), we obtain the explicit obstructions as
\begin{align}
(dF)^{a,b,c}
&=(d_AF^{\cdot,\cdot})^{a,b,c}
=\underbrace{\frac{{}^a{\left(F^{b,c}\right)} \times F^{a,bc}}{F^{ab,c}\times F^{a,b}}}_{\delta_0F_{0,2}}
=1,\\
(dF)^{a,b,\ag{1}{k}}
&=\left(d_GF^{a,b}\right)^{\ag{1}{k}}
\times (d_A F^{\cdot,\ag{1}{k}})^{a,b}
=
\underbrace{\frac{\left({}^{\ag{1}{k}}F\right)^{a,b}}{F^{a,b}}}_{\delta_1F_{0,2}}
\underbrace{\frac{{}^{a}{(F^{b,\ag{1}{k}})} \times F^{a,\ag{1}{k}}}{F^{ab,\ag{1}{k}}}}_{\delta_0F_{1,1}}
=1,\\\nonumber\label{dF21}
(dF)^{a,\ag{1}{h},\ag{1}{k}}
&=\left[\iota_{\nu(\mb{h,k})}\left({}^{\ag{1}{hk}}\!F\right)^{\cdot,\cdot}\right]^{a}
\times \left[\left(d_GF^{a,\cdot}\right)^{\ag{1}{h},\ag{1}{k}}\right]^{-1}
\times \left(d_AF^{\ag{1}{h},\ag{1}{k}}\right)^a\\
&=\underbrace{\frac{\left({}^{\ag{1}{hk}}\!F\right)^{a,\nu(\mb{h,k})}}{\left({}^{\ag{1}{hk}}\!F\right)^{\nu(\mb{h,k}),a}}}_{\delta_2F_{0,2}}
\underbrace{\frac{F^{a,\ag{1}{hk}}}{{}^{\ag{1}{h}}{\left(F^{\gia{h}{a},\ag{1}{k}}\right)} \times F^{a,\ag{1}{h}}}}_{\delta_1F_{1,1}}
\underbrace{\frac{{}^{a}{\left(F^{\ag{1}{h},\ag{1}{k}}\right)}}{F^{\ag{1}{h},\ag{1}{k}}}}_{\delta_0F_{2,0}}
=1,\\\nonumber\label{dF30}
(dF)^{\ag{1}{g},\ag{1}{h},\ag{1}{k}}\nonumber
&=
\left[\iota_{(\nu\otimes\nu)(-\Delta_1(\mb{g,h,k}))} \left({}^{\ag{1}{ghk}}\!F\right)^{\cdot,\cdot}\right]
\times \left[\iota_{-\nu(\mb{g,h})} \left({}^{\ag{1}{gh}}\!F\right)^{\cdot,\ag{1}{k}}\right]
\times (d_GF^{\cdot,\cdot})^{\ag{1}{g},\ag{1}{h},\ag{1}{k}}\\
&=
\underbrace{\frac{\left({}^{\ag{1}{ghk}}\!F\right)^{\nu(\mb{g,hk}),\ga{g}{\nu(\mb{h,k})}}}{\left({}^{\ag{1}{ghk}}\!F\right)^{\nu(\mb{gh,k}),\nu(\mb{g,h})}}}_{\delta_3F_{0,2}}
\underbrace{\frac{1}{{}^{\ag{1}{gh}}\!\left[F^{\gia{gh}{\nu(\mb{g,h})},\ag{1}{k}}\right]}}_{\delta_2F_{1,1}}
\underbrace{\frac{{}^{\ag{1}{g}}{\left(F^{\ag{1}{h},\ag{1}{k}}\right)} \times F^{\ag{1}{g},\ag{1}{hk}}}{F^{\ag{1}{gh},\ag{1}{k}} \times F^{\ag{1}{g},\ag{1}{h}}}}_{\delta_1F_{2,0}}
=1.
\end{align}
These four equations have the desired form Eq.~(\ref{E03})-(\ref{E30}), and correspond to the four figures in Fig.~\ref{fig:LHS2}, respectively.

\subsection{LHS spectral sequence for degree-3}
\label{App:H3}

Replacing the differential of 2-cocycle $dF$ in Eq.~(\ref{dF2}) by 3-cocycle $F$, we obtain the generic expression for $F^{\ag{a}{g},\ag{b}{h},\ag{c}{k}}$:
\begin{align}\nonumber\label{F3}
F^{\ag{a}{g},\ag{b}{h},\ag{c}{k}}
&=
\bigg\{
{}^{\ag{[\nu(\mb{g,h})\nu(\mb{gh,k})]}{ghk}}\! \left[F^{\gia{ghk}{a},\gia{hk}{b},\gia{k}{c}}\right]
\times {}^{\ag{\nugh{gh,k}}{ghk}}\! \left[F^{\gia{ghk}{\nugh{g,h}},\gia{ghk}{a}\gia{hk}{b},\gia{k}{c}}\right]\\\nonumber
&\quad
\underbrace{
\quad\times
\frac{{}^{\ag{\nu(\mb{g,hk})}{ghk}}\! \left[ F^{\gia{ghk}{a},\gia{hk}{\nugh{h,k}},\gia{hk}{b}\gia{k}{c}}\right]}{{}^{\ag{\nu(\mb{g,hk})}{ghk}}\! \left[ F^{\gia{hk}{\nugh{h,k}},\gia{ghk}{a},\gia{hk}{b}\gia{k}{c}}\right]}
\times
\frac{{}^{\ag{1}{ghk}}\! \left[F^{\gia{ghk}{\nugh{g,hk}},\gia{hk}{\nugh{h,k}},\gia{ghk}{a}\gia{hk}{b}\gia{k}{c}}\right]}{{}^{\ag{1}{ghk}}\! \left[F^{\gia{ghk}{\nugh{gh,k}},\gia{ghk}{\nugh{g,h}},\gia{ghk}{a}\gia{hk}{b}\gia{k}{c}}\right]}
\bigg\}
}_{E_0^{0,3}}
\\
&\quad\times
\underbrace{
{}^{\ag{\nugh{g,h}}{gh}}{\left[F^{\gia{gh}{a},\gia{h}{b},\ag{1}{k}}\right]}
\times {}^{\ag{1}{gh}}{\left[F^{\gia{gh}{\nugh{g,h}},\gia{gh}{a}\gia{h}{b},\ag{1}{k}}\right]}
}_{E_0^{1,2}}
\times
\underbrace{
{}^{\ag{1}{g}}\!\left[F^{\gia{g}{a},\ag{1}{h},\ag{1}{k}}\right]
}_{E_0^{2,1}}
\times
\underbrace{
F^{\ag{1}{g},\ag{1}{h},\ag{1}{k}}
}_{E_0^{3,0}}.
\end{align}
Tedious but straightforward calculations show that the differential $dF$ can be decomposed as
\begin{align}\label{dF3}
(dF)^{\ag{a}{g},\ag{b}{h},\ag{c}{k},\ag{d}{l}}
&=
(dF)|_{E^{0,4}}
\times (dF)|_{E_0^{1,3}}
\times (dF)|_{E_0^{2,2}}
\times (dF)|_{E_0^{3,1}}
\times (dF)|_{E_0^{4,0}},
\end{align}
where the fives terms at different locations in the LHS spectral sequence are given by
\begin{align}
(dF)|_{E_0^{0,4}}\nonumber\label{dF04}
&={}^{\ag{\nu(\mb{g,h}) \nu(\mb{gh,k}) \nu(\mb{ghk,l})}{}}\!\!\left[\left({}^{\ag{1}{ghkl}}(dF)\right)^{a,\ga{g}{b},\ga{gh}{c},\ga{ghk}{d}}\right]
\times
{}^{\ag{\nu(\mb{gh,k}) \nu(\mb{ghk,l})}{}}\!\!\left[\iota_{\nu(\mb{g,h})}\left({}^{\ag{1}{ghkl}}(dF)\right)^{\cdot,a\ga{g}{b},\ga{gh}{c},\ga{ghk}{d}}\right]
\\\nonumber
&\quad\times
{}^{\ag{\nu(\mb{g,hk}) \nu(\mb{ghk,l})}{}}\!\!\left[\left(\iota_{[\ga{g}{\nu(\mb{h,k})}]}\left({}^{\ag{1}{ghkl}}(dF)\right)^{\cdot,\cdot,\ga{g}{b}\ga{gh}{c},\ga{ghk}{d}}\right)^{\!\!a}\,\right]
\times
{}^{\ag{\nu(\mb{g,h}) \nu(\mb{gh,kl})}{}}\!\!\left[\left(\iota_{[\ga{gh}{\nu(\mb{k,l})}]}\left({}^{\ag{1}{ghkl}}(dF)\right)^{\cdot,\cdot,\cdot,\ga{gh}{c}\ga{ghk}{d}}\right)^{\!\!a,\ga{g}{b}}\right]
\\\nonumber
&\quad\times
{}^{\ag{\nu(\mb{ghk,l})}{}}\!\!\left[\iota_{(\nu\otimes\nu)(-\Delta_1(\mb{g,h,k}))}\left({}^{\ag{1}{ghkl}}(dF)\right)^{\cdot,\cdot,a\ga{g}{b}\ga{gh}{c},\ga{ghk}{d}}\right]
\times
{}^{\ag{\nu(\mb{gh,kl})}{}}\!\!\left[\left(\iota_{[\ga{gh}{\nu(\mb{k,l})}]}\left({}^{\ag{1}{ghkl}}(dF)\right)^{\cdot,\cdot,\ga{gh}{c}\ga{ghk}{d}}\right)^{\!\!\nu(\mb{g,h}),a\ga{g}{b}}\right]
\\\nonumber
&\quad\times
{}^{\ag{\nu(\mb{g,hkl})}{}}\!\!\left[\left(\iota_{(\nu\otimes\nu)(-\Delta_1(\mb{h,k,l}))}\left({}^{\ag{1}{ghkl}}(dF)\right)^{\cdot,\cdot,\cdot,\ga{g}{b}\ga{gh}{c}\ga{ghk}{d}}\right)^{\!\!a}\,\right]
\\
&\quad
\times
\left[\iota_{(\nu\otimes\nu\otimes\nu)[((\Delta_1\otimes\operatorname{id.})\Delta_1)(\mb{g,h,k,l})]}\left({}^{\ag{1}{ghkl}}(dF)\right)^{\cdot,\cdot,\cdot,a\ga{g}{b}\ga{gh}{c}\ga{ghk}{d}}\right],
\end{align}
\begin{align}\nonumber\label{dF13}
(dF)|_{E_0^{1,3}}
&={}^{\ag{[\nu(\mb{g,h}) \nu(\mb{gh,k})]}{ghk}}\!\!\left[(dF)^{\gia{ghk}{a},\gia{hk}{b},\gia{k}{c},\ag{1}{l}}\right]
\times
{}^{\ag{\nu(\mb{gh,k})}{ghk}}\!\!\left[(dF)^{\gia{ghk}{\nu(\mb{g,h})},\gia{ghk}{a}\gia{hk}{b},\gia{k}{c},\ag{1}{l}}\right]\\
&\quad\times
\frac{{}^{\ag{\nu(\mb{g,hk})}{ghk}}\!\!\left[(dF)^{\gia{ghk}{a},\gia{hk}{\nu(\mb{h,k})},\gia{hk}{b}\gia{k}{c},\ag{1}{l}}\right]}
{{}^{\ag{\nu(\mb{g,hk})}{ghk}}\!\!\left[(dF)^{\gia{hk}{\nu(\mb{h,k})},\gia{ghk}{a},\gia{hk}{b}\gia{k}{c},\ag{1}{l}}\right]}
\times
\frac{{}^{\ag{1}{ghk}}\!\!\left[(dF)^{\gia{ghk}{\nu(\mb{g,hk})},\gia{hk}{\nu(\mb{h,k})},\gia{ghk}{a}\gia{hk}{b}\gia{k}{c},\ag{1}{l}}\right]}
{{}^{\ag{1}{ghk}}\!\!\left[(dF)^{\gia{ghk}{\nu(\mb{gh,k})},\gia{ghk}{\nu(\mb{g,h})},\gia{ghk}{a}\gia{hk}{b}\gia{k}{c},\ag{1}{l}}\right]},
\end{align}
\begin{align}\label{dF22}
(dF)|_{E_0^{2,2}}
&={}^{\ag{\nu(\mb{g,h})}{gh}}\!\!\left[(dF)^{\gia{gh}{a},\gia{h}{b},\ag{1}{k},\ag{1}{l}}\right]
\times
{}^{\ag{1}{gh}}\!\!\left[(dF)^{\gia{gh}{\nu(\mb{g,h})},\gia{gh}{a}\gia{h}{b},\ag{1}{k},\ag{1}{l}}\right],
\end{align}
\begin{align}\label{dF31}
(dF)|_{E_0^{3,1}}
&={}^{\ag{1}{g}}\!\!\left[(dF)^{\gia{g}{a},\ag{1}{h},\ag{1}{k},\ag{1}{l}}\right],
\end{align}
\begin{align}\label{dF40}
(dF)|_{E_0^{4,0}}
&=(dF)^{\ag{1}{g},\ag{1}{h},\ag{1}{k},\ag{1}{l}}.
\end{align}
As introduced in Section~\ref{sec:HigherCup}, we used the notation of diagonal approximation $\Delta_1$ to simplify the expression Eq.~(\ref{dF04}). The explicit expression for the three terms containing $\Delta_1$ in Eq.~(\ref{dF04}) are
\begin{align}\nonumber
{}^{\ag{\nu(\mb{ghk,l})}{}}\!\!\left[\iota_{(\nu\otimes\nu)(-\Delta_1(\mb{g,h,k}))}\left({}^{\ag{1}{ghkl}}(dF)\right)^{\cdot,\cdot,a\ga{g}{b}\ga{gh}{c},\ga{ghk}{d}}\right]
&=
\frac{{}^{\ag{\nu(\mb{ghk,l})}{}}\!\!\left[\left({}^{\ag{1}{ghkl}}(dF)\right)^{\gia{}{\nu(\mb{g,hk})},\ga{g}{\nu(\mb{h,k})},a\ga{g}{b}\ga{gh}{c},\ga{ghk}{d}}\right]}
{{}^{\ag{\nu(\mb{ghk,l})}{}}\!\!\left[\left({}^{\ag{1}{ghkl}}(dF)\right)^{\gia{}{\nu(\mb{gh,k})},\gia{}{\nu(\mb{g,h})},a\ga{g}{b}\ga{gh}{c},\ga{ghk}{d}}\right]}
\\
&=
\frac{{}^{\ag{\nu(\mb{ghk,l})}{ghkl}}\!\!\left[(dF)^{\gia{ghkl}{\nu(\mb{g,hk})},\gia{hkl}{\nu(\mb{h,k})},\gia{ghkl}{a}\gia{hkl}{b}\gia{kl}{c},\gia{l}{d}}\right]}
{{}^{\ag{\nu(\mb{ghk,l})}{ghkl}}\!\!\left[(dF)^{\gia{ghkl}{\nu(\mb{gh,k})},\gia{ghkl}{\nu(\mb{g,h})},\gia{ghkl}{a}\gia{hkl}{b}\gia{kl}{c},\gia{l}{d}}\right]},
\end{align}
\begin{align}
{}^{\ag{\nu(\mb{g,hkl})}{}}\!\!\left[\left(\iota_{(\nu\otimes\nu)(-\Delta_1(\mb{h,k,l}))}\left({}^{\ag{1}{ghkl}}(dF)\right)^{\cdot,\cdot,\cdot,\ga{g}{b}\ga{gh}{c}\ga{ghk}{d}}\right)^{\!\!a}\right]
&=
\frac{{}^{\ag{\nu(\mb{g,hkl})}{}}\!\!\left[\left(\iota_{\nu(\mb{h,kl}),\ga{h}{\nu(\mb{k,l})}}\left({}^{\ag{1}{ghkl}}(dF)\right)^{\cdot,\cdot,\cdot,\ga{g}{b}\ga{gh}{c}\ga{ghk}{d}}\right)^{\!\!a}\right]}
{{}^{\ag{\nu(\mb{g,hkl})}{}}\!\!\left[\left(\iota_{\nu(\mb{hk,l}),\nu(\mb{h,k})}\left({}^{\ag{1}{ghkl}}(dF)\right)^{\cdot,\cdot,\cdot,\ga{g}{b}\ga{gh}{c}\ga{ghk}{d}}\right)^{\!\!a}\right]},
\end{align}
\begin{align}\nonumber
\left[\iota_{(\nu\otimes\nu\otimes\nu)[((\Delta_1\otimes\operatorname{id.})\Delta_1)(\mb{g,h,k,l})]}\left({}^{\ag{1}{ghkl}}(dF)\right)^{\cdot,\cdot,\cdot,a\ga{g}{b}\ga{gh}{c}\ga{ghk}{d}}\right]
&=
\frac{\left[{}^{\ag{1}{ghkl}}(dF)\right]^{\nu(\mb{gh,kl}),\ga{gh}{\nu(\mb{k,l})},\nu(\mb{g,h}),a\ga{g}{b}\ga{gh}{c}\ga{ghk}{d}}}
{\left[{}^{\ag{1}{ghkl}}(dF)\right]^{\nu(\mb{ghk,l}),\nu(\mb{gh,k}),\nu(\mb{g,h}),a\ga{g}{b}\ga{gh}{c}\ga{ghk}{d}}}\\\nonumber
&\quad\times
\frac{\left[{}^{\ag{1}{ghkl}}(dF)\right]^{\nu(\mb{ghk,l}),\nu(\mb{g,hk}),\ga{gh}{\nu(\mb{h,k})},a\ga{g}{b}\ga{gh}{c}\ga{ghk}{d}}}
{\left[{}^{\ag{1}{ghkl}}(dF)\right]^{\nu(\mb{g,hkl}),\ga{g}{\nu(\mb{hk,l})},\ga{g}{\nu(\mb{h,k})},a\ga{g}{b}\ga{gh}{c}\ga{ghk}{d}}}\\
&\quad\times
\frac{\left[{}^{\ag{1}{ghkl}}(dF)\right]^{\nu(\mb{g,hkl}),\ga{g}{\nu(\mb{h,kl})},\ga{gh}{\nu(\mb{k,l})},a\ga{g}{b}\ga{gh}{c}\ga{ghk}{d}}}
{\left[{}^{\ag{1}{ghkl}}(dF)\right]^{\nu(\mb{gh,kl}),\nu(\mb{g,h}),\ga{gh}{\nu(\mb{k,l})},a\ga{g}{b}\ga{gh}{c}\ga{ghk}{d}}}.
\end{align}

\begin{figure}[ht]
\begin{sseq}[grid=crossword, entrysize=6mm, labelstep=1]{0...4}{0...4}
\ssmoveto{0}{3} \ssdrop[]{F_{0,3}}
\ssmoveto{1}{2} \ssdrop[]{F_{1,2}}
\ssmoveto{2}{1} \ssdrop[]{F_{2,1}}
\ssmoveto{3}{0} \ssdrop[]{F_{3,0}}
\ssmoveto{0}{4} \ssdropbull \ssdroplabel[RD]{d_0}
\ssmoveto{1}{3} \ssdropbull
\ssmoveto{2}{2} \ssdropbull
\ssmoveto{3}{1} \ssdropbull
\ssmoveto{4}{0} \ssdropbull
\ssmoveto{0}{3} \ssarrow{0}{1}
\end{sseq}
\ 
\begin{sseq}[grid=crossword, entrysize=6mm, labelstep=1]{0...4}{0...4}
\ssmoveto{0}{3} \ssdrop[]{F_{0,3}}
\ssmoveto{1}{2} \ssdrop[]{F_{1,2}}
\ssmoveto{2}{1} \ssdrop[]{F_{2,1}}
\ssmoveto{3}{0} \ssdrop[]{F_{3,0}}
\ssmoveto{0}{4} \ssdropbull
\ssmoveto{1}{3} \ssdropbull \ssdroplabel[LU,color=red]{d_1} \ssdroplabel[RD]{d_0}
\ssmoveto{2}{2} \ssdropbull
\ssmoveto{3}{1} \ssdropbull
\ssmoveto{4}{0} \ssdropbull
\ssmoveto{0}{3} \ssarrow[color=red]{1}{0}
\ssmoveto{1}{2} \ssarrow{0}{1}
\end{sseq}
\ 
\begin{sseq}[grid=crossword, entrysize=6mm, labelstep=1]{0...4}{0...4}
\ssmoveto{0}{3} \ssdrop[]{F_{0,3}}
\ssmoveto{1}{2} \ssdrop[]{F_{1,2}}
\ssmoveto{2}{1} \ssdrop[]{F_{2,1}}
\ssmoveto{3}{0} \ssdrop[]{F_{3,0}}
\ssmoveto{0}{4} \ssdropbull
\ssmoveto{1}{3} \ssdropbull \ssdroplabel[D,color=blue]{d_2}
\ssmoveto{2}{2} \ssdropbull \ssdroplabel[LD,color=red]{d_1} \ssdroplabel[RD]{d_0}
\ssmoveto{3}{1} \ssdropbull
\ssmoveto{4}{0} \ssdropbull
\ssmoveto{0}{3} \ssarrow[color=blue]{2}{-1}
\ssmoveto{1}{2} \ssarrow[color=red]{1}{0}
\ssmoveto{2}{1} \ssarrow{0}{1}
\end{sseq}
\ 
\begin{sseq}[grid=crossword, entrysize=6mm, labelstep=1]{0...4}{0...4}
\ssmoveto{0}{3} \ssdrop[]{F_{0,3}}
\ssmoveto{1}{2} \ssdrop[]{F_{1,2}}
\ssmoveto{2}{1} \ssdrop[]{F_{2,1}}
\ssmoveto{3}{0} \ssdrop[]{F_{3,0}}
\ssmoveto{0}{4} \ssdropbull
\ssmoveto{1}{3} \ssdropbull \ssdroplabel[D,color=cyan]{d_3}
\ssmoveto{2}{2} \ssdropbull \ssdroplabel[LD,color=blue]{d_2}
\ssmoveto{3}{1} \ssdropbull \ssdroplabel[LD,color=red]{d_1}\ssdroplabel[RD]{d_0}
\ssmoveto{4}{0} \ssdropbull
\ssmoveto{0}{3} \ssarrow[color=cyan]{3}{-2}
\ssmoveto{1}{2} \ssarrow[color=blue]{2}{-1}
\ssmoveto{2}{1} \ssarrow[color=red]{1}{0}
\ssmoveto{3}{0} \ssarrow{0}{1}
\end{sseq}
\ 
\begin{sseq}[grid=crossword, entrysize=6mm, labelstep=1]{0...4}{0...4}
\ssmoveto{0}{3} \ssdrop[]{F_{0,3}}
\ssmoveto{1}{2} \ssdrop[]{F_{1,2}}
\ssmoveto{2}{1} \ssdrop[]{F_{2,1}}
\ssmoveto{3}{0} \ssdrop[]{F_{3,0}}
\ssmoveto{0}{4} \ssdropbull
\ssmoveto{1}{3} \ssdropbull \ssdroplabel[D,color=purple]{d_4}
\ssmoveto{2}{2} \ssdropbull \ssdroplabel[LD,color=cyan]{d_3}
\ssmoveto{3}{1} \ssdropbull \ssdroplabel[LD,color=blue]{d_2}
\ssmoveto{4}{0} \ssdropbull \ssdroplabel[LD,color=red]{d_1}
\ssmoveto{0}{3} \ssarrow[color=magenta]{4}{-3}
\ssmoveto{1}{2} \ssarrow[color=cyan]{3}{-2}
\ssmoveto{2}{1} \ssarrow[color=blue]{2}{-1}
\ssmoveto{3}{0} \ssarrow[color=red]{1}{0}
\end{sseq}
\caption{LHS spectral sequence for degree-3.}
\label{fig:LHS3}
\end{figure}

Similar to the degree-1 and degree-2 cases, the cocycle condition for $F^{\ag{a}{g},\ag{b}{h},\ag{c}{k}}$ has an equivalent form as a consequence of Eq.~(\ref{dF3}):
\begin{align}\label{obs3}
(dF)^{\ag{a}{g},\ag{b}{h},\ag{c}{k},\ag{d}{l}}=1 \quad (\forall \ag{a}{g},\ag{b}{h},\ag{c}{k},\ag{d}{l}\in \tilde G)
\ \Longleftrightarrow\ 
\begin{cases}
(dF)^{a,b,c,d}=1 & (\forall a,b,c,d\in A),\\
(dF)^{a,b,c,\ag{1}{l}}=1 & (\forall a,b,c\in A,\ \ \forall \mb{l}\in G),\\
(dF)^{a,b,\ag{1}{k},\ag{1}{l}}=1 & (\forall a,b\in A,\ \ \forall \mb{k,l}\in G),\\
(dF)^{a,\ag{1}{h},\ag{1}{k},\ag{1}{l}}=1 & (\forall a\in A,\ \ \forall \mb{h,k,l}\in G),\\
(dF)^{\ag{1}{g},\ag{1}{h},\ag{1}{k},\ag{1}{l}}=1 & (\forall \mb{g,h,k,l}\in G).
\end{cases}
\end{align}
These five right-hand-side equations correspond to the five figures in Fig.~\ref{fig:LHS3}, respectively. So we expect they have the form:
\begin{align}\label{E04}
E_0^{0,4}&:\quad (\delta_0 F_{0,3}) = 1,\\\label{E13}
E_0^{1,3}&:\quad (\delta_1 F_{0,3})(\delta_0 F_{1,2}) = 1,\\\label{E22}
E_0^{2,2}&:\quad (\delta_2 F_{0,3})(\delta_1F_{1,2})(\delta_0F_{2,1}) = 1,\\\label{E31}
E_0^{3,1}&:\quad (\delta_3 F_{0,3})(\delta_2F_{1,2})(\delta_1F_{2,1})(\delta_0F_{3,0}) = 1,\\\label{E40}
E_0^{4,0}&:\quad (\delta_4 F_{0,3})(\delta_3 F_{1,2})(\delta_2F_{2,1})(\delta_1F_{3,0}) = 1.
\end{align}
Indeed, we can obtain the following explicit obstruction functions from the right-hand-side of Eq.~(\ref{obs3}) by using Eq.~(\ref{F3}):
\begin{align}\label{dFabcd}
(dF)^{a,b,c,d}
&=(d_AF^{\cdot,\cdot,\cdot})^{a,b,c,d}
=\underbrace{\frac{{}^{a}\! \left(F^{b,c,d}\right) \times F^{a,bc,d} \times F^{a,b,c}}{F^{ab,c,d} \times F^{a,b,cd}}}_{\delta_0F_{0,3}}
=1,
\end{align}
\begin{align}\label{dFabcl}
(dF)^{a,b,c,\ag{1}{l}}
&=
\frac{1}{\left(d_GF^{a,b,c}\right)^{\ag{1}{l}}}
\times
\left(d_A F^{\cdot,\cdot,\ag{1}{l}}\right)^{a,b,c}
=
\underbrace{\frac{F^{a,b,c}}{({}^{\ag{1}{l}}\!F)^{a,b,c}}}_{\delta_1F_{0,3}}
\underbrace{\frac{{}^a (F^{b,c,\ag{1}{l}}) \times F^{a,bc,\ag{1}{l}}}{F^{ab,c,\ag{1}{l}} \times F^{a,b,\ag{1}{l}}}}_{\delta_0F_{1,2}}
=1,
\end{align}
\begin{align}\nonumber\label{dFabkl}
(dF)^{a,b,\ag{1}{k},\ag{1}{l}}
&=\left[\iota_{-\nu(\mb{k,l})}\left({}^{\ag{1}{kl}}\!F\right)^{\cdot,\cdot,\cdot}\right]^{a,b}
\times
\left(d_GF^{a,b,\cdot}\right)^{\ag{1}{k},\ag{1}{l}}
\times
\left(d_A F^{\cdot,\ag{1}{k},\ag{1}{l}}\right)^{a,b}\\
&=\underbrace{\frac{({}^{\ag{1}{kl}}\!F)^{a,\nugh{k,l},b}}{({}^{\ag{1}{kl}}\!F)^{a,b,\nugh{k,l}} \times ({}^{\ag{1}{kl}}\!F)^{\nugh{k,l},a,b}}}_{\delta_2F_{0,3}}
\times
\underbrace{\frac{{}^{\ag{1}{k}}\!\left(F^{\gia{k}{a},\gia{k}{b},\ag{1}{l}}\right) \times F^{a,b,\ag{1}{k}}}{F^{a,b,\ag{1}{kl}}}}_{\delta_1F_{1,2}}
\times
\underbrace{\frac{{}^{a}\!\left(F^{b,\ag{1}{k},\ag{1}{l}}\right) \times F^{a,\ag{1}{k},\ag{1}{l}}}{F^{ab,\ag{1}{k},\ag{1}{l}}}}_{\delta_0F_{2,1}}
=1,
\end{align}
\begin{align}\nonumber\label{dFahkl}
(dF)^{a,\ag{1}{h},\ag{1}{k},\ag{1}{l}}
&=
\left[\iota_{(\nu\otimes\nu)[\Delta_1(\mb{h,k,l})]} \left({}^{\ag{1}{hkl}}\!F\right)^{\cdot,\cdot,\cdot}\right]^{a}
\times
\left[\iota_{\nu(\mb{h,k})} \left({}^{\ag{1}{hk}}\!F\right)^{\cdot,\cdot,\ag{1}{l}}\right]^{a}
\times
\left[(d_GF^{a,\cdot,\cdot})^{\ag{1}{h},\ag{1}{k},\ag{1}{l}}\right]^{-1}
\times
(d_AF^{\ag{1}{h},\ag{1}{k},\ag{1}{l}})^a
\\\nonumber
&=
\frac{\left[\iota_{\nugh{hk,l},\nugh{h,k}} \left({}^{\ag{1}{hkl}}\!F\right)^{\cdot,\cdot,\cdot}\right]^a}{\left[\iota_{\nugh{h,kl},\ga{h}{\nugh{k,l}}} \left({}^{\ag{1}{hkl}}\!F\right)^{\cdot,\cdot,\cdot}\right]^a}
\times
\left[\iota_{\nu(\mb{h,k})} \left({}^{\ag{1}{hk}}\!F\right)^{\cdot,\cdot,\ag{1}{l}}\right]^{a}
\times
\left[(d_GF^{a,\cdot,\cdot})^{\ag{1}{h},\ag{1}{k},\ag{1}{l}}\right]^{-1}
\times
(d_AF^{\ag{1}{h},\ag{1}{k},\ag{1}{l}})^a
\\\nonumber
&=
\underbrace{
\frac{\left({}^{\ag{1}{hkl}}\!F\right)^{a,\nu(\mb{hk,l}),\nu(\mb{h,k})} \times \left({}^{\ag{1}{hkl}}\!F\right)^{\nu(\mb{hk,l}),\nu(\mb{h,k}),a}}
{\left({}^{\ag{1}{hkl}}\!F\right)^{\nu(\mb{hk,l}),a,\nu(\mb{h,k})}  }
\times
\frac{\left({}^{\ag{1}{hkl}}\!F\right)^{\nu(\mb{h,kl}),a,\ga{h}{\nu(\mb{k,l})}}  }
{\left({}^{\ag{1}{hkl}}\!F\right)^{a,\nu(\mb{h,kl}),\ga{h}{\nu(\mb{k,l})}} \times \left({}^{\ag{1}{hkl}}\!F\right)^{\nu(\mb{h,kl}),\ga{h}{\nu(\mb{k,l})},a}}
}_{\delta_3F_{0,3}}
\\\nonumber
&\quad\times
\underbrace{
\frac{\left({}^{\ag{1}{hk}}\!F\right)^{a,\nu(\mb{h,k}),\ag{1}{l}}}
{\left({}^{\ag{1}{hk}}\!F\right)^{\nu(\mb{h,k}),a,\ag{1}{l}}}
}_{\delta_2F_{1,2}}
\times
\underbrace{
\frac{F^{a,\ag{1}{hk},\ag{1}{l}} \times F^{a,\ag{1}{h},\ag{1}{k}}}
{{}^{\ag{1}{h}}\!\left(F^{\gia{h}{a},\ag{1}{k},\ag{1}{l}}\right) \times F^{a,\ag{1}{h},\ag{1}{kl}}}
}_{\delta_1F_{2,1}}
\times
\underbrace{
\frac{{}^{a}\!\left(F^{\ag{1}{h},\ag{1}{k},\ag{1}{l}}\right)}
{F^{\ag{1}{h},\ag{1}{k},\ag{1}{l}}}
}_{\delta_0F_{3,0}}
\\
&=1,
\end{align}
\begin{align}\nonumber\label{dFghkl}
(dF)^{\ag{1}{g},\ag{1}{h},\ag{1}{k},\ag{1}{l}}
&=\left[\iota_{(\nu\otimes\nu\otimes\nu)[(-(\Delta_1\otimes\operatorname{id.})\Delta_1)(\mb{g,h,k,l})]} ({}^{\ag{1}{ghkl}}\!F)^{\cdot,\cdot,\cdot}\right]
\times
\left[\iota_{(\nu\otimes\nu)(-\Delta_1(\mb{g,h,l}))}F^{\cdot,\cdot,\ag{1}{l}}\right]
\\\nonumber
&\quad\times
\left[\iota_{-\nu(\mb{g,h})}F^{\cdot,\ag{1}{k},\ag{1}{l}}\right]
\times
(d_GF^{\cdot,\cdot,\cdot})^{\ag{1}{g},\ag{1}{h},\ag{1}{k},\ag{1}{l}}
\\\nonumber
&=\underbrace{
\frac{({}^{\ag{1}{ghkl}}\!F)^{\nu(\mb{ghk,l}),\nu(\mb{gh,k}),\nu(\mb{g,h})}}{({}^{\ag{1}{ghkl}}\!F)^{\nu(\mb{gh,kl}),\ga{gh}{\nu(\mb{k,l})},\nu(\mb{g,h})}}
\frac{({}^{\ag{1}{ghkl}}\!F)^{\nu(\mb{g,hkl}),\ga{g}{\nu(\mb{hk,l})},\ga{g}{\nu(\mb{h,k})}}}{({}^{\ag{1}{ghkl}}\!F)^{\nu(\mb{ghk,l}),\nu(\mb{g,hk}),\ga{g}{\nu(\mb{h,k})}}}
\frac{({}^{\ag{1}{ghkl}}\!F)^{\nu(\mb{gh,kl}),\nu(\mb{g,h}),\ga{gh}{\nu(\mb{k,l})}}}{({}^{\ag{1}{ghkl}}\!F)^{\nu(\mb{g,hkl}),\ga{g}{\nu(\mb{h,kl})},\ga{gh}{\nu(\mb{k,l})}}}
}_{\delta_4F_{0,3}}
\\\nonumber
&\quad\times
\underbrace{
\frac{({}^{\ag{1}{ghk}}\!F)^{\ga{}{\nu(\mb{g,hk})},\ga{g}{\nu(\mb{h,k})},\ag{1}{l}}}{({}^{\ag{1}{ghk}}\!F)^{\ga{}{\nu(\mb{gh,k})},\ga{}{\nu(\mb{g,h})},\ag{1}{l}}}
}_{\delta_3F_{1,2}}
\times
\underbrace{
\frac{1}{({}^{\ag{1}{gh}}\!F)^{\nu(\mb{g,h}),\ag{1}{k},\ag{1}{l}}}
}_{\delta_2F_{2,1}}
\times
\underbrace{
\frac{{}^{\ag{1}{g}}\!\left(F^{\ag{1}{h},\ag{1}{k},\ag{1}{l}}\right) F^{\ag{1}{g},\ag{1}{hk},\ag{1}{l}} F^{\ag{1}{g},\ag{1}{h},\ag{1}{k}}} {F^{\ag{1}{gh},\ag{1}{k},\ag{1}{l}} F^{\ag{1}{g},\ag{1}{h},\ag{1}{kl}}}
}_{\delta_1F_{3,0}}
\\
&=1.
\end{align}
These equations can be solved one by one to obtain a 3-cocycle of $\tilde G$ in Eq.~(\ref{F3}).

\subsection{LHS spectral sequence for degree-4}
\label{App:H4}

Replacing the differential of 3-cocycle $dF$ in Eq.~(\ref{dF3}) by 4-cocycle $F$, we obtain the generic expression for $F^{\ag{a}{g},\ag{b}{h},\ag{c}{k},\ag{d}{l}}$:
\begin{align}
F\nonumber
&=\bigg\{{}^{\ag{\nu(\mb{g,h}) \nu(\mb{gh,k}) \nu(\mb{ghk,l})}{}}\!\!\left[\left({}^{\ag{1}{ghkl}}F\right)^{a,\ga{g}{b},\ga{gh}{c},\ga{ghk}{d}}\right]
\times
{}^{\ag{\nu(\mb{gh,k}) \nu(\mb{ghk,l})}{}}\!\!\left[\iota_{\nu(\mb{g,h})}\left({}^{\ag{1}{ghkl}}F\right)^{\cdot,a\ga{g}{b},\ga{gh}{c},\ga{ghk}{d}}\right]
\\\nonumber
&\quad\quad\times
{}^{\ag{\nu(\mb{g,hk}) \nu(\mb{ghk,l})}{}}\!\!\left[\left(\iota_{[\ga{g}{\nu(\mb{h,k})}]}\left({}^{\ag{1}{ghkl}}F\right)^{\cdot,\cdot,\ga{g}{b}\ga{gh}{c},\ga{ghk}{d}}\right)^{\!\!a}\right]
\times
{}^{\ag{\nu(\mb{g,h}) \nu(\mb{gh,kl})}{}}\!\!\left[\left(\iota_{[\ga{gh}{\nu(\mb{k,l})}]}\left({}^{\ag{1}{ghkl}}F\right)^{\cdot,\cdot,\cdot,\ga{gh}{c}\ga{ghk}{d}}\right)^{\!\!a,\ga{g}{b}}\right]
\\\nonumber
&\quad\quad\times
{}^{\ag{\nu(\mb{ghk,l})}{}}\!\!\left[\iota_{(\nu\otimes\nu)(-\Delta_1(\mb{g,h,k}))}\left({}^{\ag{1}{ghkl}}F\right)^{\cdot,\cdot,a\ga{g}{b}\ga{gh}{c},\ga{ghk}{d}}\right]
\times
{}^{\ag{\nu(\mb{gh,kl})}{}}\!\!\left[\left(\iota_{[\ga{gh}{\nu(\mb{k,l})}]}\left({}^{\ag{1}{ghkl}}F\right)^{\cdot,\cdot,\cdot,\ga{gh}{c}\ga{ghk}{d}}\right)^{\!\!\nu(\mb{g,h}),a\ga{g}{b}}\right]
\\\nonumber
&\quad
\underbrace{
\quad\times
{}^{\ag{\nu(\mb{g,hkl})}{}}\!\!\left[\left(\iota_{(\nu\otimes\nu)(-\Delta_1(\mb{h,k,l}))}\left({}^{\ag{1}{ghkl}}F\right)^{\cdot,\cdot,\cdot,\ga{g}{b}\ga{gh}{c}\ga{ghk}{d}}\right)^{\!\!a}\right]
\times
\left[\iota_{(\nu\otimes\nu\otimes\nu)[((\Delta_1\otimes\operatorname{id.})\Delta_1)(\mb{g,h,k,l})]}\left({}^{\ag{1}{ghkl}}F\right)^{\cdot,\cdot,\cdot,a\ga{g}{b}\ga{gh}{c}\ga{ghk}{d}}\right]\bigg\}
}_{E_0^{0,4}}
\\
\nonumber
&\quad\times
\bigg\{{}^{\ag{[\nu(\mb{g,h}) \nu(\mb{gh,k})]}{ghk}}\!\!\left[F^{\gia{ghk}{a},\gia{hk}{b},\gia{k}{c},\ag{1}{l}}\right]
\times
{}^{\ag{\nu(\mb{gh,k})}{ghk}}\!\!\left[F^{\gia{ghk}{\nu(\mb{g,h})},\gia{ghk}{a}\gia{hk}{b},\gia{k}{c},\ag{1}{l}}\right]\\\nonumber
&\quad\quad
\underbrace{
\quad\times
\frac{{}^{\ag{\nu(\mb{g,hk})}{ghk}}\!\!\left[F^{\gia{ghk}{a},\gia{hk}{\nu(\mb{h,k})},\gia{hk}{b}\gia{k}{c},\ag{1}{l}}\right]}
{{}^{\ag{\nu(\mb{g,hk})}{ghk}}\!\!\left[F^{\gia{hk}{\nu(\mb{h,k})},\gia{ghk}{a},\gia{hk}{b}\gia{k}{c},\ag{1}{l}}\right]}
\times
\frac{{}^{\ag{1}{ghk}}\!\!\left[F^{\gia{ghk}{\nu(\mb{g,hk})},\gia{hk}{\nu(\mb{h,k})},\gia{ghk}{a}\gia{hk}{b}\gia{k}{c},\ag{1}{l}}\right]}
{{}^{\ag{1}{ghk}}\!\!\left[F^{\gia{ghk}{\nu(\mb{gh,k})},\gia{ghk}{\nu(\mb{g,h})},\gia{ghk}{a}\gia{hk}{b}\gia{k}{c},\ag{1}{l}}\right]}\bigg\}
}_{E_0^{1,3}}
\\
&\quad\times
\underbrace{
\bigg\{{}^{\ag{\nu(\mb{g,h})}{gh}}\!\!\left[F^{\gia{gh}{a},\gia{h}{b},\ag{1}{k},\ag{1}{l}}\right]
\times
{}^{\ag{1}{gh}}\!\!\left[F^{\gia{gh}{\nu(\mb{g,h})},\gia{gh}{a}\gia{h}{b},\ag{1}{k},\ag{1}{l}}\right]\bigg\}
}_{E_0^{2,2}}
\times
\underbrace{
{}^{\ag{1}{g}}\!\!\left[F^{\gia{g}{a},\ag{1}{h},\ag{1}{k},\ag{1}{l}}\right]
}_{E_0^{3,1}}
\times
\underbrace{
F^{\ag{1}{g},\ag{1}{h},\ag{1}{k},\ag{1}{l}}
}_{E_0^{4,0}}
.
\end{align}
Now we can use the above equation to show that the differential of $F$ has the form
\begin{align}\label{dF4}
(dF)^{\ag{a}{g},\ag{b}{h},\ag{c}{k},\ag{d}{l},\ag{e}{m}}
&=
(dF)|_{E_0^{0,5}}
\times (dF)|_{E_0^{1,4}}
\times (dF)|_{E_0^{2,3}}
\times (dF)|_{E_0^{3,2}}
\times (dF)|_{E_0^{4,1}}
\times (dF)|_{E_0^{5,0}}.
\end{align}
Since we are interested in systems in 3+1 or lower dimensions, we will stop here without showing the explicit form of $(dF)|_{E_0^{i,5-i}}$, which is only useful to obtain the expression of 5-cocycles.

\begin{figure}[ht]
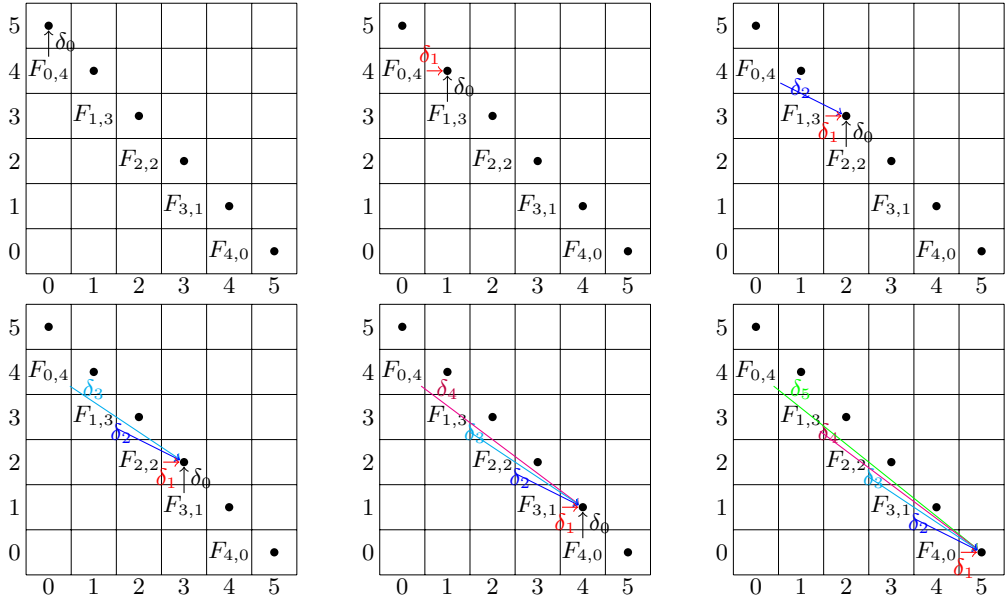

\begin{sseq}[grid=crossword, entrysize=6mm, labelstep=1]{0...5}{0...5}
\ssmoveto{0}{4} \ssdrop[]{F_{0,4}}
\ssmoveto{1}{3} \ssdrop[]{F_{1,3}}
\ssmoveto{2}{2} \ssdrop[]{F_{2,2}}
\ssmoveto{3}{1} \ssdrop[]{F_{3,1}}
\ssmoveto{4}{0} \ssdrop[]{F_{4,0}}
\ssmoveto{0}{5} \ssdropbull \ssdroplabel[RD]{\delta_0}
\ssmoveto{1}{4} \ssdropbull
\ssmoveto{2}{3} \ssdropbull
\ssmoveto{3}{2} \ssdropbull
\ssmoveto{4}{1} \ssdropbull
\ssmoveto{5}{0} \ssdropbull
\ssmoveto{0}{4} \ssarrow{0}{1}
\end{sseq}
\quad\quad
\begin{sseq}[grid=crossword, entrysize=6mm, labelstep=1]{0...5}{0...5}
\ssmoveto{0}{4} \ssdrop[]{F_{0,4}}
\ssmoveto{1}{3} \ssdrop[]{F_{1,3}}
\ssmoveto{2}{2} \ssdrop[]{F_{2,2}}
\ssmoveto{3}{1} \ssdrop[]{F_{3,1}}
\ssmoveto{4}{0} \ssdrop[]{F_{4,0}}
\ssmoveto{0}{5} \ssdropbull
\ssmoveto{1}{4} \ssdropbull \ssdroplabel[LU,color=red]{\delta_1} \ssdroplabel[RD]{\delta_0}
\ssmoveto{2}{3} \ssdropbull
\ssmoveto{3}{2} \ssdropbull
\ssmoveto{4}{1} \ssdropbull
\ssmoveto{5}{0} \ssdropbull
\ssmoveto{0}{4} \ssarrow[color=red]{1}{0}
\ssmoveto{1}{3} \ssarrow{0}{1}
\end{sseq}
\quad\quad
\begin{sseq}[grid=crossword, entrysize=6mm, labelstep=1]{0...5}{0...5}
\ssmoveto{0}{4} \ssdrop[]{F_{0,4}}
\ssmoveto{1}{3} \ssdrop[]{F_{1,3}}
\ssmoveto{2}{2} \ssdrop[]{F_{2,2}}
\ssmoveto{3}{1} \ssdrop[]{F_{3,1}}
\ssmoveto{4}{0} \ssdrop[]{F_{4,0}}
\ssmoveto{0}{5} \ssdropbull
\ssmoveto{1}{4} \ssdropbull \ssdroplabel[D,color=blue]{\delta_2}
\ssmoveto{2}{3} \ssdropbull \ssdroplabel[LD,color=red]{\delta_1} \ssdroplabel[RD]{\delta_0}
\ssmoveto{3}{2} \ssdropbull
\ssmoveto{4}{1} \ssdropbull
\ssmoveto{5}{0} \ssdropbull
\ssmoveto{0}{4} \ssarrow[color=blue]{2}{-1}
\ssmoveto{1}{3} \ssarrow[color=red]{1}{0}
\ssmoveto{2}{2} \ssarrow{0}{1}
\end{sseq}
\\\ 
\begin{sseq}[grid=crossword, entrysize=6mm, labelstep=1]{0...5}{0...5}
\ssmoveto{0}{4} \ssdrop[]{F_{0,4}}
\ssmoveto{1}{3} \ssdrop[]{F_{1,3}}
\ssmoveto{2}{2} \ssdrop[]{F_{2,2}}
\ssmoveto{3}{1} \ssdrop[]{F_{3,1}}
\ssmoveto{4}{0} \ssdrop[]{F_{4,0}}
\ssmoveto{0}{5} \ssdropbull
\ssmoveto{1}{4} \ssdropbull \ssdroplabel[D,color=cyan]{\delta_3}
\ssmoveto{2}{3} \ssdropbull \ssdroplabel[LD,color=blue]{\delta_2}
\ssmoveto{3}{2} \ssdropbull \ssdroplabel[LD,color=red]{\delta_1}\ssdroplabel[RD]{\delta_0}
\ssmoveto{4}{1} \ssdropbull
\ssmoveto{5}{0} \ssdropbull
\ssmoveto{0}{4} \ssarrow[color=cyan]{3}{-2}
\ssmoveto{1}{3} \ssarrow[color=blue]{2}{-1}
\ssmoveto{2}{2} \ssarrow[color=red]{1}{0}
\ssmoveto{3}{1} \ssarrow{0}{1}
\end{sseq}
\quad\quad
\begin{sseq}[grid=crossword, entrysize=6mm, labelstep=1]{0...5}{0...5}
\ssmoveto{0}{4} \ssdrop[]{F_{0,4}}
\ssmoveto{1}{3} \ssdrop[]{F_{1,3}}
\ssmoveto{2}{2} \ssdrop[]{F_{2,2}}
\ssmoveto{3}{1} \ssdrop[]{F_{3,1}}
\ssmoveto{4}{0} \ssdrop[]{F_{4,0}}
\ssmoveto{0}{5} \ssdropbull
\ssmoveto{1}{4} \ssdropbull \ssdroplabel[D,color=purple]{\delta_4}
\ssmoveto{2}{3} \ssdropbull \ssdroplabel[LD,color=cyan]{\delta_3}
\ssmoveto{3}{2} \ssdropbull \ssdroplabel[LD,color=blue]{\delta_2}
\ssmoveto{4}{1} \ssdropbull \ssdroplabel[LD,color=red]{\delta_1}\ssdroplabel[RD]{\delta_0}
\ssmoveto{5}{0} \ssdropbull
\ssmoveto{0}{4} \ssarrow[color=magenta]{4}{-3}
\ssmoveto{1}{3} \ssarrow[color=cyan]{3}{-2}
\ssmoveto{2}{2} \ssarrow[color=blue]{2}{-1}
\ssmoveto{3}{1} \ssarrow[color=red]{1}{0}
\ssmoveto{4}{0} \ssarrow[color=black]{0}{1}
\end{sseq}
\quad\quad
\begin{sseq}[grid=crossword, entrysize=6mm, labelstep=1]{0...5}{0...5}
\ssmoveto{0}{4} \ssdrop[]{F_{0,4}}
\ssmoveto{1}{3} \ssdrop[]{F_{1,3}}
\ssmoveto{2}{2} \ssdrop[]{F_{2,2}}
\ssmoveto{3}{1} \ssdrop[]{F_{3,1}}
\ssmoveto{4}{0} \ssdrop[]{F_{4,0}}
\ssmoveto{0}{5} \ssdropbull
\ssmoveto{1}{4} \ssdropbull \ssdroplabel[D,color=green]{\delta_5}
\ssmoveto{2}{3} \ssdropbull \ssdroplabel[LD,color=purple]{\delta_4}
\ssmoveto{3}{2} \ssdropbull \ssdroplabel[LD,color=cyan]{\delta_3}
\ssmoveto{4}{1} \ssdropbull \ssdroplabel[LD,color=blue]{\delta_2}
\ssmoveto{5}{0} \ssdropbull \ssdroplabel[LD,color=red]{\delta_1}
\ssmoveto{0}{4} \ssarrow[color=green]{5}{-4}
\ssmoveto{1}{3} \ssarrow[color=magenta]{4}{-3}
\ssmoveto{2}{2} \ssarrow[color=cyan]{3}{-2}
\ssmoveto{3}{1} \ssarrow[color=blue]{2}{-1}
\ssmoveto{4}{0} \ssarrow[color=red]{1}{0}
\end{sseq}
\caption{LHS spectral sequence for degree-4.}
\label{fig:LHS4}
\end{figure}

To derive the obstruction functions, we again use the trick that the cocycle condition for $F^{\ag{a}{g},\ag{b}{h},\ag{c}{k},\ag{d}{l}}$ can be equivalently expressed as:
\begin{align}\label{obs4}
(dF)^{\ag{a}{g},\ag{b}{h},\ag{c}{k},\ag{d}{l},\ag{e}{m}}=1 \quad (\forall \ag{a}{g},\ag{b}{h},\ag{c}{k},\ag{d}{l},\ag{e}{m}\in \tilde G)
\ \Longleftrightarrow\ 
\begin{cases}
(dF)^{a,b,c,d,e}=1 & (\forall a,b,c,d,e\in A),\\
(dF)^{a,b,c,d,\ag{1}{m}}=1 & (\forall a,b,c,d\in A, \ \ \forall \mb{m}\in G),\\
(dF)^{a,b,c,\ag{1}{l},\ag{1}{m}}=1 & (\forall a,b,c\in A, \ \ \forall \mb{l,m}\in G),\\
(dF)^{a,b,\ag{1}{k},\ag{1}{l},\ag{1}{m}}=1 & (\forall a,b\in A, \ \ \forall \mb{k,l,m}\in G),\\
(dF)^{a,\ag{1}{h},\ag{1}{k},\ag{1}{l},\ag{1}{m}}=1 & (\forall a\in A, \ \ \forall \mb{h,k,l,m}\in G),\\
(dF)^{\ag{1}{g},\ag{1}{h},\ag{1}{k},\ag{1}{l},\ag{1}{m}}=1 & (\forall \mb{g,h,k,l,m}\in G).
\end{cases}
\end{align}
These six equations correspond to the six figures in Fig.~\ref{fig:LHS4}, respectively. And they can all expressed as product of several differentials at different locations:
\begin{align}\label{E05}
E_0^{0,5}&:\quad (\delta_0 F_{0,4}) = 1,\\\label{E14}
E_0^{1,4}&:\quad (\delta_1 F_{0,4})(\delta_0 F_{1,3}) = 1,\\\label{E23}
E_0^{2,3}&:\quad (\delta_2 F_{0,4})(\delta_1F_{1,3})(\delta_0F_{2,2}) = 1,\\\label{E32}
E_0^{3,2}&:\quad (\delta_3 F_{0,4})(\delta_2F_{1,3})(\delta_1F_{2,2})(\delta_0F_{3,1}) = 1,\\\label{E41}
E_0^{4,1}&:\quad (\delta_4 F_{0,4})(\delta_3 F_{1,3})(\delta_2 F_{2,2})(\delta_1 F_{3,1})(\delta_0 F_{4,0}) = 1,\\\label{E50}
E_0^{5,0}&:\quad (\delta_5 F_{0,4})(\delta_4 F_{1,3})(\delta_3 F_{2,2})(\delta_2 F_{3,1})(\delta_1 F_{4,0}) = 1.
\end{align}
The explicit obstruction functions on the right-hand-side are:
\begin{align}
(dF)^{a,b,c,d,e}
&=\underbrace{\frac{{}^{a}\!\left(F^{b,c,d,e}\right)F^{a,bc,d,e}F^{a,b,c,de}}{F^{ab,c,d,e}F^{a,b,cd,e}F^{a,b,c,d}}}_{\delta_0F_{0,4}}
=1,
\end{align}
\begin{align}
(dF)^{a,b,c,d,\ag{1}{m}}
=
\left(d_GF^{a,b,c,d}\right)^{\!\ag{1}{m}}
\left(d_AF^{\cdot,\cdot,\cdot,\ag{1}{m}}\right)^{\!a,b,c,d}
=
\underbrace{\frac{\left({}^{\ag{1}{m}}\!F\right)^{a,b,c,d}}{F^{a,b,c,d}}}_{\delta_1F_{0,4}}
\underbrace{
\frac{{}^{a}\!\left(F^{b,c,d,\ag{1}{m}}\right) F^{a,bc,d,\ag{1}{m}} F^{a,b,c,\ag{1}{m}}}
{F^{ab,c,d,\ag{1}{m}} F^{a,b,cd,\ag{1}{m}}}
}_{\delta_0F_{1,3}}
=1,
\end{align}
\begin{align}\nonumber
(dF)^{a,b,c,\ag{1}{l},\ag{1}{m}}
&=
\left[\iota_{\nu(\mb{l,m})}\left({}^{\ag{1}{lm}}F\right)^{\cdot,\cdot,\cdot,\cdot}\right]^{a,b,c}
\times
\left[\left(d_GF^{a,b,c,\cdot}\right)^{\ag{1}{l},\ag{1}{m}}\right]^{-1}
\times
\left(d_AF^{\cdot,\cdot,\ag{1}{l},\ag{1}{m}}\right)^{a,b,c}\\
&=
\underbrace{
\frac{\left({}^{\ag{1}{lm}}F\right)^{a,b,c,\nu(\mb{l,m})}\left({}^{\ag{1}{lm}}F\right)^{a,\nu(\mb{l,m}),b,c}}
{\left({}^{\ag{1}{lm}}F\right)^{a,b,\nu(\mb{l,m}),c}\left({}^{\ag{1}{lm}}F\right)^{\nu(\mb{l,m}),a,b,c}}
}_{\delta_2F_{0,4}}
\underbrace{
\frac{F^{a,b,c,\ag{1}{lm}}}
{({}^{\ag{1}{l}}F)^{a,b,c,\ag{1}{m}} F^{a,b,c,\ag{1}{l}}}
}_{\delta_1F_{1,3}}
\underbrace{
\frac{{}^a(F^{b,c,\ag{1}{l},\ag{1}{m}}) F^{a,bc,\ag{1}{l},\ag{1}{m}}}
{F^{ab,c,\ag{1}{l},\ag{1}{m}} F^{a,b,\ag{1}{l},\ag{1}{m}}}
}_{\delta_0F_{2,2}}=1,
\end{align}
\begin{align}\nonumber\label{dFabklm}
(dF)^{a,b,\ag{1}{k},\ag{1}{l},\ag{1}{m}}
&=
\left[\iota_{(\nu\otimes\nu)(-\Delta_1(\mb{k,l,m}))}\left({}^{\ag{1}{klm}}F\right)^{\cdot,\cdot,\cdot,\cdot}\right]^{a,b}
\left[\iota_{-\nu(\mb{k,l})}\left({}^{\ag{1}{kl}}F\right)^{\cdot,\cdot,\cdot,\ag{1}{m}}\right]^{a,b}
\left(d_GF^{a,b,\cdot,\cdot}\right)^{\ag{1}{k},\ag{1}{l},\ag{1}{m}}
\left(d_AF^{\cdot,\ag{1}{k},\ag{1}{l},\ag{1}{m}}\right)^{a,b}\\\nonumber
&=
\underbrace{
\frac{\left[\iota_{\nu(\mb{k,lm}),[\ga{k}{\nu(\mb{l,m})}]}\left({}^{\ag{1}{klm}}F\right)^{\cdot,\cdot,\cdot,\cdot}\right]^{a,b}}
{\left[\iota_{\nu(\mb{kl,m}),\nu(\mb{k,l})}\left({}^{\ag{1}{klm}}F\right)^{\cdot,\cdot,\cdot,\cdot}\right]^{a,b}}
}_{\delta_3 F_{0,4}}
\times
\underbrace{
\frac{\left({}^{\ag{1}{kl}}F\right)^{a,\nu(\mb{k,l}),b,\ag{1}{m}} }
{\left({}^{\ag{1}{kl}}F\right)^{a,b,\nu(\mb{k,l}),\ag{1}{m}} \left({}^{\ag{1}{kl}}F\right)^{\nu(\mb{k,l}),a,b,\ag{1}{m}}}
}_{\delta_2F_{1,3}}\\\nonumber
&\quad\times
\underbrace{
\frac{{}^{\ag{1}{k}}\!\left(F^{\gia{k}{a},\gia{k}{b},\ag{1}{l},\ag{1}{m}}\right) F^{a,b,\ag{1}{k},\ag{1}{lm}}}
{F^{a,b,\ag{1}{kl},\ag{1}{m}} F^{a,b,\ag{1}{k},\ag{1}{l}}}
}_{\delta_1F_{2,2}}
\times
\underbrace{
\frac{{}^a\!\left(F^{b,\ag{1}{k},\ag{1}{l},\ag{1}{m}}\right) F^{a,\ag{1}{k},\ag{1}{l},\ag{1}{m}}}
{F^{ab,\ag{1}{k},\ag{1}{l},\ag{1}{m}}}
}_{\delta_0F_{3,1}}\\
&=1,
\end{align}
\begin{align}\nonumber
(dF)^{a,\ag{1}{h},\ag{1}{k},\ag{1}{l},\ag{1}{m}}
&=
\left[\iota_{(\nu\otimes\nu\otimes\nu)[((\Delta_1\otimes\operatorname{id.})\Delta_1)(\mb{h,k,l,m})]}\left({}^{\ag{1}{hklm}}F\right)^{\cdot,\cdot,\cdot,\cdot}\right]^{a}
\times
\left[\iota_{(\nu\otimes\nu)(\Delta_1(\mb{h,k,l}))} \left({}^{\ag{1}{hkl}}F\right)^{\cdot,\cdot,\cdot,\ag{1}{m}}\right]^a
\\\nonumber
&\quad\times
\left[\iota_{\nu(\mb{h,k})} \left({}^{\ag{1}{hk}}F\right)^{\cdot,\cdot,\ag{1}{l},\ag{1}{m}}\right]^a
\times
\left[\left(d_GF^{a,\cdot,\cdot,\cdot}\right)^{\ag{1}{h},\ag{1}{k},\ag{1}{l},\ag{1}{m}}\right]^{-1}
\times
\left(d_AF^{\ag{1}{h},\ag{1}{k},\ag{1}{l},\ag{1}{m}}\right)^a
\\\nonumber
&=
\bigg\{
\frac{\left[\iota_{\nu(\mb{hk,lm}),\ga{hk}{\nu(\mb{l,m})},\nu(\mb{h,k})}\left({}^{\ag{1}{hklm}}F\right)^{\cdot,\cdot,\cdot,\cdot}\right]^{a}}{\left[\iota_{\nu(\mb{hkl,m}),\nu(\mb{hk,l}),\nu(\mb{h,k})}\left({}^{\ag{1}{hklm}}F\right)^{\cdot,\cdot,\cdot,\cdot}\right]^{a}}
\times
\frac{\left[\iota_{\nu(\mb{hkl,m}),\nu(\mb{h,kl}),\ga{h}{\nu(\mb{k,l})}}\left({}^{\ag{1}{hklm}}F\right)^{\cdot,\cdot,\cdot,\cdot}\right]^{a}}{\left[\iota_{\nu(\mb{h,klm}),\ga{h}{\nu(\mb{kl,m})},\ga{h}{\nu(\mb{k,l})}}\left({}^{\ag{1}{hklm}}F\right)^{\cdot,\cdot,\cdot,\cdot}\right]^{a}}
\\\nonumber
&\quad
\underbrace{
\quad\times
\frac{\left[\iota_{\nu(\mb{h,klm}),\ga{h}{\nu(\mb{k,lm})},\ga{hk}{\nu(\mb{l,m})}}\left({}^{\ag{1}{hklm}}F\right)^{\cdot,\cdot,\cdot,\cdot}\right]^{a}}{\left[\iota_{\nu(\mb{hk,lm}),\nu(\mb{h,k}),\ga{hk}{\nu(\mb{l,m})}}\left({}^{\ag{1}{hklm}}F\right)^{\cdot,\cdot,\cdot,\cdot}\right]^{a}}
\bigg\}
}_{\delta_4F_{0,4}}
\\\nonumber
&\quad\times
\underbrace{
\frac{\left[\iota_{\ga{}{\ga{}{\nu(\mb{hk,l})},\nu(\mb{h,k})}}\left({}^{\ag{1}{hkl}}F\right)^{\cdot,\cdot,\cdot,\ag{1}{m}}\right]^{a}}{\left[\iota_{\ga{}{\nu(\mb{h,kl})},[\ga{h}{\nu(\mb{k,l})}]}\left({}^{\ag{1}{hkl}}F\right)^{\cdot,\cdot,\cdot,\ag{1}{m}}\right]^{a}}
}_{\delta_3F_{1,3}}
\times
\underbrace{
\frac{\left({}^{\ag{1}{hk}}F\right)^{a,\nu(\mb{h,k}),\ag{1}{l},\ag{1}{m}}}{\left({}^{\ag{1}{hk}}F\right)^{\nu(\mb{h,k}),a,\ag{1}{l},\ag{1}{m}}}
}_{\delta_2F_{2,2}}
\\\nonumber
&\quad
\times
\underbrace{
\frac{F^{a,\ag{1}{hk},\ag{1}{l},\ag{1}{m}} F^{a,\ag{1}{h},\ag{1}{k},\ag{1}{lm}}}{{}^{\ag{1}{h}}\!\left(F^{\gia{h}{a},\ag{1}{k},\ag{1}{l},\ag{1}{m}}\right) F^{a,\ag{1}{h},\ag{1}{kl},\ag{1}{m}} F^{a,\ag{1}{h},\ag{1}{k},\ag{1}{l}}}
}_{\delta_1F_{3,1}}
\times
\underbrace{
\frac{{}^{a}\!\left(F^{\ag{1}{h},\ag{1}{k},\ag{1}{l},\ag{1}{m}}\right)}{F^{\ag{1}{h},\ag{1}{k},\ag{1}{l},\ag{1}{m}}}
}_{\delta_0F_{4,0}}
\\
&=1,
\end{align}
\begin{align}\nonumber\label{dFghklm}
(dF)^{\ag{1}{g},\ag{1}{h},\ag{1}{k},\ag{1}{l},\ag{1}{m}}
&=
\left[\iota_{(\nu\otimes\nu\otimes\nu\otimes\nu)[-((\Delta_1\otimes\operatorname{id.}\otimes\operatorname{id.})(\Delta_1\otimes\operatorname{id.})\Delta_1)(\mb{g,h,k,l,m})]}\left({}^{\ag{1}{ghklm}}F\right)^{\cdot,\cdot,\cdot,\cdot}\right]
\\\nonumber
&\quad\times
\left[\iota_{(\nu\otimes\nu\otimes\nu)[-((\Delta_1\otimes\operatorname{id.})\Delta_1)(\mb{g,h,k,l})]}\left({}^{\ag{1}{ghkl}}F\right)^{\cdot,\cdot,\cdot,\ag{1}{m}}\right]
\times
\left[\iota_{(\nu\otimes\nu)(-\Delta_1(\mb{g,h,k}))}\left({}^{\ag{1}{ghk}}F\right)^{\cdot,\cdot,\ag{1}{l},\ag{1}{m}}\right]
\\\nonumber
&\quad\times
\left[\iota_{-\nu(\mb{g,h})}\left({}^{\ag{1}{gh}}F\right)^{\cdot,\ag{1}{k},\ag{1}{l},\ag{1}{m}}\right]
\times
\left(d_GF^{\cdot,\cdot,\cdot,\cdot}\right)^{\ag{1}{g},\ag{1}{h},\ag{1}{k},\ag{1}{l},\ag{1}{m}}
\\\nonumber
&=
\underbrace{
\left[\iota_{(\nu\otimes\nu\otimes\nu\otimes\nu)[-((\Delta_1\otimes\operatorname{id.}\otimes\operatorname{id.})(\Delta_1\otimes\operatorname{id.})\Delta_1)(\mb{g,h,k,l,m})]}\left({}^{\ag{1}{ghklm}}F\right)^{\cdot,\cdot,\cdot,\cdot}\right]
}_{\delta_5F_{0,4}}
\\\nonumber
&\quad\times
\underbrace{
\frac{\left({}^{\ag{1}{ghkl}}\!F\right)^{\nu(\mb{ghk,l}),\nu(\mb{gh,k}),\nu(\mb{g,h}),\ag{1}{m}}}{\left({}^{\ag{1}{ghkl}}\!F\right)^{\nu(\mb{gh,kl}),\ga{gh}{\nu(\mb{k,l})},\nu(\mb{g,h}),\ag{1}{m}}}
\frac{\left({}^{\ag{1}{ghkl}}\!F\right)^{\nu(\mb{g,hkl}),\ga{g}{\nu(\mb{hk,l})},\ga{g}{\nu(\mb{h,k})},\ag{1}{m}}}{\left({}^{\ag{1}{ghkl}}\!F\right)^{\nu(\mb{ghk,l}),\nu(\mb{g,hk}),\ga{g}{\nu(\mb{h,k})},\ag{1}{m}}}
\frac{\left({}^{\ag{1}{ghkl}}\!F\right)^{\nu(\mb{gh,kl}),\nu(\mb{g,h}),\ga{gh}{\nu(\mb{k,l})},\ag{1}{m}}}{\left({}^{\ag{1}{ghkl}}\!F\right)^{\nu(\mb{g,hkl}),\ga{g}{\nu(\mb{h,kl})},\ga{gh}{\nu(\mb{k,l})},\ag{1}{m}}}
}_{\delta_4F_{1,3}}
\\\nonumber
&\quad\times
\underbrace{
\frac{\left({}^{\ag{1}{ghk}}\!F\right)^{\nu(\mb{g,hk}),\ga{g}{\nu(\mb{h,k})},\ag{1}{l},\ag{1}{m}}}{\left({}^{\ag{1}{ghk}}\!F\right)^{\nu(\mb{gh,k}),\nu(\mb{g,h}),\ag{1}{l},\ag{1}{m}}}
}_{\delta_3F_{2,2}}
\times
\underbrace{
\frac{1}{\left({}^{\ag{1}{gh}}F\right)^{\nu(\mb{g,h}),\ag{1}{k},\ag{1}{l},\ag{1}{m}}}
}_{\delta_2F_{3,1}}
\times
\underbrace{
\frac{{}^{\ag{1}{g}}\!\left(F^{\ag{1}{h},\ag{1}{k},\ag{1}{l},\ag{1}{m}}\right) F^{\ag{1}{g},\ag{1}{hk},\ag{1}{l},\ag{1}{m}} F^{\ag{1}{g},\ag{1}{h},\ag{1}{k},\ag{1}{lm}}} {F^{\ag{1}{gh},\ag{1}{k},\ag{1}{l},\ag{1}{m}} F^{\ag{1}{g},\ag{1}{h},\ag{1}{kl},\ag{1}{m}} F^{\ag{1}{g},\ag{1}{h},\ag{1}{k},\ag{1}{l}}}
}_{\delta_1F_{4,0}}
\\
&=1.
\end{align}
We note that the first term $\delta_5F_{0,4}$ of $(dF)^{\ag{1}{g},\ag{1}{h},\ag{1}{k},\ag{1}{l},\ag{1}{m}}$ is the product of 24 $F$ symbols given by Eq.~(\ref{d5F04}).

In principle, we can obtain the expressions for higher degree cocycles in the similar way order by order. We first decompose the differential of $n$-cocycle into different terms located at different positions in the LHS spectral sequence. By replacing $(n+1)$-coboundary $dF$ by $(n+1)$-cocycle $F$, we then have the expression of cocycles in one higher degree. Similarly, the cocycle condition $dF=1$ can be also decomposed into several different obstruction equations. And the differentials in LHS spectral sequence are summarized in Section~\ref{App:diff}.

\subsection{Example}
\label{app:example}
In this subsection, we will use results in the previous subsections to determine cocycle-level differentials of the LHS spectral sequence for $A=\Z_n$ in degree 3. The procedure is to calculate the cochain-level obstruction conditions in Eqs.~\eqref{E04}-\eqref{E40} one after another. The explicit expressions for these equations are given in Eqs.~\eqref{dFabcd}-\eqref{dFghkl}.

For simplicity, we assume that $G$ has trivial actions on $A$ and the $\U$ coefficient. The central extension 2-cocycle is denoted as $\nu\in\H^2[G,A]$.
We will identify the coefficient $\U$ with the the group $\RZ=[0,1)$. So all the obstruction conditions in the previous calculations should be changed into the corresponding additive notations, and all the formulas are understood as mod 1 equations.

In degree 3, the relevant cohomology groups on page 2 of the spectral sequence are $E_2^{0,3}=\H^3[A,\RZ]=\Z_n$, $E_2^{1,2}=\H^1[G,\H^2[A,\RZ]]=0$, $E_2^{2,1}=\H^2[G,\Z]$, and $E_2^{3,0}=\H^3[G,\RZ]$.

We can choose the standard 3-cocycle of $\Z_n$ at $E_2^{0,3}=\H^3[\Z_n,\RZ]$ to be
\begin{align}\label{F03}
F_{0,3}(a,b,c)=\frac{p}{n^2} a \left(b+c-[b+c]_n\right),
\end{align}
for $p=0,1,\dots, n-1$. It satisfies the obstruction condition $\delta_0F_{0,3}=d_AF_{0,3}=0$ in Eq.~\eqref{dFabcd}. Since we assumed the trivial action of $G$ on $A$ and the coefficient, the second obstruction condition Eq.~\eqref{dFabcl} becomes $\delta_0F_{1,2}=0$. As $E_2^{0,2}=0$, we can simply choose
\begin{align}
F_{1,2}(a,b,\ag{1}{k})=0.
\end{align}
The next level obstruction is Eq.~\eqref{dFabkl} which means $\delta_2 F_{0,3}+\delta_0F_{2,1} = 0$ additively. Since $\delta_0F_{2,1}$ is a coboundary in $E_2^{2,2}$, the cocycle-level differential of $F_{0,3}$ is $d_2F_{0,3}:=\delta_2F_{0,3}$. Using the explicit expression Eq.~\eqref{F03}, this term becomes
\begin{align}
(d_2F_{0,3})(a,b,\ag{1}{k},\ag{1}{l}) := (\delta_2F_{0,3})(a,b,\ag{1}{k},\ag{1}{l}) = - \frac{p}{n^2} \nu(\mb{k,l}) \left(a+b-[a+b]_n\right).
\end{align}
Therefore, the equation $\delta_0F_{2,1}=-\delta_2F_{0,3}$ of Eq.~\eqref{dFabkl} has a general solution for the cochain $F_{2,1}$ as
\begin{align}
F_{2,1}(a,\ag{1}{h},\ag{1}{k}) = \frac{p}{n^2} a\nu(\mb{h,k}) + f_{2,1}(a,\ag{1}{h},\ag{1}{k}).
\end{align}
Here, $f_{2,1}$ is some cochain in $\mathcal{C}^2[G,\mathcal{Z}^1[A,\RZ]]$ which is a homogeneous solution with $\delta_0f_{2,1}=0$. The explicit form of it will be determined by subsequent obstruction conditions.

Now let us turn to the obstruction condition Eq.~\eqref{dFahkl}, which means $\delta_3 F_{0,3}+\delta_1F_{2,1} = 0$ additively (both $\delta_2F_{1,2}$ and $\delta_0F_{3,0}$ are zero). Using the explicit expressions of cochains $F_{0,3}$ and $F_{2,1}$, we have
\begin{align}
(\delta_3F_{0,3})(a,\ag{1}{h},\ag{1}{k},\ag{1}{l})
&=-\frac{p}{n} a (\beta_n\nu)(\mb{h,k,l}),\\
(\delta_1F_{2,1})(a,\ag{1}{h},\ag{1}{k},\ag{1}{l})
&=-\frac{p}{n} a (\beta_n\nu)(\mb{h,k,l}) - (d_Gf_{2,1})(a,\ag{1}{h},\ag{1}{k},\ag{1}{l}),
\end{align}
where $\beta_n\nu = \frac{1}{n} d_G\tilde\nu$ is the Bockstein homomorphism of $\nu$ and $\tilde\nu$ is an integral lift of the $\Z_n$-valued $\nu$. The cochain-level differentials $\delta_3F_{0,3}$ and (the first term of) $\delta_1F_{2,1}$ are combined into a cocycle-level differential of $F_{0,3}$ as
\begin{align}\label{d3F03}
(d_3F_{0,3})(a,\ag{1}{h},\ag{1}{k},\ag{1}{l})
=-\frac{2p}{n} a (\beta_n\nu)(\mb{h,k,l}),
\end{align}
which doubles the single result of $\delta_3F_{0,3}$. Now we can solve the obstruction condition $\delta_3F_{0,3}+\delta_1F_{2,1}=0$, which is equivalent to $d_3F_{0,3}=d_Gf_{2,1}$. Since $f_{2,1}$ is a cochain in $\mathcal{C}^2[G,\mathcal{Z}^1[A,\RZ]]$, $d_3F_{0,3}=d_Gf_{2,1}$ should be a coboundary in $\mathcal{B}^3[G,\mathcal{Z}^1[A,\RZ]]$. $-2p\nu$ should be a coboundary in $\H^2[G, A]$, which means $-2p\tilde{\nu} = d_G \tilde{\lambda} + n\tilde{\mu}$.

From the expression Eq.~\eqref{d3F03}, we assume that $\mu:=-2p\tilde\nu/n$ is a 2-cochain $\mathcal{C}^2[G,\Z]$. In this way, we can choose
\begin{align}
f_{2,1}(a,\ag{1}{h},\ag{1}{k})=\frac{a\mu(\mb{h,k})}{n}
=-\frac{2p}{n^2}a\nu(\mb{h,k}),
\end{align}
to satisfy the obstruction condition Eq.~\eqref{dFahkl}. Here and later, we choose the integral lift $\tilde\nu$ to be always valued in $\{0,1,...,n-1\}$, and denote it also by $\nu$ abusively.

The last obstruction condition is Eq.~\eqref{dFghkl}, i.e., $\delta_4 F_{0,3} + \delta_2F_{2,1} + \delta_1F_{3,0} = 0$ additively. Using the expressions of $F_{0,3}$ and $F_{2,1}$, we have
\begin{align}\nonumber
(\delta_4 F_{0,3})(\mb{g,h,k,l})
&=-\frac{p}{n}(\nu\cup_1\beta_n\nu)(\mb{g,h,k,l})\\
&=-\frac{p}{n}\nu(\mb{ghk,l})(\beta_n\nu)(\mb{g,h,k})
-\frac{p}{n}\nu(\mb{g,hkl})(\beta_n\nu)(\mb{h,k,l}),\\
(\delta_2F_{2,1})(\mb{g,h,k,l})
&=-F_{2,1}(\nu(\mb{g,h}),\ag{1}{k},\ag{1}{l})=\frac{p}{n^2}\nu(\mb{g,h})\nu(\mb{k,l}).
\end{align}
The condition $\delta_4 F_{0,3} + \delta_2F_{2,1} =- d_GF_{3,0}$ is equivalent to the statement that the cocycle-level differential
\begin{align}
d_4F_{0,3}
:=\delta_4 F_{0,3} + \delta_2F_{2,1}
=-\frac{p}{n}\nu\cup_1\beta_n\nu + \frac{p}{n^2}\nu\cup\nu
\end{align}
is trivial in $\H^4[G,\RZ]$. We note again that $\nu$ is understood as an integral lift valued in $\{0,1,...,n-1\}$.

We can also derive the cocycle-level differential $d_2$ for $F_{2,1}\in E_2^{2,1}$. The procedure is to assume $F_{0,3}=F_{1,2}=0$, and try to solve the obstruction conditions. The conditions Eqs.~\eqref{dFabkl} and \eqref{dFahkl} become the cocycle conditions $d_AF_{2,1}=d_GF_{2,1}=0$, indicating $F_{2,1}\in E_2^{2,1}$.
Since $E_2^{2,1}=\H^2[G,\H^1[\Z_n,\RZ]]=\H^2[G,\Z_n]$, $F_{2,1}$ can be characterized by $m\in \H^2[G,\Z_n]$ as
\begin{align}
F_{2,1}(a,\ag{1}{h},\ag{1}{k}) = \frac{a}{n} m(\mb{h,k}).
\end{align}
The last obstruction condition Eq.~\eqref{dFghkl} is $\delta_2F_{2,1}+\delta_GF_{3,0}=0$. It means that $\delta_2F_{2,1}$ is a coboundary in $\mathcal B^4[G,\RZ]$. So the cocycle-level differential $d_2$ for $F_{2,1}\in E_2^{2,1}$ is
\begin{align}
(d_2F_{2,1})(\mb{g,h,k,l})
:= (\delta_2F_{2,1})(\mb{g,h,k,l})
= -F_{2,1}(\nu(\mb{g,h}),\ag{1}{k},\ag{1}{l})
= -\frac{1}{n} \nu(\mb{g,h}) m(\mb{k,l}).
\end{align}

The $A=\Z_n$ example above can be generalized to $A=\U$ straightforwardly. The cocycle level differentials are summarized in Eq.~\eqref{dU1} of the main text.

\section{Extracting domain wall decorations by Lyndon's algorithm}
\label{sec:lyndon}

Most of this paper is focus on how to obtain the cocycles of $\tilde G$ by domain wall decorations using cocycles of $A$ and $G$. In this appendix, we want to answer the inverse problem: How to extract the domain wall decoration data given a cocycle of $\tilde G$? Equivalently, we want to decompose a cocycle of $\tilde G$ into different pieces located at different positions of the LHS spectral sequence. In fact, Lyndon already proposed an algorithm in Ref.~\onlinecite{Lyndon} to achieve the goal.

In this appendix, we will review Lyndon's algorithm and list the explicit formulas for degrees $\le 4$. We note that this appendix will use the symmetry action convention in Ref.~\onlinecite{HS}, which is different from Ref.~\onlinecite{Lyndon}. So we have a different order of variable $\ag{a}{g}$ ($A$ is in front of $G$) in, for example, Eq.~(\ref{normal}). The explicit formulas in the algorithm are also a little bit different from those in Ref.~\onlinecite{Lyndon}. But it is merely a gauge convention and would not change the physical results.

\subsection{Lyndon's algorithm and decomposition of cocycles}

We first introduce the notational abbreviation $s_{h,k}$ ($2\le h<k\le n+1$) for the argument set
\begin{align}
s_{h,k} = \left( \ag{(a_1)}{g_1},...,\ag{(a_{h-2})}{g_{h-2}},
\ag{1}{g_{h-1}},
a_h,...,a_{k-1},
\ag{1}{g_k},...,\ag{1}{g_n} \right),
\end{align}
which is a special case of the generic argument $\left(\ag{(a_1)}{g_1},\ag{(a_2)}{g_2},...,\ag{(a_n)}{g_n}\right) \in \tilde G^{n}$. We can define an order of all $s_{h,k}$. Two arguments are said to be $(h,k)<(h',k')$ if $k<k'$, or $h<h'$ and $k=k'$. So the set of $s_{h,k}$ are ordered as $(2,3)<(2,4)<(3,4)<(2,5)<...<(n,n+1)$. And in the legal range of $h$ and $k$ ($2\le h<k\le n+1$), the first and the last $s_{h,k}$ are
\begin{align}
s_{2,3} &= \left(
\ag{1}{g_{1}},
a_2,
\ag{1}{g_3},...,\ag{1}{g_n} \right),\\
s_{n,n+1} &= \left( \ag{(a_1)}{g_1},...,\ag{(a_{h-2})}{g_{n-2}},
\ag{1}{g_{n-1}},
a_n
\right).
\end{align}
We note that the order of $s_{h,k}$ defined here is different from the lexicographical order defined in Ref.~\onlinecite{Lyndon}. This difference comes from the different conventions on the symmetry actions.

Following Ref.~\onlinecite{Lyndon}, an $n$-cochain $\omega_n\in C[\tilde G,\U]$ is called \emph{normal} if it can be expressed as the product of its partial cochains:
\begin{align}\label{normal}
\omega_n\!\left[\ag{(a_1)}{g_1},...,\ag{(a_n)}{g_n}\right]
=\prod_{q=0}^n \omega_n(a_1,...,a_q,\ag{1}{g_{q+1}},...,\ag{1}{g_{n}}),
\end{align}
where the first $q$ and the last $n-q$ variables in the argument on the right-hand side come from $A$ and $G$, respectively. In this way, we decompose a normal $n$-cochain $\omega_n$ into different domain wall decoration terms in the LHS spectral sequence. By Lemma 5.4 of Ref.~\onlinecite{Lyndon}, if a cocycle $\omega_n$ satisfies
\begin{align}\label{ws=0}
\omega_n(s_{h,k})=1,\quad \forall s_{h,k},
\end{align}
it is a normal cocycle satisfying Eq.~(\ref{normal}).

Now the task of extracting domain wall decoration data is reduced to turning an $n$-cocycle to another $n$-cocycle (by adding coboundaries) satisfying Eq.~(\ref{ws=0}). For this purpose, we introduce the $(n-1)$-cochain $p_{h,k}$:
\begin{align}
p_{h,k} \left[\ag{(a_1)}{g_1},...,\ag{(a_n)}{g_n}\right]
:=
\omega_n
\left[\ag{(a_1)}{g_1},...,\ag{(a_{h})}{g_{h}},
\ag{1}{g_{h+1}...g_{k}},
a_{h+1}',...,a_{k}',
\ag{1}{g_{k+1}},...,\ag{1}{g_{n-1}}
\right]
\end{align}
where $a_i'$ ($h+1\le i\le k$) in the argument is defined as
\begin{align}
a_i' := \gia{g_i...g_{k}}{\left[a_i\nu(\mb{g_i},\mb{g_{i+1}...g_{k}})\right]}.
\end{align}
The differential of this $(n-1)$-cochain $p_{h,k}$ is very special (see Lemma 5.1 and 5.2 of Ref.~\onlinecite{Lyndon}):
\begin{align}
\begin{cases}
(dp_{h,k})(s_{h',k'}) = 1, & \mathrm{if}\ (h',k')>(h,k),\\
(dp_{h,k})(s_{h',k'}) = \left[\omega_n(s_{h',k'})\right]^{(-1)^{h+1}}, & \mathrm{if}\ (h',k')=(h,k).
\end{cases}
\end{align}
It means that $dp_{h,k}$ is trivial if the argument label $(h',k')$ is bigger than $(h,k)$, and is $(\omega_n)^{\pm 1}$ if the argument label is exactly $(h,k)$.

With the $(n-1)$-cochain $p_{h,k}$, we can now transform the $n$-cochain $\omega_n$ into a normal cochain. It is done using the following algorithm iteratively. Suppose $\omega_n$ already satisfies $\omega_n(s_{h',k'})=1$ for all $(h',k')>(h,k)$, then we can replace $\omega_n$ by $\omega_n/ (dp_{h,k})^{(-1)^{h+1}}$. In this way, the new $\omega_n$ would satisfies $\omega_n(s_{h',k'})=1$ for all $(h',k')\ge (h,k)$. After using this algorithm repeatedly, we obtain an $n$-cocycle $\tilde\omega_n$ satisfying $\tilde\omega_n(s_{h,k})=1$ for all $s_{h,k}$. It is automatically a normal cocycle with the property Eq.~(\ref{normal}). So we have
\begin{align}
\omega_n \times (\mathrm{coboundaries}) = \tilde \omega_n = \prod_{p+q=n} \tilde\omega_n^{p,q} = \tilde\omega_n^{0,n} \times \tilde\omega_n^{1,n-1} \times ... \times \tilde\omega_n^{n,0}.
\end{align}
We note that this is a cochain level decomposition, meaning that $\tilde\omega_n^{p,q}\in E_0^{p,q}$ may be a nontrivial torsor even when $E_2^{p,q}=0$.
The domain wall decoration data $\tilde\omega_n^{p,q}\in E_0^{p,q}$ is obtained as
\begin{align}
\tilde\omega_n^{p,q}(a_1,...,a_q,{\mb{g}_{q+1}},...,{\mb{g}_{n}}) := \tilde\omega_n(a_1,...,a_q,1_{\mb{g}_{q+1}},...,1_{\mb{g}_{n}}).
\end{align}
Since the coboundaries $dp$ added to $\omega_n$ are expressed as $\omega_n$ itself, the decoration data $\tilde\omega_n^{p,q}$ should also be expressed as a product of $\omega_n$'s with different arguments in the end.

In the remaining part of this appendix, we will list the explicit domain wall decoration data in degrees $\le 4$ obtained using the Lyndon's algorithm.

\subsection{Domain wall decorations for degree-1}

For arbitrary 1-cocycle $\omega_1\in\H^1[\tilde G,\U]$, the domain wall decoration data $\tilde\omega_1^{p,q} \in E_0^{p,q}=\mathcal{C}^p[G,\mathcal{C}^q[A,\U]]$ are simply
\begin{align}
\tilde\omega_1^{0,1}(a) &= \omega_1(a),\\
\tilde\omega_1^{1,0}(\mb{g}) &= \omega_1(\ag{1}{g}).
\end{align}

\subsection{Domain wall decorations for degree-2}

For 2-cocycle $\omega_2\in\H^2[\tilde G,\U]$, we can extract $\tilde\omega_2^{p,q} \in E_0^{p,q}=\mathcal{C}^p[G,\mathcal{C}^q[A,\U]]$ as
\begin{align}
\tilde\omega_2^{0,2}(a,b) &= \omega_2(a,b),\\
\tilde\omega_2^{1,1}(a,\mb{h}) &= \left[\iota_{(\gia{h}{a})} \omega_2\right](\ag{1}{h}) = \frac{\omega_2(a,\ag{1}{h})}{\omega_2(\ag{1}{h},\gia{h}{a})},\\
\tilde\omega_2^{2,0}(\mb{g,h}) &= \frac{\omega_2(\ag{1}{g},\ag{1}{h})}{\omega_2(\ag{1}{gh},\gia{gh}{\nu(\mb{g,h})})}.
\end{align}

\subsection{Domain wall decorations for degree-3}

For $\omega_3\in\H^3[\tilde G,\U]$, we can extract $\tilde\omega_3^{p,q} \in E_0^{p,q}=\mathcal{C}^p[G,\mathcal{C}^q[A,\U]]$ as
\begin{align}
\tilde\omega_3^{0,3}(a,b,c) &= \omega_3(a,b,c),\\
\tilde\omega_3^{1,2}(a,b,\mb{k}) &= \left[\iota_{(\ag{1}{k})} \omega_3\right](a,b)
=\frac{\omega_3(a,b,\ag{1}{k}) \ \omega_3(\ag{1}{k},\gia{k}{a},\gia{k}{b})}{\omega_3(a,\ag{1}{k},\gia{k}{b})},\\\nonumber
\tilde\omega_3^{2,1}(a,\mb{h,k}) &= \frac{\left[\iota_{(\gia{hk}{a})} \omega_3\right](\ag{1}{h},\ag{1}{k})}{\left[\iota_{(\gia{hk}{a})} \omega_3\right](\ag{1}{hk},\gia{hk}{\nu(\mb{h,k})})}\\
&=\frac{\omega_3(\ag{1}{h},\ag{1}{k},\gia{hk}{a}) \ \omega_3(a,\ag{1}{h},\ag{1}{k})}{\omega_3(\ag{1}{h},\gia{h}{a},\ag{1}{k})}
\times
\frac{\omega_3(\ag{1}{hk},\gia{hk}{a},\gia{hk}{\nu(\mb{h,k})})  }{\omega_3(\ag{1}{hk},\gia{hk}{\nu(\mb{h,k})},\gia{hk}{a}) \ \omega_3(a,\ag{1}{hk},\gia{hk}{\nu(\mb{h,k})})},\\\nonumber
\tilde\omega_3^{3,0}(\mb{g,h,k}) &= \omega_3(\ag{1}{g},\ag{1}{h},\ag{1}{k})
\times
\frac{\omega_3(\ag{1}{gh},\ag{1}{k},\gia{ghk}{\nu(\mb{g,h})})   }{\omega_3(\ag{1}{gh},\gia{gh}{\nu(\mb{g,h})},\ag{1}{k}) \ \omega_3(\ag{1}{g},\ag{1}{hk},\gia{hk}{\nu(\mb{h,k})})}\\
&\quad\times
\frac{\omega_3(\ag{1}{ghk},\gia{ghk}{\nu(\mb{g,hk})},\gia{hk}{\nu(\mb{h,k})})}{\omega_3(\ag{1}{ghk},\gia{ghk}{\nu(\mb{gh,k})},\gia{ghk}{\nu(\mb{g,h})})}.
\end{align}

\subsection{Domain wall decorations for degree-4}

For $\omega_4\in\H^4[\tilde G,\U]$, the domain wall decoration data $\tilde\omega_4^{p,q} \in E_0^{p,q}=\mathcal{C}^p[G,\mathcal{C}^q[A,\U]]$ are much more complicated:
\begin{align}\label{Lyn_w04}
\tilde\omega_4^{0,4}(a,b,c,d) &= \omega_4(a,b,c,d),
\end{align}
\begin{align}\label{Lyn_w13}
\tilde\omega_4^{1,3}(a,b,c,\mb{l}) &= \left[\iota_{(\ag{1}{l})}\right] (a,b,c)
=\frac{\omega_4(a,b,c,\ag{1}{l}) \ 
\omega_4(a,\ag{1}{l},\gia{l}{b},\gia{l}{c})  }
{\omega_4(a,b,\ag{1}{l},\gia{l}{c}) \ 
\omega_4(\ag{1}{l},\gia{l}{a},\gia{l}{b},\gia{l}{c})},
\end{align}
\begin{align}\label{Lyn_w22}
\tilde\omega_4^{2,2}(a,b,\mb{k,l}) &= \frac{\left[\iota_{(\gia{kl}{a}),(\gia{kl}{b})} \omega_4 \right] (\ag{1}{k},\ag{1}{l})}{\left[\iota_{(\gia{kl}{a}),(\gia{kl}{b})} \omega_4 \right] (\ag{1}{kl},\gia{kl}{\nu(\mb{k,l})})}\\\nonumber
&=\frac{\omega_4(\ag{1}{k},\ag{1}{l},\gia{kl}{a},\gia{kl}{b}) \ 
\omega_4(a,\ag{1}{k},\ag{1}{l},\gia{kl}{b}) \ 
\omega_4(\ag{1}{k},\gia{k}{a},\gia{k}{b},\ag{1}{l}) \ 
\omega_4(a,b,\ag{1}{k},\ag{1}{l}) }{\omega_4(\ag{1}{k},\gia{k}{a},\ag{1}{l},\gia{kl}{b}) \ \omega_4(a,\ag{1}{k},\gia{k}{b},\ag{1}{l})}
\\\nonumber
&\quad \times
\frac{\omega_4(\ag{1}{kl},\gia{kl}{a},\gia{kl}{\nu(\mb{k,l})},\gia{kl}{b}) \ 
\omega_4(a,\ag{1}{kl},\gia{kl}{b},\gia{kl}{\nu(\mb{k,l})}) }
{\omega_4(\ag{1}{kl},\gia{kl}{\nu(\mb{k,l})},\gia{kl}{a},\gia{kl}{b}) \ 
\omega_4(a,\ag{1}{kl},\gia{kl}{\nu(\mb{k,l})},\gia{kl}{b}) \ 
\omega_4(\ag{1}{kl},\gia{kl}{a},\gia{kl}{b},\gia{kl}{\nu(\mb{k,l})}) \ 
\omega_4(a,b,\ag{1}{kl},\gia{kl}{\nu(\mb{k,l})})},
\end{align}
\begin{align}\nonumber
\tilde\omega_4^{3,1}(a,\mb{h,k,l}) &=
\frac{
\left[\iota_{(\gia{hkl}{a})}\omega_4\right](\ag{1}{h},\ag{1}{kl},\gia{kl}{\nu(\mb{k,l})})
\times
\left[\iota_{(\gia{hkl}{a})}\omega_4\right](\ag{1}{hk},\gia{hk}{\nu(\mb{h,k})},\ag{1}{l})
\times
\left[\iota_{(\gia{hkl}{a})}\omega_4\right](\ag{1}{hkl},\gia{hkl}{\nu(\mb{hk,l})},\gia{hkl}{\nu(\mb{h,k})})
}
{\left[\iota_{(\gia{hkl}{a})}\omega_4\right] (\ag{1}{h},\ag{1}{k},\ag{1}{l})
\times
\left[\iota_{(\gia{hkl}{a})}\omega_4\right](\ag{1}{hk},\ag{1}{l},\gia{ghk}{\nu(\mb{h,k})})
\times
\left[\iota_{(\gia{hkl}{a})}\omega_4\right](\ag{1}{hkl},\gia{hkl}{\nu(\mb{h,kl})},\gia{kl}{\nu(\mb{k,l})})
}
\\\nonumber
&= \frac{\omega_4(a,\ag{1}{h},\ag{1}{k},\ag{1}{l}) \ 
\omega_4(\ag{1}{h},\ag{1}{k},\gia{hk}{a},\ag{1}{l})}
{\omega_4(\ag{1}{h},\gia{h}{a},\ag{1}{k},\ag{1}{l}) \ 
\omega_4(\ag{1}{h},\ag{1}{k},\ag{1}{l},\gia{hkl}{a})}
\times
\frac{\omega_4(\ag{1}{h},\gia{h}{a},\ag{1}{kl},\gia{kl}{\nu(\mb{k,l})}) \ \omega_4(\ag{1}{h},\ag{1}{kl},\gia{kl}{\nu(\mb{k,l})},\gia{hkl}{a})}
{\omega_4(a,\ag{1}{h},\ag{1}{kl},\gia{kl}{\nu(\mb{k,l})}) \ \omega_4(\ag{1}{h},\ag{1}{kl},\gia{hkl}{a},\gia{kl}{\nu(\mb{k,l})})}
\\\nonumber
&\quad\times
\frac{\omega_4(a,\ag{1}{hk},\ag{1}{l},\gia{ghk}{\nu(\mb{h,k})}) \ \omega_4(\ag{1}{hk},\ag{1}{l},\gia{hkl}{a},\gia{ghk}{\nu(\mb{h,k})})}
{\omega_4(\ag{1}{hk},\gia{hk}{a},\ag{1}{l},\gia{ghk}{\nu(\mb{h,k})}) \ \omega_4(\ag{1}{hk},\ag{1}{l},\gia{ghk}{\nu(\mb{h,k})},\gia{hkl}{a})}
\\\nonumber
&\quad\times
\frac{\omega_4(\ag{1}{hk},\gia{hk}{a},\gia{hk}{\nu(\mb{h,k})},\ag{1}{l}) \ \omega_4(\ag{1}{hk},\gia{hk}{\nu(\mb{h,k})},\ag{1}{l},\gia{hkl}{a})}
{\omega_4(a,\ag{1}{hk},\gia{hk}{\nu(\mb{h,k})},\ag{1}{l}) \ \omega_4(\ag{1}{hk},\gia{hk}{\nu(\mb{h,k})},\gia{hk}{a},\ag{1}{l})}
\\\nonumber
&\quad\times
\frac{\omega_4(a,\ag{1}{hkl},\gia{hkl}{\nu(\mb{h,kl})},\gia{kl}{\nu(\mb{k,l})}) \ \omega_4(\ag{1}{hkl},\gia{hkl}{\nu(\mb{h,kl})},\gia{hkl}{a},\gia{kl}{\nu(\mb{k,l})})}
{\omega_4(\ag{1}{hkl},\gia{hkl}{a},\gia{hkl}{\nu(\mb{h,kl})},\gia{kl}{\nu(\mb{k,l})}) \ \omega_4(\ag{1}{hkl},\gia{hkl}{\nu(\mb{h,kl})},\gia{kl}{\nu(\mb{k,l})},\gia{hkl}{a})}
\\
&\quad\times
\frac{\omega_4(\ag{1}{hkl},\gia{hkl}{a},\gia{hkl}{\nu(\mb{hk,l})},\gia{hkl}{\nu(\mb{h,k})}) \ \omega_4(\ag{1}{hkl},\gia{hkl}{\nu(\mb{hk,l})},\gia{hkl}{\nu(\mb{h,k})},\gia{hkl}{a})}
{\omega_4(a,\ag{1}{hkl},\gia{hkl}{\nu(\mb{hk,l})},\gia{hkl}{\nu(\mb{h,k})}) \ \omega_4(\ag{1}{hkl},\gia{hkl}{\nu(\mb{hk,l})},\gia{hkl}{a},\gia{hkl}{\nu(\mb{h,k})})},
\end{align}
\begin{align}\nonumber
\tilde\omega_4^{4,0}(\mb{g,h,k,l})
&=
\omega_4(\ag{1}{g},\ag{1}{h},\ag{1}{k},\ag{1}{l})
\times
\frac{1}{\omega_4(\ag{1}{g},\ag{1}{h},\ag{1}{kl},\gia{kl}{\nu(\mb{k,l})})}
\times
\frac{\omega_4(\ag{1}{g},\ag{1}{hk},\ag{1}{l},\gia{hkl}{\nu(\mb{h,k})})}{\omega_4(\ag{1}{g},\ag{1}{hk},\gia{hk}{\nu(\mb{h,k})},\ag{1}{l})}
\\\nonumber
&\quad\times
\frac{\omega_4(\ag{1}{gh},\ag{1}{k},\gia{ghk}{\nu(\mb{g,h})},\ag{1}{l})}{\omega_4(\ag{1}{gh},\ag{1}{k},\ag{1}{l},\gia{ghkl}{\nu(\mb{g,h})}) \ \omega_4(\ag{1}{gh},\gia{gh}{\nu(\mb{g,h})},\ag{1}{k},\ag{1}{l})}
\times
\frac{\omega_4(\ag{1}{g},\ag{1}{hkl},\gia{hkl}{\nu(\mb{h,kl})},\gia{kl}{\nu(\mb{k,l})})}{\omega_4(\ag{1}{g},\ag{1}{hkl},\gia{hkl}{\nu(\mb{hk,l})},\gia{hkl}{\nu(\mb{h,k})})}
\\\nonumber
&\quad\times
\frac{\omega_4(\ag{1}{gh},\gia{gh}{\nu(\mb{g,h})},\ag{1}{kl},\gia{kl}{\nu(\mb{k,l})}) \ \omega_4(\ag{1}{gh},\ag{1}{kl},\gia{kl}{\nu(\mb{k,l})},\gia{ghkl}{\nu(\mb{g,h})})}{\omega_4(\ag{1}{gh},\ag{1}{kl},\gia{ghkl}{\nu(\mb{g,h})},\gia{kl}{\nu(\mb{k,l})})}
\\\nonumber
&\quad\times
\frac{\omega_4(\ag{1}{ghk},\ag{1}{l},\gia{ghkl}{\nu(\mb{g,hk})},\gia{hkl}{\nu(\mb{h,k})}) \ \omega_4(\ag{1}{ghk},\gia{ghk}{\nu(\mb{g,hk})},\gia{hk}{\nu(\mb{h,k})},\ag{1}{l}) }
{\omega_4(\ag{1}{ghk},\gia{ghk}{\nu(\mb{g,hk})},\ag{1}{l},\gia{hkl}{\nu(\mb{h,k})}) }
\\\nonumber
&\quad\times
\frac{\omega_4(\ag{1}{ghk},\gia{ghk}{\nu(\mb{gh,k})},\ag{1}{l},\gia{ghkl}{\nu(\mb{g,h})})}
{\omega_4(\ag{1}{ghk},\ag{1}{l},\gia{ghkl}{\nu(\mb{gh,k})},\gia{ghkl}{\nu(\mb{g,h})}) \ \omega_4(\ag{1}{ghk},\gia{ghk}{\nu(\mb{gh,k})},\gia{ghk}{\nu(\mb{g,h})},\ag{1}{l})}
\\\nonumber
&\quad\times
\frac{\omega_4(\ag{1}{ghkl},\gia{ghkl}{\nu(\mb{g,hkl})},\gia{hkl}{\nu(\mb{hk,l})},\gia{hkl}{\nu(\mb{h,k})})}
{\omega_4(\ag{1}{ghkl},\gia{ghkl}{\nu(\mb{g,hkl})},\gia{hkl}{\nu(\mb{h,kl})},\gia{kl}{\nu(\mb{k,l})})}
\times
\frac{\omega_4(\ag{1}{ghkl},\gia{ghkl}{\nu(\mb{gh,kl})},\gia{ghkl}{\nu(\mb{g,h})},\gia{kl}{\nu(\mb{k,l})})}
{\omega_4(\ag{1}{ghkl},\gia{ghkl}{\nu(\mb{gh,kl})},\gia{kl}{\nu(\mb{k,l})},\gia{ghkl}{\nu(\mb{g,h})})}
\\
&\quad\times
\frac{\omega_4(\ag{1}{ghkl},\gia{ghkl}{\nu(\mb{ghk,l})},\gia{ghkl}{\nu(\mb{gh,k})},\gia{ghkl}{\nu(\mb{g,h})})}
{\omega_4(\ag{1}{ghkl},\gia{ghkl}{\nu(\mb{ghk,l})},\gia{ghkl}{\nu(\mb{g,hk})},\gia{hkl}{\nu(\mb{h,k})})}.
\end{align}

\twocolumngrid
\end{document}